\documentclass[11pt]{article}
\usepackage{lmodern}
\usepackage{dsfont}
\usepackage{environ}
\input{definitions.sty}

\begin{document}

\title{Measuring CEX-DEX Extracted Value and Searcher Profitability: \\ The Darkest of the MEV Dark Forest}

\author{
  Fei Wu$^*$,
  Danning Sui$^\dagger$,
  Thomas Thiery$^\ddagger$, and
  Mallesh Pai$^\S$
  \medskip
  \\
  \small
  $^*$King's College London \\
  \small
  $^\dagger$Flashbots \\
  \small
  $^\ddagger$Ethereum Foundation \\
  \small
  $^\S$Rice University, Consensys, and Paradigm \\
  \texttt{}
}

\date{Initial Version: May 29, 2025 \\ Current version: August 3, 2025
}
\maketitle

\thispagestyle{empty}

\begin{abstract}
This paper provides a comprehensive empirical analysis of the economics and dynamics behind arbitrages between centralized and decentralized exchanges (CEX-DEX) on Ethereum. We refine heuristics to identify arbitrage transactions from on-chain data and introduce a robust empirical framework to estimate arbitrage revenue without knowing traders' actual behaviors on CEX. Leveraging an extensive dataset spanning 19 months from August 2023 to March 2025, we estimate a total of 233.8M USD extracted by 19 major CEX-DEX searchers from 7,203,560 identified CEX-DEX arbitrages. Our analysis reveals increasing centralization trends as three searchers captured three-quarters of both volume and extracted value. We also demonstrate that searchers' profitability is tied to their integration level with block builders and uncover exclusive searcher-builder relationships and their market impact. Finally, we correct the previously underestimated profitability of block builders who vertically integrate with a searcher. These insights illuminate the darkest corner of the MEV landscape and highlight the critical implications of CEX-DEX arbitrages for Ethereum's decentralization. 

\end{abstract}

\hspace{0.5cm}

\section{Introduction}
Over recent years, the focus of study around Maximal Extractable Value (MEV) extraction has broadened beyond on-chain atomic strategies to a richer landscape of more sophisticated strategies across different venues. Early studies primarily investigated MEV extracted solely from the blockchain’s internal state and mempool, termed atomic MEV \cite{daian2019flash,qin2021liqui,qin2022quantifying}. Subsequent research has turned attention towards non-atomic MEV, strategies that leverage external price information alongside on-chain execution, most notably arbitrage between centralized exchanges (CEX) and decentralized exchanges (DEX) \cite{obadia2021unity,nonatomic,burakcrosschainarb}. These CEX-DEX arbitrages have emerged as an economically substantial type of MEV.

CEX-DEX arbitrageurs (searchers) capitalize on temporary price discrepancies arising from asynchronous price discovery across venues. While centralized exchanges provide high liquidity and near-instantaneous trade execution, decentralized exchanges experience liquidity constraints and inherent latency due to blockchain settlement delays. These structural frictions regularly produce significant, short-lived deviations in asset prices on DEXes from their fair value, offering lucrative arbitrage opportunities for sophisticated market participants.

Despite high entry barriers such as capital requirements, low-latency infrastructure, inventory risk, and uncertainty of block inclusion, CEX-DEX arbitrage remains remarkably profitable \cite{frontier2}. Such trades often take up less than 2\% of block space but contribute more than 15\% of the total block value \cite{nonatomic,whowinsandwhy}. Consequently, under Ethereum’s Proposer-Builder Separation (PBS) framework, block builders with privileged access to such arbitrage flows benefit from a decisive advantage in winning block auctions. At the time of writing, three builders \texttt{beaverbuild}, \texttt{Titan}, and \texttt{rsync} dominate the Ethereum builder market, two of which vertically integrate their own CEX-DEX searchers \cite{whowinsandwhy,nonatomic,yang2024decentralization}. Such vertical integration raises important concerns for Ethereum's decentralization and security: it fosters economies of scale that strengthen dominant players, enables monopoly pricing that causes proposer loss \cite{yang2024decentralization}, and increases vulnerability to censorship and commitment attacks \cite{toni2024censorship,heimbach2023ethereum,fox2023censorshipresistanceonchainauctions}.

Although theoretical models \cite{lvr,lvr2,nezlobin2025} and prior empirical studies \cite{nonatomic,whowinsandwhy,yang2024decentralization,frontier2} have illuminated the existence and broader impact of CEX-DEX arbitrage on the Ethereum builder market, the economic details remain unclear. Accurately identifying such transactions on-chain and estimating realized revenue without knowing searchers' off-chain behaviors is particularly challenging, rendering CEX-DEX arbitrage arguably ``the darkest part of the MEV dark forest." In addition, key questions persist about how arbitrage revenue is distributed between searchers and builders, how searchers' profitability varies, and how exclusive deals between searchers and builders influence both the searcher and builder market. 

This paper provides a rigorous empirical investigation into value extraction, searcher profitability, and market structure effects of CEX-DEX arbitrage on Ethereum. Integrating DEX trades data, CEX pricing, and MEV-Boost data over 19 months from August 2023 to March 2025 and quantifying \emph{who} earns \emph{what}, our results supply the missing inputs needed to reason about decentralization guarantees. Our contributions are summarized as follows:
\begin{enumerate}[topsep=1pt, itemsep=1pt, parsep=0pt]
    \item We refine existing heuristics for identifying CEX-DEX arbitrage transactions \cite{nonatomic,whowinsandwhy}, significantly expanding coverage and accuracy.
    \item Without directly observing the CEX leg of searchers' CEX-DEX arbitrages, we develop an empirical framework to infer the likely CEX execution by tracking when each searcher's information advantage maximizes and begins to erode, thereby inferring from each on-chain swap a realized arbitrage revenue. This data-driven proxy matches theoretical price-impact intuition and enables consistent comparisons between CEX-DEX searchers.
    \item We demonstrate that token liquidity influences how searchers extract value and hedge: high-liquidity tokens enable searchers to place large orders at tight spreads with minimal market impact, whereas low-liquidity tokens present wider spreads but restrict them to smaller trades that incur larger price impact.
    \item We estimate that 19 major searchers extracted a total value of 233.8M USD from 7,203,560 CEX-DEX arbitrages during the observed period and uncover an increasing centralization trend, with the three leading searchers capturing approximately three-quarters of the total arbitrage volume and extracted value.
    \item Our results reveal a clear profit-sharing pattern tied to searchers' integration levels with block builders. Neutral searchers, distributing flow among multiple builders, retain higher profit margins. Conversely, exclusive searchers maintain lower margins, sharing most of their extracted value with affiliated builders, sometimes even operating at negative net profit. Additionally, we find that the searcher-builder exclusivity deal is likely to mutually reinforce competitive positions for both the searcher and the builder in their respective markets.
    \item By accounting for searcher profits, we correct previously understated estimates of integrated builder profits and subsidies and better illuminate their profitability.

\end{enumerate}

\section{Background and Related Works}

\subparagraph*{Ethereum, Proposer-Builder Separation, and MEV-Boost.} Ethereum advances in discrete 12-second slots. At each slot, a randomly selected validator—the beacon proposer—publishes a new block \cite{pos}. To enhance validator decentralization and censorship resistance, Proposer-Builder Separation (PBS) allows proposers to outsource block construction tasks to specialized entities called \emph{builders}. PBS is currently implemented via MEV-Boost \cite{mevboost}, an out-of-protocol solution where builders compete in MEV-Boost auctions by submitting blocks alongside bids to trusted intermediaries known as \emph{relays}. Builders submit bids for slot $n$ starting from the start of slot $n-1$, funded by tips and payments alongside transactions from users and MEV searchers. At the start of slot $n$, the proposer signs the highest-bid block header from any relay to which the proposer is connected, after which the winning relay publishes the block.

\subparagraph*{Maximal Extractable Value and its supply chain.} 
Maximal Extractable Value (MEV) refers to profits obtainable through strategic transaction ordering, inclusion, or censorship within a block. MEV extraction can rely solely on internal blockchain state (atomic MEV) \cite{daian2019flash,qin2022quantifying,zhou2021high,qin2021liqui}, or incorporate external market information (non-atomic MEV), such as prices from centralized exchanges and decentralized exchanges on other blockchains \cite{nonatomic,burakcrosschainarb,obadia2021unity}.

Under PBS, MEV extraction is typically performed by specialized searchers. The competitive nature of MEV extraction gives rise to a structured MEV supply chain: searchers share with builders part of their extracted value to ensure prioritized inclusion of their transactions. Builders subsequently share a portion of these MEV revenues with proposers through competitive bids in MEV-Boost auctions. Such intense competition incentivizes \emph{vertical integration}, where entities optimize MEV extraction by simultaneously controlling upstream (searcher-level strategies) and downstream (builder and relay infrastructure) activities, thus securing a strategic advantage in capturing valuable MEV opportunities.

\begin{figure}[t]
    \centering
    \includegraphics[width=0.7\linewidth]{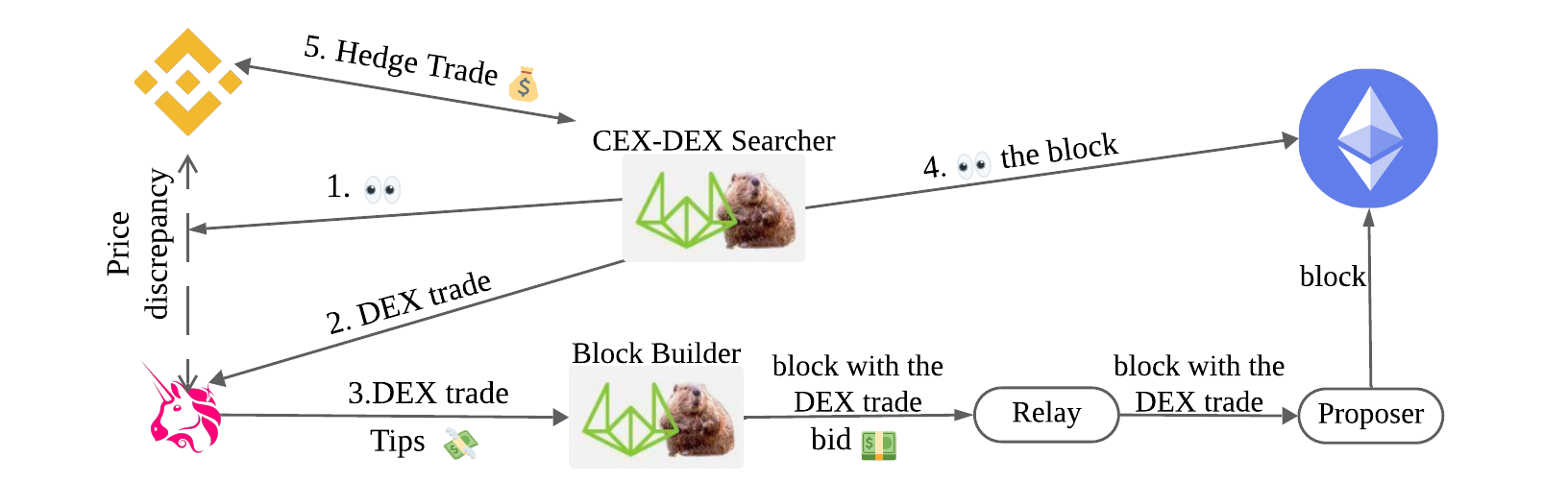}
    \caption{An example of a CEX-DEX arbitrage operation.}
    \label{fig:cex-dex_arb}
\end{figure}

\subparagraph*{CEX-DEX Arbitrages.} 
CEX-DEX arbitrages represent a prominent category of non-atomic MEV. Typically, these arbitrages involve two complementary trades: one executed on-chain via a decentralized exchange, and the other executed off-chain via a centralized exchange, in this order. \Cref{fig:cex-dex_arb} illustrates an example of a CEX-DEX operation. First, searchers identify price discrepancies between assets listed on a DEX and a CEX. Based on the observed price differences, searchers form an \emph{ex-ante expectation} of the extractable value (i.e., arbitrage revenue) once hedged. They then initiate the arbitrage by submitting their DEX trade to builders, offering a fraction of their expected arbitrage revenue as payments to the builder to secure prioritized block inclusion. Once the DEX trade is confirmed on-chain, searchers swiftly execute an offsetting hedge trade on the CEX to realize the profit.

\subparagraph*{Related Works.} \cite{lvr,lvr2,nezlobin2025} introduced the concept of Loss-Versus-Rebalancing (LVR), modeling losses of DEX liquidity providers due to informed arbitrageurs, establishing a theoretical benchmark for CEX-DEX arbitrage revenue. \cite{frontier1} characterized the extractable value by informed searchers and showed that MEV extraction grows sophisticated with more informed trading, notably CEX-DEX arbitrages. The authors further elaborated on the risks and entry barriers of CEX-DEX arbitrage operation and observed that CEX-DEX arbitrage primarily involves trading high-liquidity tokens \cite{frontier2}. \cite{colinpost} provided initial insights into CEX-DEX arbitrages and searcher-builder relationships, and estimated searcher profits from trades in the Uniswap V2 WETH-USDC pool. \cite{greenfieldpost} further investigated CEX-DEX arbitrage activities in all Uniswap V2 and V3 pools, highlighting searcher-builder integration and economy of scale in the Ethereum builder market. \cite{nonatomic} conducted an extensive empirical study of CEX-DEX arbitrages identified by comprehensive heuristics they introduced, highlighting a significant correlation between arbitrage volumes and token price volatility, and demonstrated that builders integrated with CEX-DEX searchers have greater odds in MEV-Boost auctions at times of high volatility. Additionally, this work provided empirical evidence for vertical integration between specific searchers and builders, such as \texttt{beaverbuild} and \texttt{rsync}. Similarly, \cite{gupta2023centralizing} revealed how builders operated by high-frequency trading firms leverage access to CEX-DEX arbitrage flows to gain competitive advantages in block building, a finding further empirically validated by \cite{cefidefiofa}. \cite{whowinsandwhy} empirically demonstrated that builders with exclusive orderflow providers, such as CEX-DEX searchers, enjoy increased market share and profitability. Similar to \cite{nonatomic}, the work found that CEX-DEX arbitrage flow contributes to a prominent share of the block value. \cite{wu2024strategic, wu2024competition} conducted simulations and empirical game-theoretic analysis to show builders' strategic advantages in block auctions with exclusive access to high-value orderflow. \cite{yang2024decentralization} further analyzed how vertical integration adversely impacts proposer revenue, block optimality, and censorship resistance. Lastly, longitudinal studies by \cite{wahrstatter2023time, heimbach2023ethereum} provided empirical overviews of Ethereum’s evolving builder market, highlighting centralization and censorship pressures under PBS.

\section{Identifying CEX-DEX Arbitrage Transactions}
CEX-DEX arbitrage transactions are typically more challenging to identify than other types of MEV transactions \cite{zhou2021high, qin2022quantifying, qin2021liqui}, primarily because only the DEX leg of such arbitrages is explicitly observable on-chain. Nonetheless, by exploiting certain properties of the transaction and behavioral regularities of CEX-DEX MEV searchers, we can infer whether a given DEX trade can be part of a CEX-DEX arbitrage. Specifically, we assume that CEX-DEX arbitrage transactions are private and prioritized trades in the block executed by specialized MEV searchers, who generally do not interact with regular users.

Our methodology builds upon and extends previous approaches \cite{nonatomic, whowinsandwhy, colinpost}, refining existing heuristics and introducing additional filtering mechanisms to detect CEX-DEX arbitrage transactions from the broader set of DEX trades. We implement these detection methods using Dune Analytics \cite{dune}, leveraging established datasets including \cite{dunedextrades, hildobbyatomic, mempooldata, orderflowart}.

A significant advancement in our approach is the expansion beyond the constraints of previous research, which typically limited CEX-DEX trades to transactions that contained exactly one on-chain swap (or two ERC-20 token transfers) \cite{nonatomic, whowinsandwhy}. Our investigation reveals that certain CEX-DEX searchers often execute multiple swaps across different DEX pools within a single transaction to optimize on-chain liquidity utilization.%
\footnote{Consider for example searcher contract address: \texttt{0x767c8bb1574bee5d4fe35e27e0003c89d43c5121}, with sample transaction: \href{https://etherscan.io/tx/0x0488cad7726ed6947be6af09d17b012fd7c986cb32a9d59cd7285b9a4e0926be}{\texttt{0x0488cad7726ed6947be6af09d17b012fd7c986cb32a9d59cd7285b9a4e0926be}.}}
These transactions do not fit into the traditional heuristic of one swap and two ERC-20 token transfers. Our methodology accommodates these multi-swap transactions by aggregating sequential swaps and reconstructing the final effective trading token pair with corresponding aggregate traded amounts. Specifically, we apply the following set of formulated heuristics:


\begin{description} [topsep=1pt, itemsep=1pt, parsep=0pt]
\item[\textbf{Heuristic 1.}] The transaction is \emph{private}, i.e., it is not observed in any public mempool \cite{mempooldata} prior to its on-chain inclusion. 
\item[\textbf{Heuristic 2.}] At least one of the swaps included in the transaction is the first swap executed in the respective direction and DEX pool for the given block.
\item[\textbf{Heuristic 3.}] The transaction is not categorized by existing frameworks \cite{zeromev, hildobbyatomic} or our own Dune query as any type of atomic MEV activity (e.g., sandwich attacks or atomic arbitrages). The transaction does not include a liquidation event.
\item[\textbf{Heuristic 4.}] The transaction is not a backrun transaction associated with an Orderflow Auction (OFA) bundle.
\item[\textbf{Heuristic 5.}] The transaction is not submitted to any known router smart contract identified in \cite{orderflowart}, nor to any trading bot that is labeled in \cite{dextradingbot} or controls an Externally Owned Account (EOA) with an Ethereum Name Service (ENS) name attached. 
\item[\textbf{Heuristic 6.}] The transaction contains no ERC-721 token transfer and settles as a swap between two tokens that are both listed on a major centralized exchange (e.g., Binance) after all intermediate swaps.
\end{description}

Heuristic 1 and 2 ensure transaction prioritization within blocks and favorable execution pricing in at least one DEX pool. Heuristics 3, 4, and 5 eliminate potential confusion with other MEV activities or retail trades executed by non-MEV bots. In particular, Heuristic 5 recognizes that bot contracts controlling EOAs with attached ENS names typically interact with regular users, suggesting they likely function as telegram bots or trading front-ends rather than specialized CEX-DEX MEV bots. For Heuristic 6, we cross-verified ERC-20 contract addresses for 287 Binance-listed tokens collected from Etherscan \cite{etherscan} and CoinMarketCap \cite{coinmarketcap} with the Dune DEX trades dataset \cite{dunedextrades}, ensuring that we filtered out meme tokens that share the same symbols with major tokens but are not listed on a CEX.

Applying the above criteria, we curate an empirical dataset comprising 8,723,233 transactions 
that are likely to be CEX-DEX arbitrages across multiple DEXes \cite{dunedextrades} from block 17866488 to block 21998438 spanning the period from August 8, 2023, to March 8, 2025 with their detailed information, including searcher contract address, fees, payments to the builder, trade volume, and the precise tokens and amounts traded. To the best of our knowledge, this constitutes the most comprehensive empirical dataset to date for analyzing CEX-DEX arbitrages in terms of observation duration, transaction count, and token coverage. We open source the Dune query for reproducibility \cite{cexdexdune}.

\subsection{Current CEX-DEX Arbitrage Landscape}
Utilizing the curated dataset, we analyze temporal trends in CEX-DEX arbitrage activity over the observed period. To enhance readability and streamline the visualization, we assign intuitive labels to the 23 major searcher entities identified in the dataset with the highest total volume in place of their contract addresses, except for \texttt{Wintermute} and \texttt{SCP}, whose identities are retained as-is. Our subsequent analysis will focus on these 23 searcher entities, who collectively contribute to 99\% of total volume. A full mapping of contract addresses to labels is available in \Cref{appendix:searcher_label}.
\begin{figure}[t]      
  \centering\
  \includegraphics[width=1\linewidth]{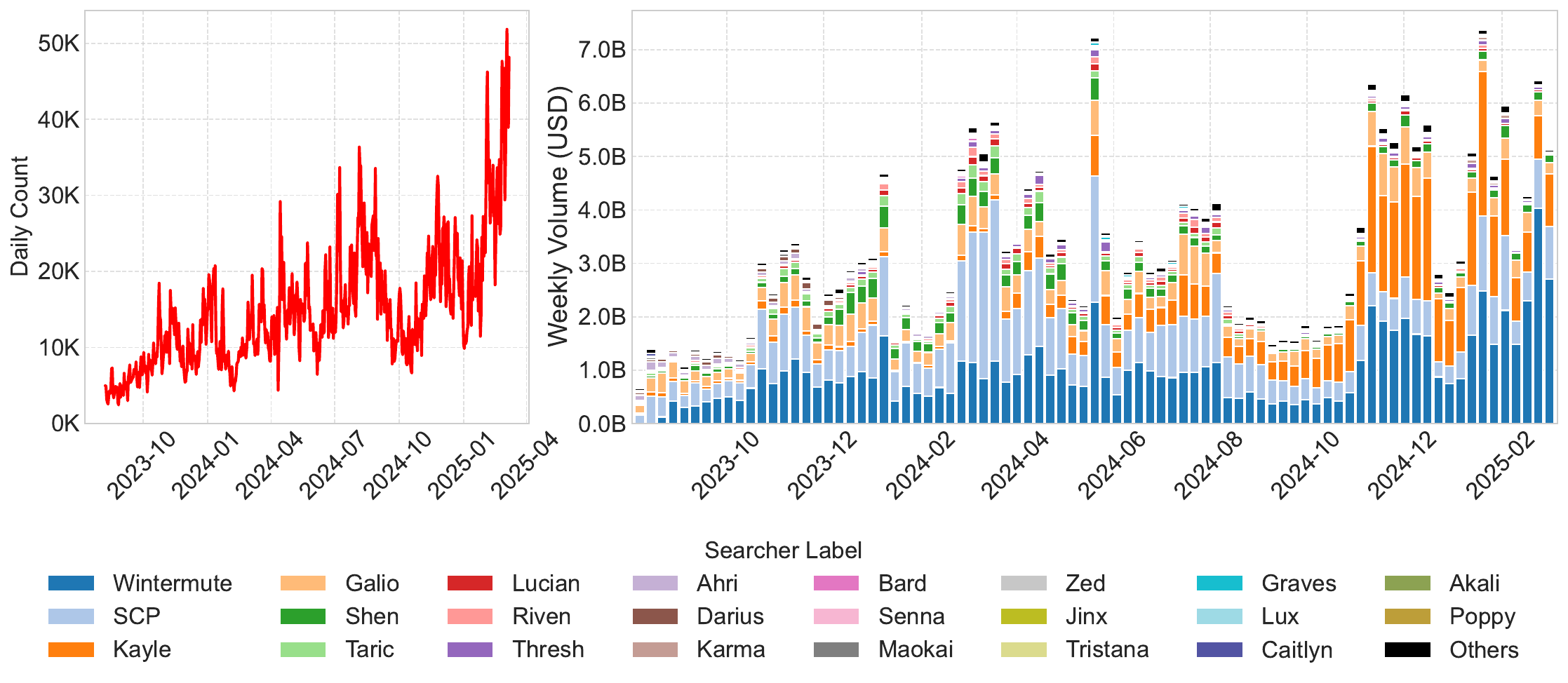}
  \caption{Daily count and weekly volume of detected CEX-DEX trades between August 2023 and March 2025.}
  
  \label{fig:cexdex_landscape}
\end{figure}

Figure 2 presents the daily CEX-DEX transaction counts (left panel) alongside the weekly volumes of each searcher entity (right panel). Note that all trades flagged by our heuristics are shown; transactions later excluded by the stricter profitability filter (cf. \Cref{sec:pnl_definition}), which roughly contribute to roughly 15\% of the total count and 7\% of the total volume, are still included here for landscape context. The daily count of CEX-DEX arbitrage transactions exhibits a clear upward trend over the 19-month period, despite occasional fluctuations. Notably, in Q3 2023, daily transaction counts averaged approximately 5,000-6,000. By Q1 2025, this figure increased by 7.2 times. 

Despite the overall increase in CEX-DEX arbitrage activity, we observe reduced diversity of successful searchers over time. Before October 2024, 23 labeled searchers collectively accounted for nearly 99\% of the total CEX-DEX arbitrage volume on-chain. This landscape transformed substantially afterward, with just 14 labeled searchers remaining active and successful. The figure further reduced to 11 by Q1 2025, with three leading searchers capturing 90\% of the total volume (cf. \Cref{appendix:mev_compare} \Cref{fig:volume_hhi}). Among all searchers, \texttt{Wintermute} and \texttt{SCP} have maintained dominant positions throughout the entire period. Interestingly, \texttt{Kayle}, initially a relatively minor searcher, began to rise significantly in influence around June 2024 and emerged as a leading searcher.

\section{Estimating CEX-DEX Arbitrage Revenue and Searcher Profit}
\label{sec:estimate}
We next proceed to introduce our methodology of estimating CEX-DEX arbitrage revenue and searcher profit. We assume that searchers first establish token inventory positions through on-chain DEX trades before subsequently hedging these accumulated positions on off-chain CEX venues. This sequential execution assumption stems from the inherently higher execution uncertainty on-chain compared to centralized venues. When multiple searchers target the same arbitrage opportunity in a specific DEX pool, the probability of any searcher's on-chain transaction being included in a block is uncertain. Consequently, searchers initiate off-chain hedging operations \emph{only after} receiving confirmation that their on-chain trades have been successfully executed, thereby avoiding potential unhedged inventory risk \cite{frontier2}. 

In addition, we assume that searchers execute the hedge for \emph{both directions} on CEX within a short timeframe or even instantaneously after confirming successful on-chain trade. Concretely, if a searcher buys BTC and sells ETH on the DEX, they will—within a very short window—sell the newly acquired BTC and repurchase ETH on the CEX. This behavior is economically rational given the inherent volatility risk of token inventory left unhedged, particularly for low-liquidity tokens that exhibit higher price volatility \cite{frontier2}.


Since searchers' actual hedging details on CEX are unobservable, we present a methodology to estimate their realized arbitrage revenues based on empirical observations.

\subsection{Searcher-Specific Optimal Execution Horizons}
\label{sec:pnl_definition}
A key challenge in estimating realized revenue is identifying the timing when searchers typically hedge their DEX positions off-chain. Searchers exhibit notable variations in the timing of their observations of DEX trade confirmations and subsequent CEX hedge, contrasting with standardized \texttt{slot time} reported by platforms such as Dune \cite{dune} and Etherscan \cite{etherscan}. 
These timing variations stem from several factors:

\begin{enumerate}[topsep=2pt, itemsep=2pt, parsep=2pt]
    \item \textbf{Block confirmation:} Searchers observe the block and on-chain transaction executions at varying times due to differences in network connectivity and block propagation path.
    \item \textbf{Information asymmetry:} Searchers possess different levels of market information (i.e., \emph{alpha}) affecting their execution timing decisions.
    \item \textbf{Execution infrastructure:} Variations in the infrastructure of different searchers result in different processing speed. Certain integrated searchers have privileged access to the block building process and can execute their strategy closer to the actual block time.
\end{enumerate}

To capture these searcher-specific timing characteristics, we analyze each searcher's historical trades by examining when their information advantage maximizes, reflected in the maximum achievable spread. To facilitate fair comparisons across different trade sizes, we define the \emph{gross return} ($\text{GR}$) at time $t$ (the \emph{markout horizon}) as the markout revenue per unit of trade volume, i.e., the spread captured per dollar deployed in the arbitrage.

Formally, consider a CEX-DEX arbitrage transaction $i$ with volume $V_i$ that purchases $x$ amount of token A and sells $y$ amount of token B on the DEX, net of liquidity provider fees. Let $P_A(t)$ and $P_B(t)$ represent the respective CEX USD prices at markout horizon $t$. The markout revenue ($\text{MR}$) and gross return ($\text{GR}$) achievable by flattening the DEX-acquired inventory on CEX at time $t$ can be estimated as:
\[
\text{MR}_i(t) = x \cdot P_A(t) - y \cdot P_B(t) - \text{CEX taker fees.}\footnotemark\
\]
\footnotetext{We assume CEX-DEX searchers operate at the highest user tier on Binance, thereby benefiting from the lowest applicable fee rate of 0.01725\% \cite{cexfees}.}
The gross return at time $t$ is thus:
\[\text{GR}_i(t) = \frac{\text{MR}_i(t)}{V_i}.\]

We value all tokens using USD to ensure consistent and accurate pricing by leveraging the superior liquidity of USDT pairs on Binance. To derive accurate USD prices for each token at specific markout horizons, we collect Binance historical quote data for USDT pairs from Tardis.dev \cite{tardisdata} and employ the \emph{mid-price} at each relevant markout timestamp.

\begin{remark}
We assume 1 USDT = 1 USD, as most tokens are priced in USDT on Binance. In contrast, other USD stablecoins (e.g., USDC) may not equal 1 USD due to potential arbitrage opportunities between stablecoins.
\end{remark}



We then calculate the gross return for all trades across a time window spanning from $\texttt{slot time}-1$ second to $\texttt{slot time}+10$ seconds, measured at an interval of 0.5 seconds. This interval range was chosen to represent searchers' typical submission time of DEX trades to builders, suggested by builders' bidding timing in the MEV-Boost auctions \cite{bidtiming}, and a reasonable upper bound for immediate hedging activity following DEX execution without introducing further noise caused by unrelated trades on CEX.




We exclude trades whose token price data is not recorded on Tardis on certain dates \cite{tardistoken} and transactions whose markout revenue persistently fails to cover the base fees throughout the interval, indicative of inventory adjustments rather than genuine arbitrage opportunities.\footnote{We observe searchers pay much lower tips to builders for these non-arbitrage transactions. For instance, \texttt{SCP} pays on average only 0.0021 ETH for these transactions compared to 0.0081 ETH for arbitrage transactions; similarly, \texttt{Wintermute} pays 0.0011 ETH versus 0.0089 ETH for arbitrage transactions.} 


\begin{figure}[t]
  \centering
  \begin{subfigure}[t]{0.33\linewidth}
    \centering
    \includegraphics[width=\linewidth]{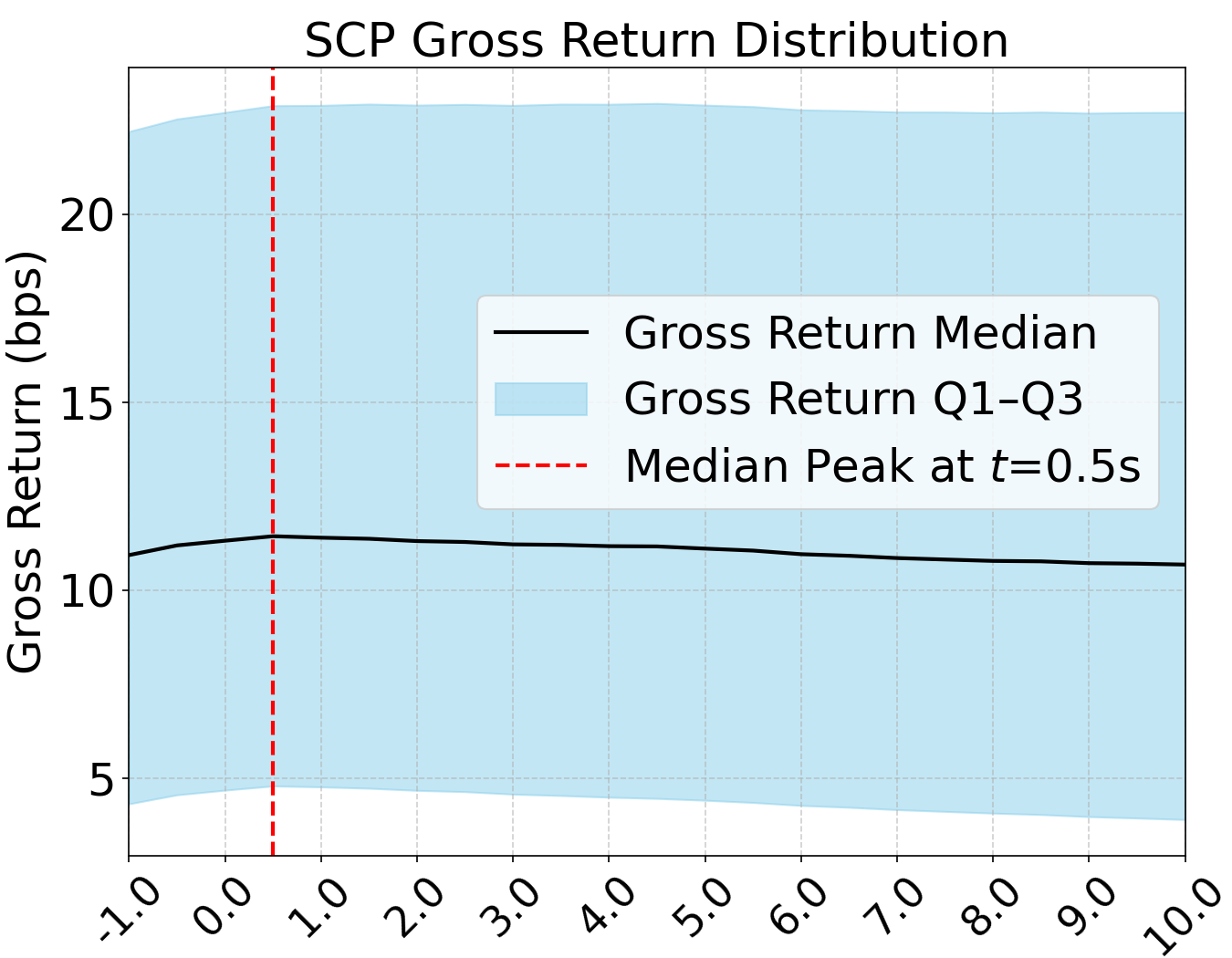}
  \end{subfigure}\hfill
  \begin{subfigure}[t]{0.33\linewidth}
    \centering
    \includegraphics[width=\linewidth]{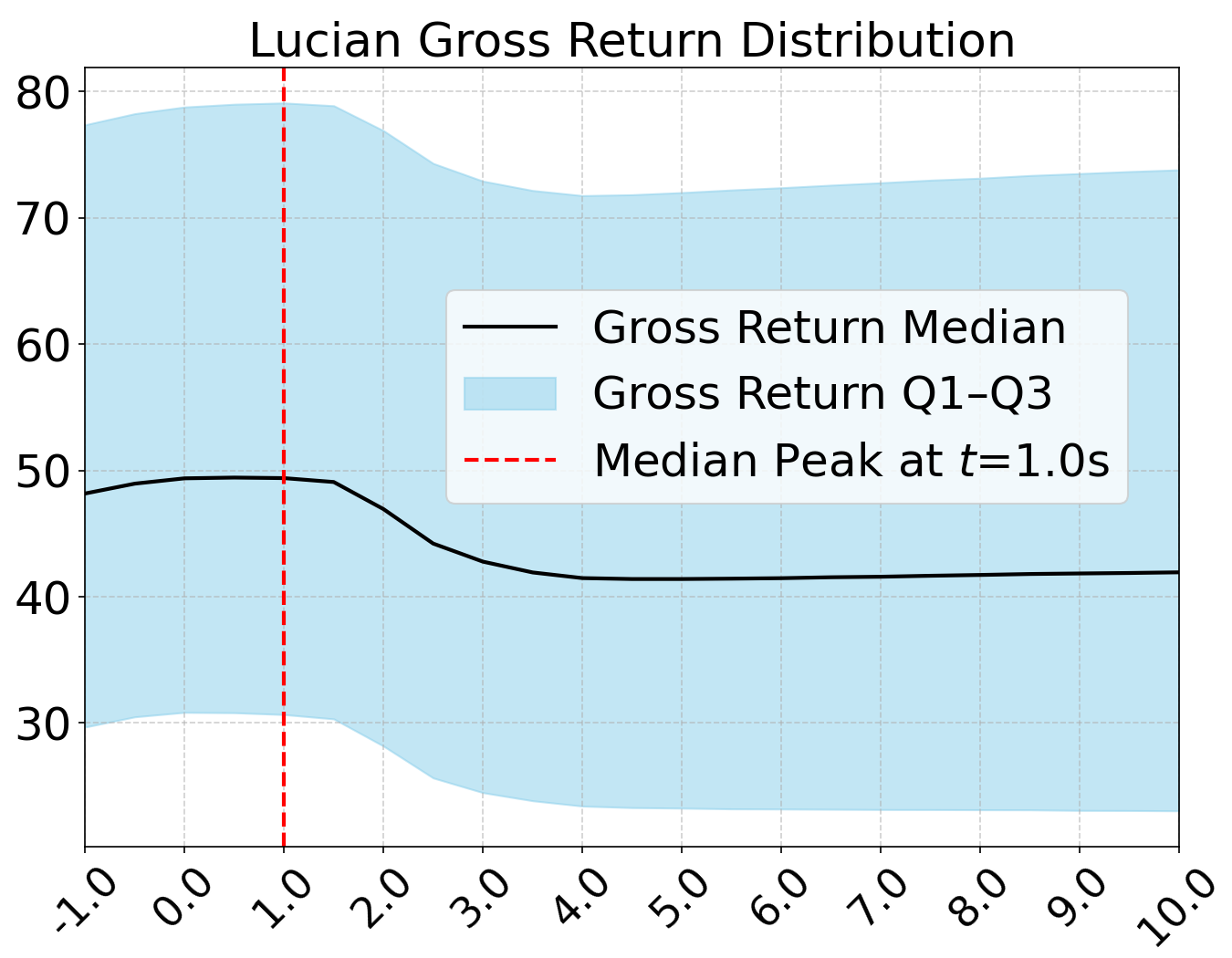}
  \end{subfigure}\hfill
  \begin{subfigure}[t]{0.33\linewidth}
    \centering
    \includegraphics[width=\linewidth]{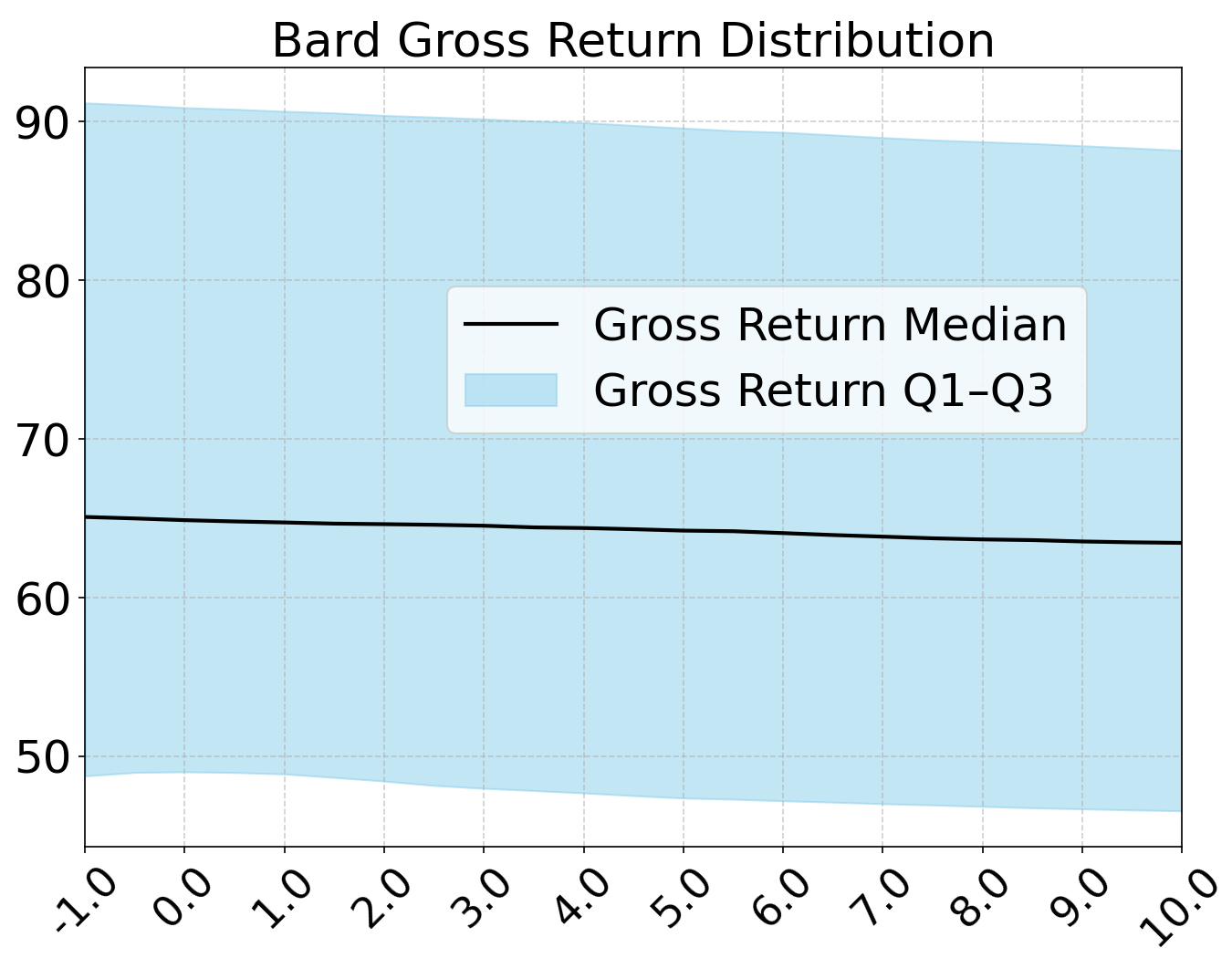}
  \end{subfigure}
  \\[0.3em] 
  \begin{subfigure}[t]{0.33\linewidth}
    \centering
    \captionsetup{justification=centering}
    \includegraphics[width=\linewidth]{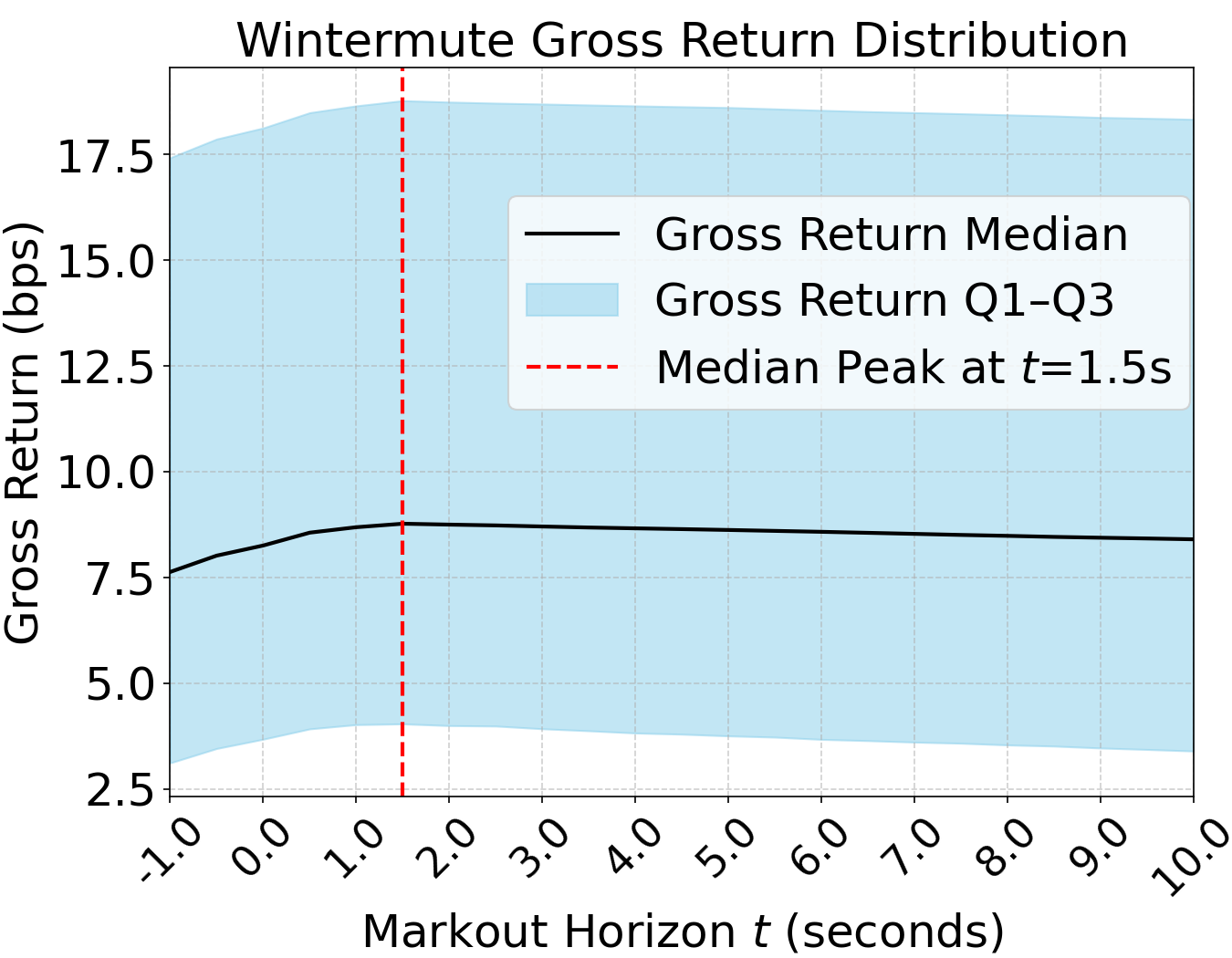}
    \caption{Pattern 1: \texttt{SCP} \& \texttt{Wintermute}}
    \label{fig:pattern1}
  \end{subfigure}\hfill
  \begin{subfigure}[t]{0.33\linewidth}
    \centering
    \captionsetup{justification=centering}
    \includegraphics[width=\linewidth]{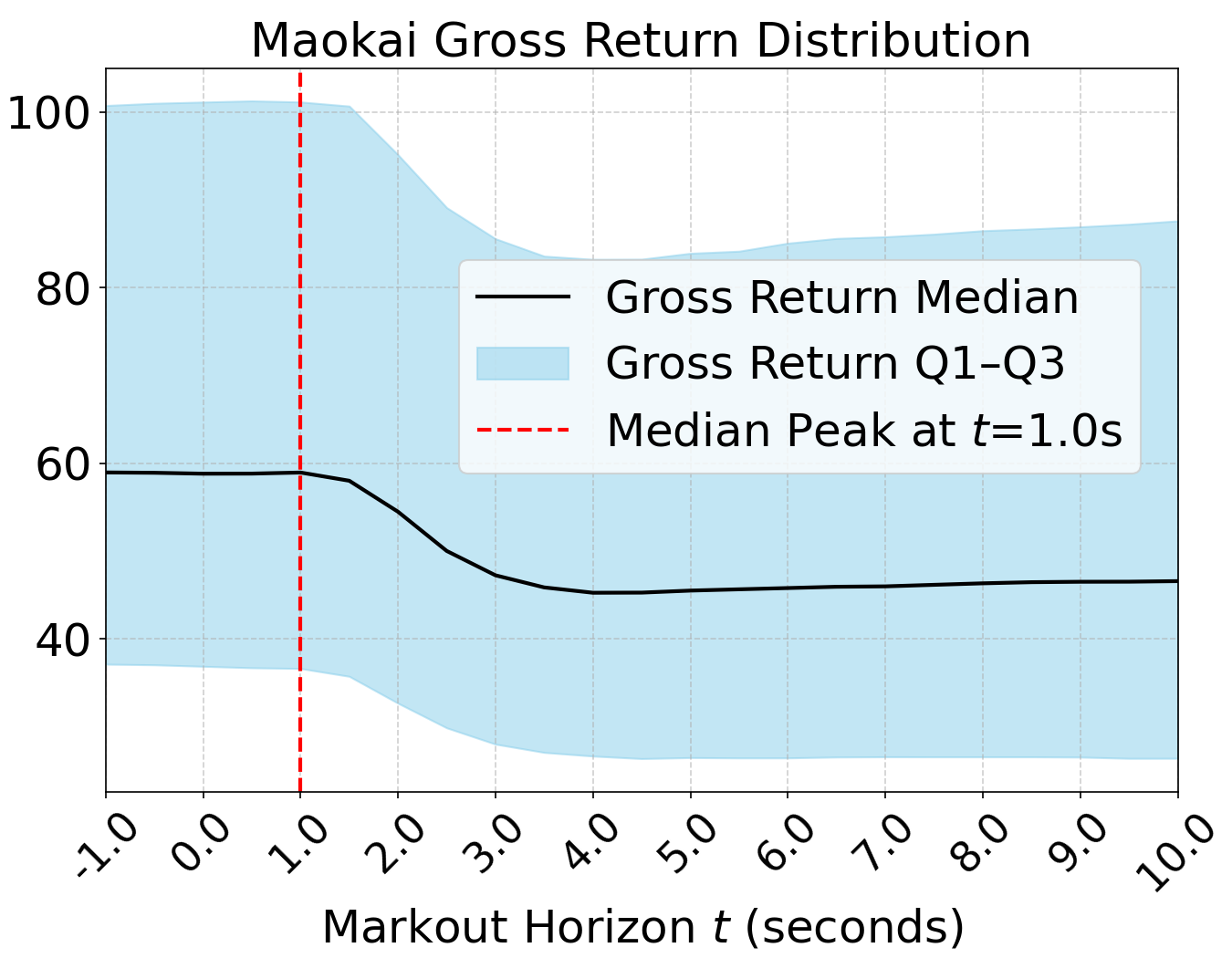}
    \caption{Pattern 2: \texttt{Lucian} \& \texttt{Maokai}}
    \label{fig:pattern2}
  \end{subfigure}\hfill
  \begin{subfigure}[t]{0.33\linewidth}
    \centering
    \captionsetup{justification=centering}
    \includegraphics[width=\linewidth]{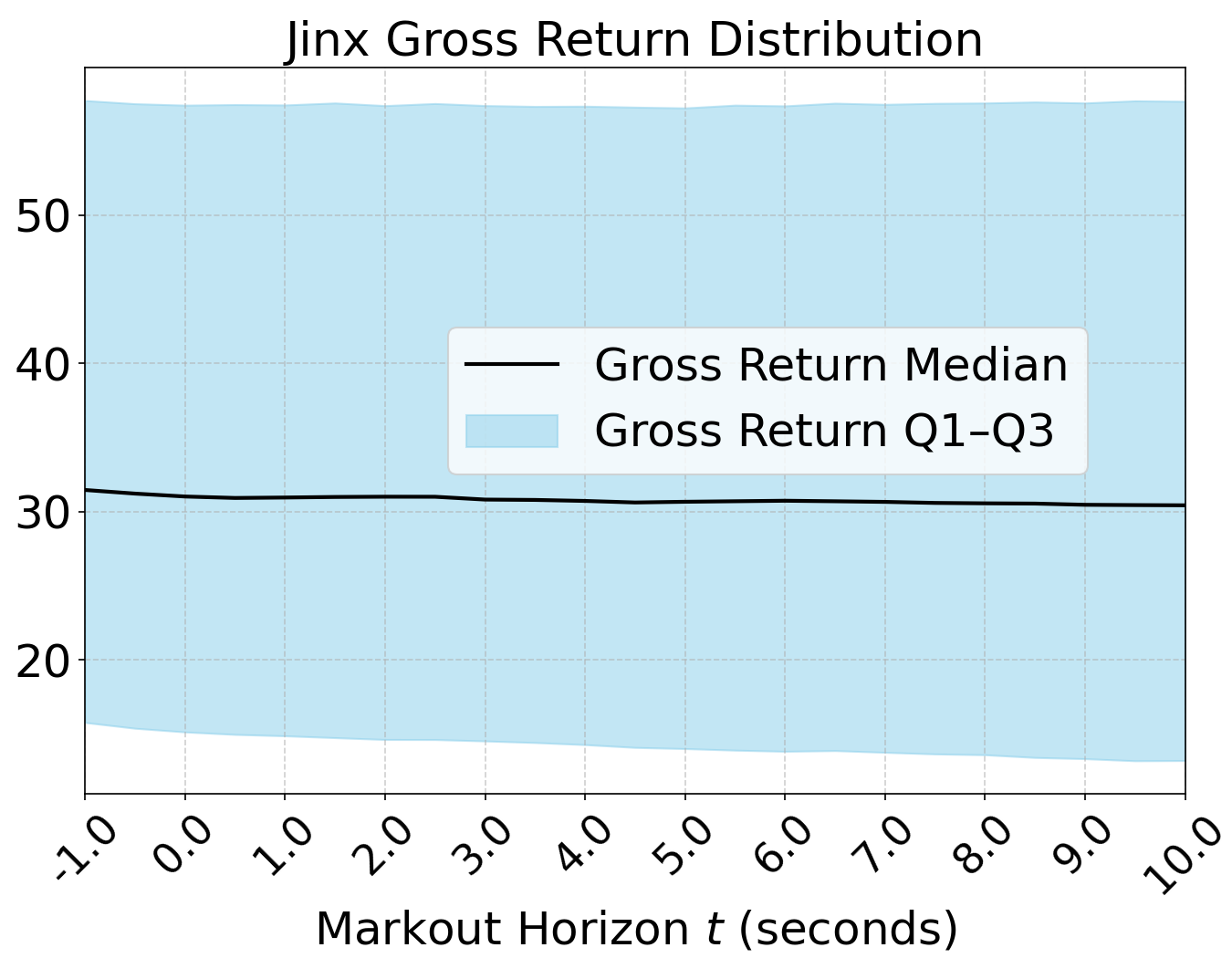}
    \caption{Pattern 3: \texttt{Bard} \& \texttt{Jinx}}
    \label{fig:pattern3}    
  \end{subfigure}

  \caption{Three empirically observed patterns of gross return distributions across all labeled searchers. Each \emph{column} corresponds to one pattern that contains two representative searchers that exhibit that pattern. The black line is the median, the shaded band the interquartile range, and the dashed red line indicates the markout horizon where the median gross return peaks.  $t=0$ second denotes the \texttt{slot time}. The labels of the x-axis and y-axis are shared between plots.}
  \label{fig:searcher_best_markout}
\end{figure}

The results reveal distinct temporal patterns in gross return across different searchers. \Cref{fig:searcher_best_markout} presents the gross return distributions for all trades of 6 representative searchers over the interval of $t\in[-1,10]$, where $t=0$ denotes the \texttt{slot time}, and reveals 3 different patterns that span all 23 labeled searchers (see \Cref{appendix:revenue_dist} for the remaining searchers). We here summarize the three patterns observed:

\begin{description} [topsep=2pt, itemsep=2pt, parsep=2pt]
\item[\textbf{Pattern 1 — Rising or plateau then gentle decay.}]
Gross return climbs (or maintains) at the peak then tapers off gradually (\Cref{fig:pattern1}). Representative searchers: \texttt{SCP}, \texttt{Wintermute}.

\item[\textbf{Pattern 2 — Rising or plateau then abrupt drop.}]
Gross return climbs (or maintains) at the peak, followed by a rapid decline and a flat tail (\Cref{fig:pattern2}). Representative searchers: \texttt{Lucian}, \texttt{Maokai}.

\item[\textbf{Pattern 3 — Flat or Minimal decline.}]
Gross return remains flat or declines slightly across the interval with no discernible peak (\Cref{fig:pattern3}).
Representative searchers: \texttt{Bard}, \texttt{Jinx}.
\end{description}
For simplicity, we refer to searchers whose gross return behaviors follow Pattern 1, 2, or 3 as Pattern 1, 2, or 3 searchers, respectively.

\subparagraph*{Pattern 1\&2.} 19 out of 23 labeled searchers follow the first two patterns and exhibit a characteristic \emph{optimal execution horizon}, identified by their (median) gross return rising to or maintaining at its maximum before declining. These patterns align with theoretical expectations: before hedge execution begins, the spread typically increases as the trader's information advantage manifests in favorable price movements. Once hedging commences, gross return declines as the trade's market impact pushes prices against their position. For instance, we observe that the median gross return of \texttt{SCP}'s trades reaches its maximum $0.5$ seconds after the \texttt{slot time}, while the optimal horizon for \texttt{Lucian} and \texttt{Maokai} occurs at $1.0$ second, for \texttt{Wintermute} at $1.5$ seconds.

We interpret the \emph{optimal execution horizon} as the most plausible empirical proxy for when searchers initiate their CEX hedge—the instant at which their information advantage is maximized and before the profit opportunity begins to erode due to delaying execution and market impact. Accordingly, we assume each searcher initiates their hedge at their optimal execution horizon empirically identified from all their historical trades. Although a single horizon cannot capture the exact timing of every trade, it summarizes the dominant behavior across all historical trades for each searcher: the cross-trade median peaks are sharp, and the interquartile bands around them are flat, indicating the estimate is not driven by outliers. 

Formally, For each searcher \(j\) we define the optimal execution horizon \(t^{*}_{j}\) as the markout horizon that maximizes the cross-trade median of gross return:
\[
  t^{*}_{j} =
  \underset{t\in\mathcal{T}}{\arg\max}\;
  \operatorname{Median}_{\,i\in\mathcal{I}_{j}}
  \bigl\{\operatorname{GR}_{i}(t)\bigr\},
\]
where $\mathcal{T}=\{-1.0,-0.5,0,\dots,10.0\}\,\text{seconds}$ is set of markout horizons around the \texttt{slot time} and $\mathcal{I}_{j}$ denotes the full set of historical trades of searcher \(j\). If multiple markout horizons attain the same maximum before the median gross return decreases, we select the largest $t$. Given \(t^{*}_{j}\), the revenue (i.e., extractable value) and net \emph{profit and loss} (PnL) for any trade $i\in\mathcal{I}_{j}$ are estimated as
\begin{align}
  \widehat{\operatorname{EV}}_{i} = \operatorname{MR}_{i}\!\bigl(t^{*}_{j}\bigr) - \text{base\_fees}_i, \text{ and }
  \widehat{\operatorname{PnL}}_{i} = \widehat{\operatorname{EV}}_{i}-\text{builder\_tips}_i,
\end{align}
where builder tips include priority fees and coinbase transfer.\footnote{For calculation, we convert the price unit of base fees, priority fees, and coinbase transfer from ETH to USD using the ETH-USDT mid-price at the \texttt{slot time}.} The \emph{profit margin} is thus:
\[\text{PM}_{i} = \frac{\widehat{\operatorname{PnL}}_{i}}{\widehat{\operatorname{EV}}_{i}}, \widehat{\operatorname{EV}}_{i} > 0.\]
We compute the profit margin only for trades whose estimated revenue is positive. 
Trades with $\widehat{\operatorname{EV}}_{i} \leq 0$ are labeled as "N/A" for margin calculations and excluded from profit margin analysis, but they are retained in all other analyses related to PnL directly.\footnote{For example, consider a trade with an estimated revenue of -1 USD, for which the searcher pays 10 USD to the builder. This results in a net PnL of -11 USD. Although the trade incurs a loss, computing the profit margin would yield an illogical value of +1100\%, which motivates its exclusion from margin-based analysis.}

Our approach to inferring CEX-DEX searchers' optimal execution horizon builds on methodologies from informed trading and market microstructure literature. Specifically, to identify this timing for each searcher, we adapt strategies from \cite{barclay_warner_1993, korajczyk_murphy_2018}, who empirically recover hidden execution timing of informed traders from observable post-trade price movements, by observing how CEX price movements affect the spread of their trades around the \texttt{slot time}. Moreover, our data-driven framework aligns with theoretical predictions by \cite{almgren_chriss_2000,bertsimas1998optimal}, suggesting each trader possesses a unique, liquidity-driven optimal execution horizon characterized by a distinct ``peak-then-decay" pattern in returns (cf. \Cref{sec:liquidity_hedge}). While our method necessarily provides a simplified approximation due to unobservable hedging price-impact costs, it nonetheless yields a consistent, empirically grounded upper-bound estimate of extracted value and PnL across searchers.



\subparagraph*{Pattern 3.} However, for four searchers—\texttt{Bard}, \texttt{Jinx}, \texttt{Tristana} and \texttt{Lux}—their gross return curve is essentially flat (or drifts only marginally downward) across the $[-1, 10]$ seconds window, so no clear "peak-then-decay" point emerges. The most plausible interpretation is that these searchers hold their inventory and do not hedge their positions on CEX within this window: if they were unwinding inventory promptly, the resulting price impact on the CEX leg should depress the gross return. More explanations can be found in~\Cref{sec:liquidity_hedge}. 

While extending the window further than +10 seconds might eventually reveal such a decline, that comes at the cost of substantial noise—price moves unrelated to the original trade. As our revenue and PnL estimation relies on identifying an optimal execution horizon, it cannot be applied reliably to these searchers. We therefore exclude these four searchers from subsequent revenue and PnL calculations, whose trades together account only for 2.7\% of total transactions detected; they remain in the descriptive landscape figures but are omitted from revenue and PnL analyses (see \Cref{appendix:tx_count} for summary statistics). 

\section{Token Liquidity as the Driver}
\label{sec:liquidity}
In this section, we demonstrate that the distinct patterns of the gross return curve observed across searchers are associated with the difference in the token liquidity they trade, which shapes searchers' value extraction regimes and subsequent hedge execution.

\subsection{Liquidity and Value Extraction Regimes}
\label{sec:liquidity_revenue}

As shown in Figures~\ref{fig:searcher_best_markout} and~\ref{fig:gross_returns_10s_14searchers}, Pattern 2 searchers earn markedly higher gross returns compared to Pattern 1 searchers. To highlight the role of token liquidity, we group tokens into "Major"—large-cap assets and commonly traded stablecoins (WETH, WBTC, USDT, USDC, TUSD, FDUSD, BUSD, DAI)—and "ALT", the remaining long-tail assets, and compare the trading pairs between these searchers. 

\begin{figure}[t]
    \centering
    \begin{subfigure}{0.66\textwidth}
    \captionsetup{justification=centering}
    \includegraphics[width=\linewidth]{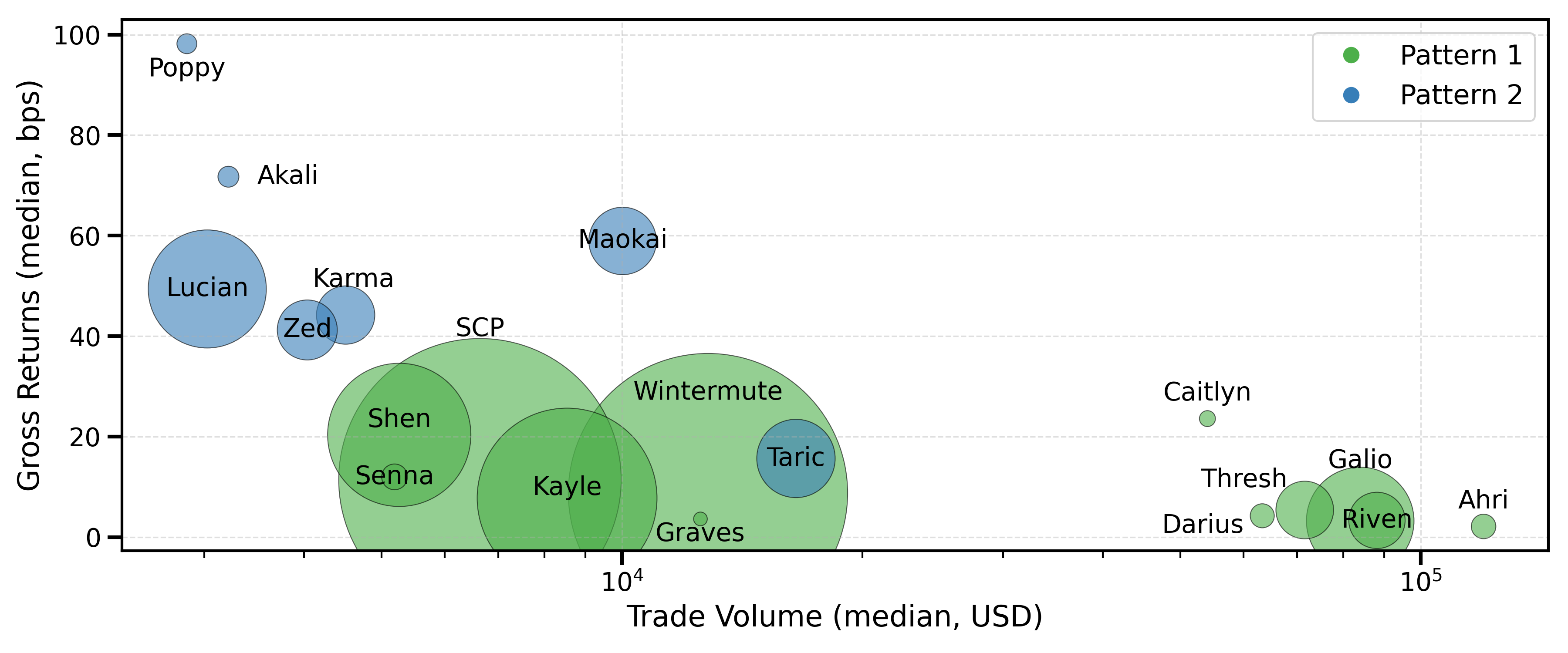}
    \caption{}
    \label{fig:bubble_return}
    \end{subfigure}%
    \begin{subfigure}{0.34\textwidth}
    \includegraphics[width=\linewidth]{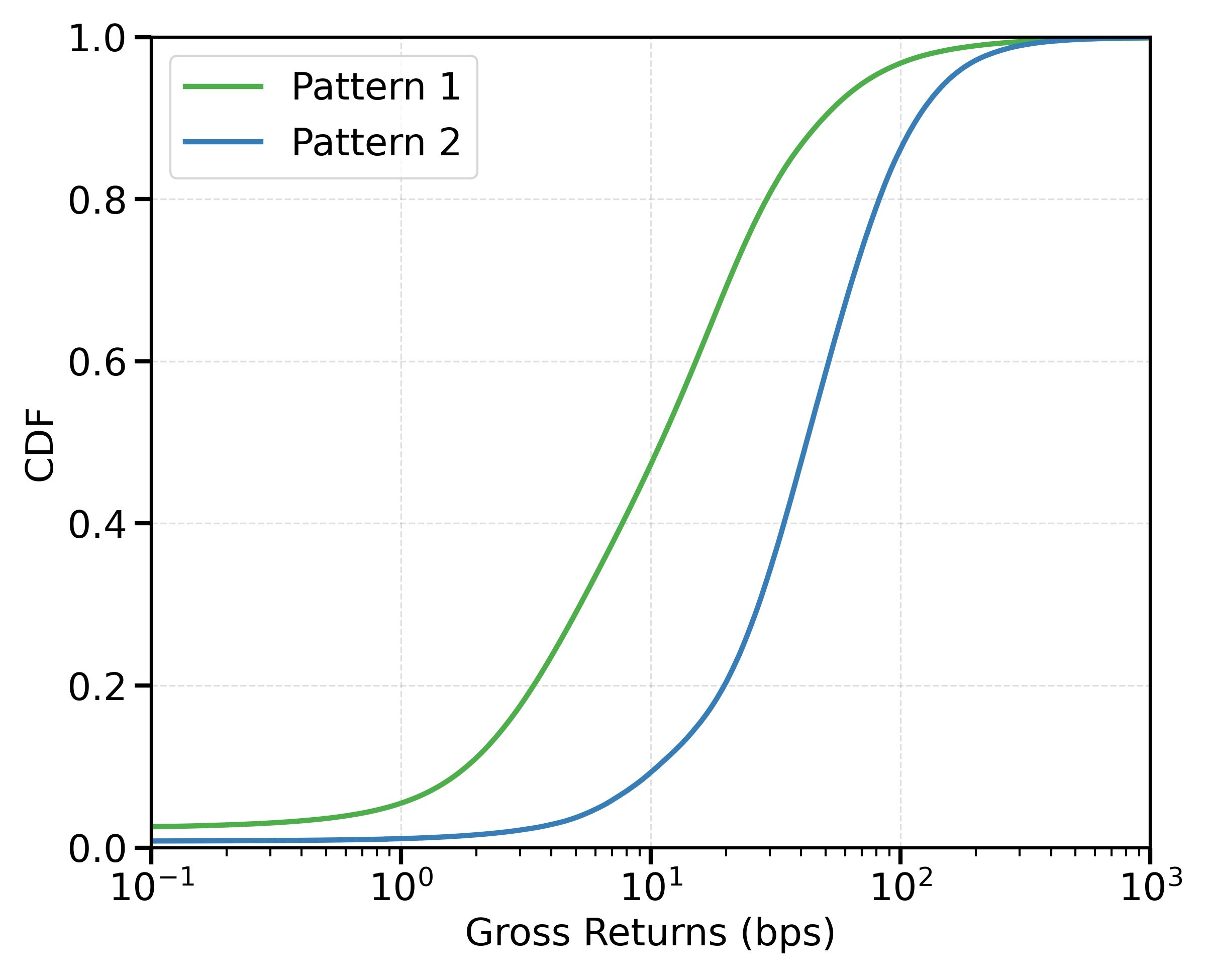}
    \captionsetup{justification=centering}
    \caption{}
    \label{fig:ecdf_return}
    \end{subfigure}
    \caption{\textbf{a)} The relationship between median trade volume and gross returns for each searcher, with bubble size indicating searchers' total estimated revenue and color encoding their execution pattern. \textbf{b)} Cumulative Distribution Function (CDF) of gross returns across all trades between Pattern 1 and 2 searchers.}
    \label{fig:gross_return}
\end{figure}

We find that Pattern 1 searchers such as \texttt{SCP} and \texttt{Wintermute} predominantly engage with highly liquid major tokens characterized by deep order books. In contrast, Pattern 2 searchers—typically smaller and less competitive—tend to focus more on lower-liquidity ALT tokens. Prices of these assets show higher volatility and suffer significant slippage even with modest trade volumes. This suggests that differences in token liquidity may influence both the achievable spread and the revenue strategies adopted by different searchers. We illustrate the empirical relationship between trade size and gross return in \Cref{fig:gross_return}. Detailed counts and volumes of token pairs traded by each searcher are available in \Cref{appendix:token}.

From \Cref{fig:bubble_return}, we observe that Pattern 1 searchers frequently execute trades exceeding 10,000 USD, leveraging the liquidity depth of major tokens to deploy substantial capital without significant price impact. However, they only earn single-digit basis-point gross returns. \Cref{fig:ecdf_return} confirms that nearly 90\% of their trades earn less than 20 bps gross return. Despite lower spreads per trade, these searchers accumulate substantial total revenue by leveraging large-scale execution. This pattern is particularly evident for leading searchers.

Conversely, Pattern 2 searchers face constrained liquidity conditions, typically restricting median trade sizes below 5,000 USD to avoid significant self-induced slippage. Nonetheless, limited liquidity frequently creates greater mispricing opportunities for long-tail assets, resulting in markedly higher gross returns ranging from 20 to 120 bps per trade. \Cref{fig:ecdf_return} confirms that 50\% of these searchers' trades exceed 40 bps in gross return.

This divergence reveals two distinct strategies to extract value: large-volume trade with narrow-spread targeting major tokens (Pattern 1) versus small-volume trade with wider-spread targeting long-tail tokens (Pattern 2). Interestingly, certain searchers exhibit hybrid strategies: \texttt{Maokai} and \texttt{Taric}, while categorized as Pattern 2, sustain relatively large trading volumes by maintaining significant activity in Major-Major pairs. \texttt{Shen} and \texttt{Senna}, categorized as Pattern 1, engage in notable ALT token trading, resulting in comparatively smaller volumes.

\subsection{Liquidity and Hedge Execution}
\label{sec:liquidity_hedge}

Variations in the speed and magnitude of spread closure after peaking between Pattern 1 and 2 observed in \Cref{fig:searcher_best_markout} can now be better understood as the value extraction regimes above manifest mechanically in how quickly spreads collapse once the hedge is executed.

As described above, Pattern 1 searchers focus primarily on deeply liquid major tokens. Consequently, after the initial peak, their gross returns decline smoothly and gradually, indicating that their hedge executions generate limited market impact. Conversely, Pattern 2 searchers, trading primarily lower-liquidity ALT tokens, experience rapid, pronounced closures of spreads immediately after peak returns. The thin liquidity conditions mean their hedge executions will quickly push prices against their positions, collapsing the spread.

\begin{figure}[t]
    \centering
    \includegraphics[width=0.9\linewidth]{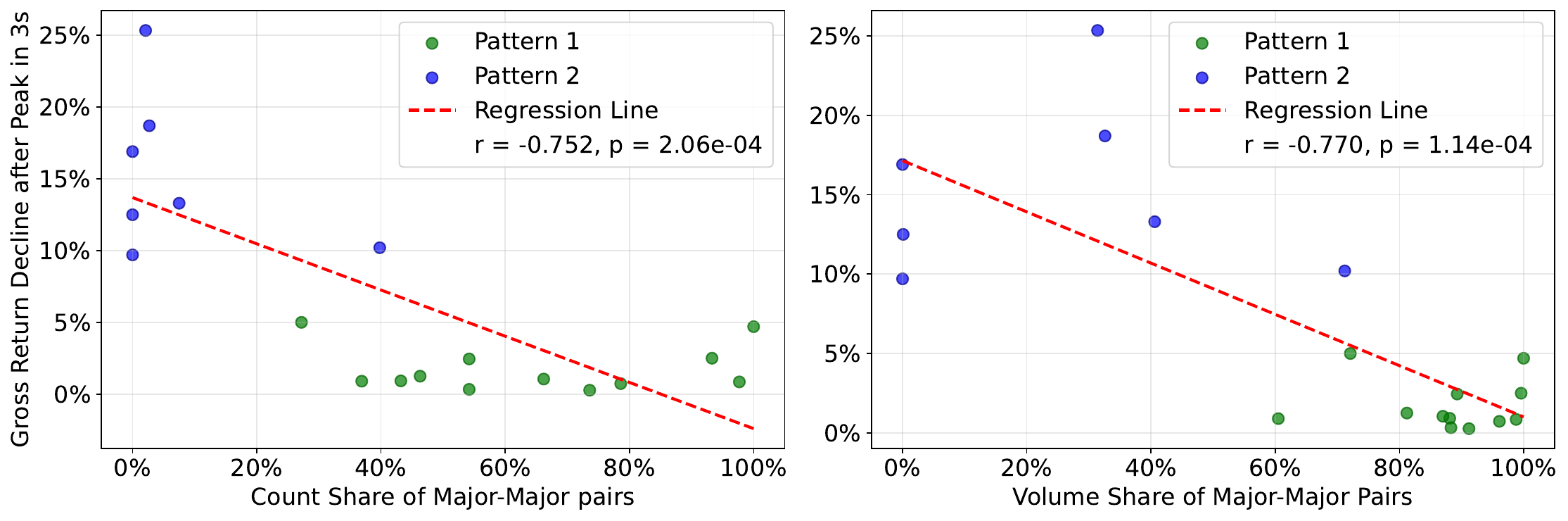}
    \caption{Correlation between the decline in median gross returns within 3 seconds after peak and the proportion of trade count (left panel) and trade volume (right panel) involving Major–Major token pairs for each searcher. Each scatter represents one searcher, with color indicating their pattern. The y-axis label is shared between the two subplots.}
    \label{fig:liquidity_corr}
\end{figure}

To quantitatively validate this link, we examine the correlation between token liquidity and how rapidly and significantly gross return declines. Specifically, \Cref{fig:liquidity_corr} plots the correlation between the proportion of major token trades and the percentage decline in median gross returns within 3 seconds after peaking for each searcher. We find a strong negative correlation, confirming that searchers predominantly trading major tokens experience slower and smaller gross return declines, reflective of reduced price impact. These findings align with the theoretical predictions in \cite{almgren_chriss_2000, bertsimas1998optimal}.

This rapid and significant spread closing after the peak for Pattern 2 searchers strongly indicates that they prioritize immediate position neutralization despite significant price impact, as they face a trade-off in less liquid markets: extending hedge execution may increase inventory risk compared to potential hedge cost savings. The observed flat tail of their gross return after the drop further confirms their emphasis on execution speed and inventory risk reduction over slippage minimization. In contrast, the muted, gradual spread compression for Pattern 1 searchers is consistent with deeper liquidity markets: even if hedging immediately with substantial volume, high liquid markets typically absorb size with only limited price impact. These searchers are further enabled to fragment large-size hedge trade into smaller pieces to minimize price impact at the cost of holding the inventory longer. 

Finally, we note that Pattern 3 searchers exclusively trade ALT tokens, yet exhibit minimal to no spread closure within the observed interval. This reinforces our earlier interpretation that these searchers likely delay hedging or choose not to hedge within our observation window, as immediate hedging would otherwise generate pronounced declines in their gross returns similar to those of Pattern 2.

Overall, our analysis empirically highlights how token liquidity profiles shape distinct value extraction and hedging regimes for CEX-DEX searchers. While token liquidity appears central to explaining observed patterns, we acknowledge other factors not presented here—such as individual searcher strategies or risk preferences—may also influence these outcomes.

\section{Extracted Value from CEX-DEX Arbitrages}
\label{sec:ev}

\begin{figure}[t]
    \centering
    \includegraphics[width=0.9\linewidth]{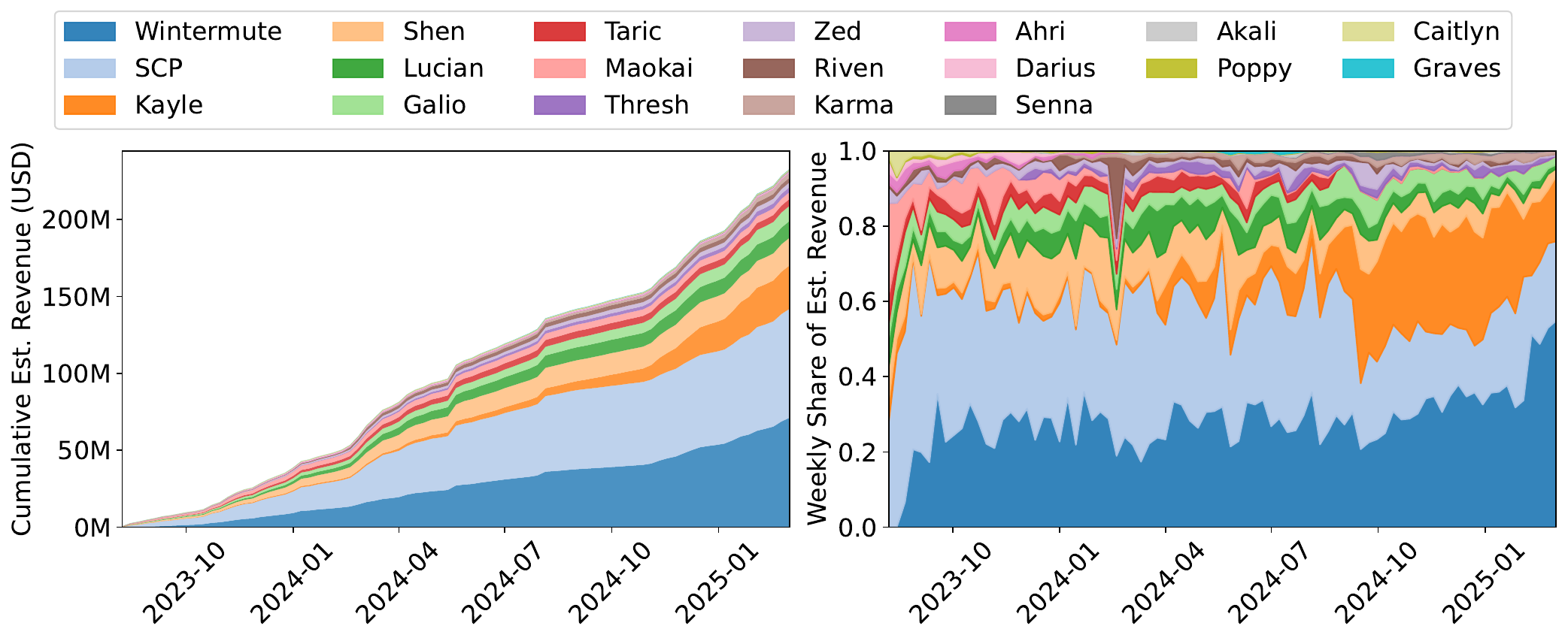}
    \caption{Cumulative CEX-DEX arbitrage revenue (left panel) and daily share of revenue (right panel) by 19 labeled searchers. A total of 233.8M USD is extracted from 7,203,560 CEX-DEX arbitrages by these 19 searchers between August 8, 2023 and March 8, 2025.}
    \label{fig:ev}
\end{figure}

Using the estimation methodology outlined in \Cref{sec:estimate}, we analyze a total of 7,203,560 CEX-DEX arbitrages with a total volume of 241.7B USD executed by 19 labeled searchers between August 8, 2023 and March 8, 2025. These figures significantly surpass atomic MEV activities during the same timeframe (cf. \Cref{appendix:mev_compare}, \Cref{fig:mev_compare}). \Cref{fig:ev} illustrates the cumulative CEX-DEX arbitrage revenue (i.e., extracted value) and weekly distribution of revenue among searchers over these 19 months. According to our estimation, a total of 233.8M USD is extracted by these 19 searchers through CEX-DEX arbitrages. At the time of writing, this figure is comparable to the total value extracted from atomic arbitrages since \emph{The Merge} \cite{libmev}. 

Beyond the revenue figures, our result highlights a structural shift towards market centralization. Both the volume of CEX-DEX arbitrages and the associated extracted value have become increasingly concentrated among a smaller subset of leading searchers. To quantify this trend, we compute the Herfindahl-Hirschman Index (HHI) for both trade volume and extracted value, revealing consistently rising concentration and a highly centralized market by the end of the observed period (cf. \Cref{appendix:mev_compare}, Figures~\ref{fig:volume_hhi} and~\ref{fig:ev_hhi}).

The largest two searchers, \texttt{Wintermute} and \texttt{SCP} consistently extracted over 50\% of total revenue throughout the entire observation period. A notable turning point occurred around June 2024, as the previously smaller searcher \texttt{Kayle} rapidly expanded its market presence and altered the competitive landscape. By October 2024, we observe a sharp reduction in searcher participation, with only 12 out of 19 labeled searchers remaining active. The market thus resembles a clear trident, with three searchers—\texttt{Wintermute}, \texttt{SCP}, and \texttt{Kayle}—collectively capturing 170.8M USD, representing approximately 73\% of the total cumulative extracted value. Furthermore, these top three players dominate arbitrage opportunities even more decisively, jointly accounting for around 90\% of the extracted value by Q1 2025. In contrast, smaller searchers' volume and revenue share have been compressed over time, and some searchers eventually stopped their operations and retreated from the market. 

\section{Searcher-Builder Integration and Profitability}
Searchers' net profitability ultimately depends on the fraction of revenue shared with the block builder. In this section, we examine searchers' integration level with block builders, and analyze its impact on searchers' PnL and profit margin. 

\subsection{Searcher-Builder Integration}
\label{sec:integration}
\begin{figure}
    \centering
    \includegraphics[width=0.9\linewidth]{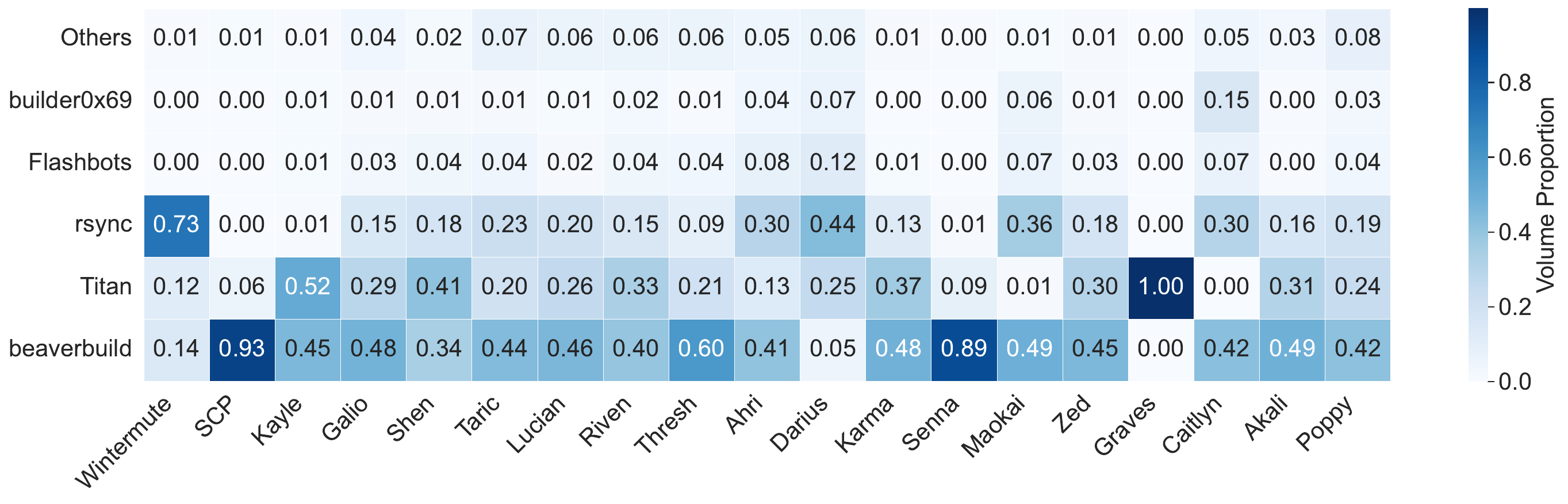}
    \caption{The proportion of CEX-DEX arbitrage volume from each labeled searcher (x-axis, sorted by total volume from left to right) in the blocks of top 5 builders (y-axis, sorted by market share from bottom to top).}
    \label{fig:heatmap}
\end{figure}

We start by assessing the level of integration between each of the 19 labeled searchers and major block builders by measuring the proportion of CEX-DEX arbitrage volume that each searcher directs to specific builders. Figure~\ref{fig:heatmap} visualizes these relationships through color intensity—darker cells indicate higher volume proportions and thus stronger integration between two entities. It is well-established that searcher \texttt{SCP} and builder \texttt{beaverbuild} are vertically integrated, as do \texttt{Wintermute} and \texttt{rsync} \cite{searcherbuilderpics, nonatomic, whowinsandwhy, yang2024decentralization}.

Our analysis further reveals exclusivity patterns between specific searcher-builder pairs: 
\begin{itemize}[topsep=2pt, itemsep=2pt, parsep=2pt]
    \item \texttt{Thresh} and \texttt{Senna} direct 60\% and 89\% of their volume to \texttt{beaverbuild}, respectively.
    \item \texttt{Kayle} and \texttt{Graves} send 52\% and 100\% of their volume to \texttt{Titan}, respectively.
\end{itemize}
We refer to these searchers, who send more than 50\% of their volume to one builder, as the \emph{exclusive searcher} (including \emph{integrated searchers} \texttt{SCP} and \texttt{Wintermute}) for that builder, and correspondingly label their primary recipient builder as their \emph{exclusive builder}. All other searchers, who spread their trades evenly among multiple builders, we classify as \emph{neutral searchers}. Notably, although \texttt{Kayle} primarily splits trades between \texttt{Titan} (52\%) and \texttt{beaverbuild} (45\%), subsequent analysis will demonstrate a higher integration level with \texttt{Titan} than with \texttt{beaverbuild}.

\subsection{Searcher Profitability Analysis}

\label{sec:searcher_profitability}
\begin{figure}[t]
    \centering
    \begin{subfigure}{0.7\textwidth}
    \captionsetup{justification=centering}
    \includegraphics[width=\linewidth]{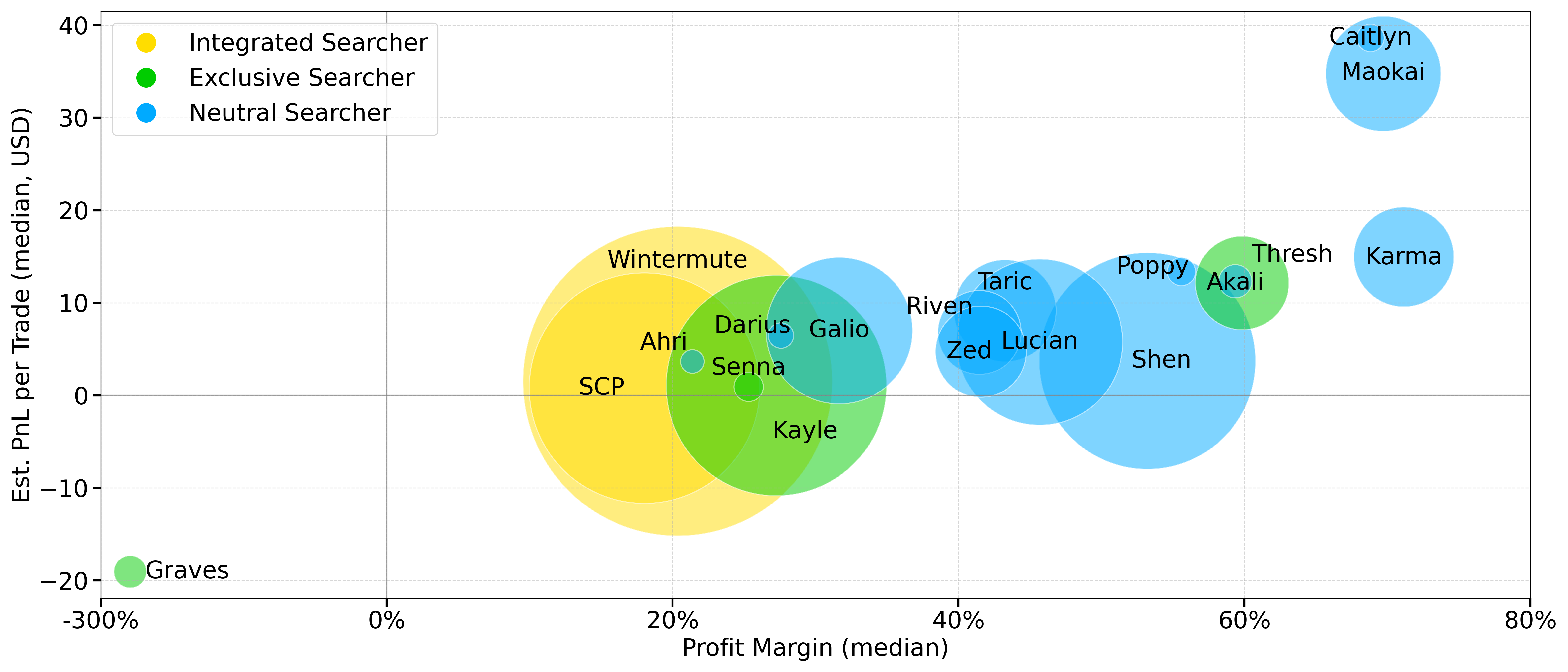}
    \caption{}
    \label{fig:bubble_pnl}
    \end{subfigure}%
    \begin{subfigure}{0.3\textwidth}
    \includegraphics[width=\linewidth]{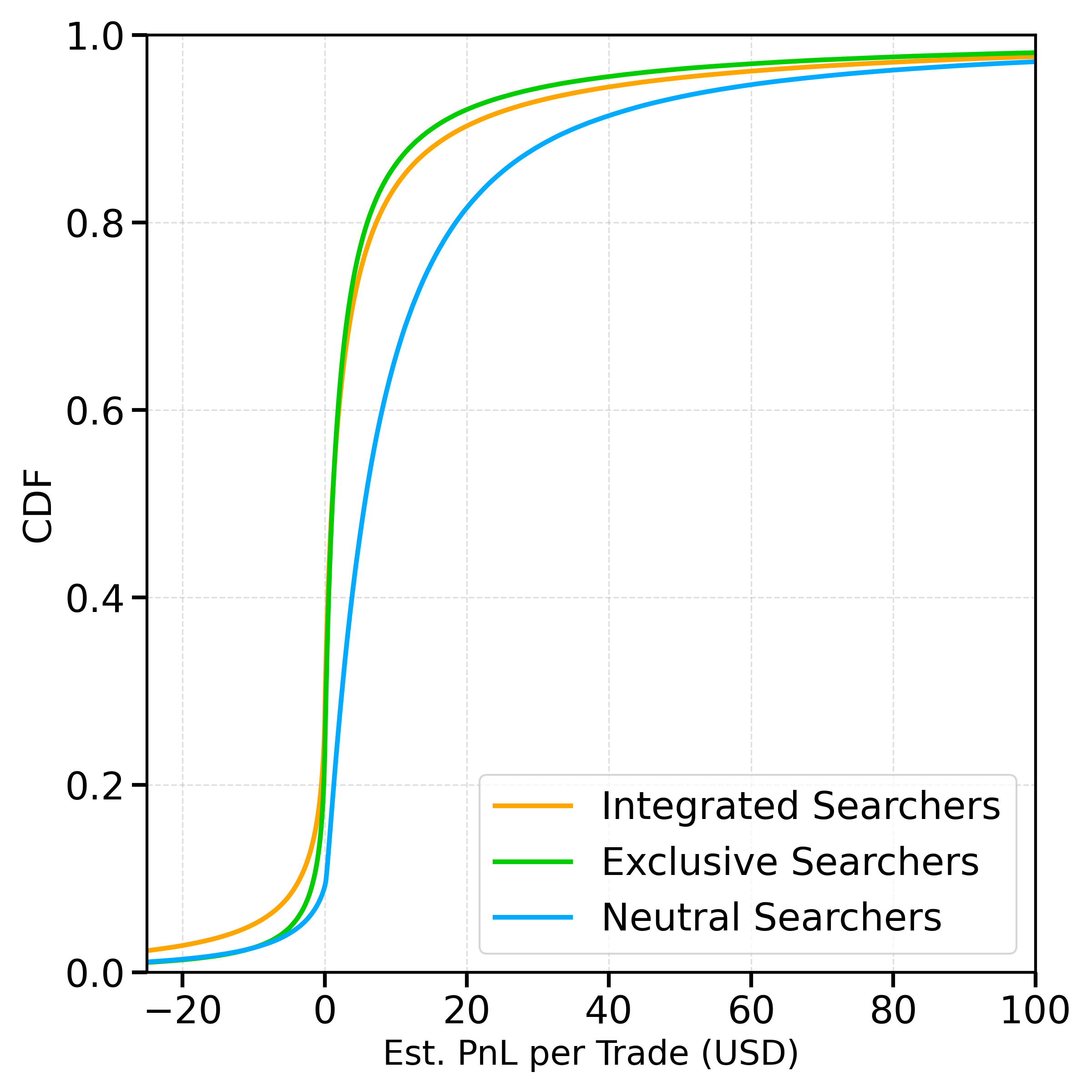}
    \captionsetup{justification=centering}
    \caption{}
    \label{fig:ecdf_pnl}
    \end{subfigure}
    \caption{\textbf{a)} Bubble plot illustrating searchers' median profit margin (x-axis) and median estimated PnL per trade (y-axis). The color shows the searcher's integration level. The bubble size represents the searcher's total PnL. Due to \texttt{Graves} having a negative total PnL, we show its absolute value. The negative tick of the x-axis (-300\%) is not proportionally spaced. \textbf{b)} Cumulative Distribution Function (CDF) of estimated PnL across all trades between the three integration levels.}
    \label{fig:profitability}
\end{figure}

We then proceed to analyze the profitability of the searchers. In \Cref{fig:bubble_pnl}, we find that neutral searchers generally have much higher profitability than most exclusive searchers: their median profit margins cluster around 30\% - 70\%, whereas the median profit margins of most exclusive searchers are within 20\% - 40\%. Neutral searchers' median PnL per trade often exceeds 10 USD, while the median PnL per trade of most exclusive searchers rarely exceeds 5 USD. \Cref{fig:ecdf_pnl} confirms the edge: 80\% of exclusive and integrated searchers' trades net fewer than 10 USD, whereas nearly half of neutral searchers' trades exceed this threshold.

\Cref{tab:profitability} summarizes the profitability status of the searchers. Despite being market leaders, \texttt{Wintermute}, \texttt{SCP}, and \texttt{Kayle} clear the largest notional volume and earn the highest total revenues, yet their profit margin is among the lowest and their median PnL per trade mostly hovers only a few dollars above zero. \texttt{SCP} and \texttt{Kayle}'s total PnL are barely larger than that of neutral searcher \texttt{Shen}, whose trade volume is much lower. Moreover, \texttt{Graves}, who sends almost all their trades to \texttt{Titan}, exhibits negative profitability according to our estimation, indicating payments to the builder often exceed their arbitrage revenue. We report the medians rather than the cumulative means for metrics of revenue per trade, PnL per trade, and profit margin to avoid the disproportionate influence of outlier trades that skew the means; the medians therefore represent the typical trade more faithfully. 

\begin{table}[t]
\centering
\caption{Searcher profitability summary sorted by total volume.}
\label{tab:profitability}

\resizebox{\textwidth}{!}{%
\begin{tabular}{lcccccccc}
\toprule
\textbf{Searcher} & \textbf{Total}  & \textbf{Total} & \textbf{Total} & \textbf{Total} & \textbf{Median} & \textbf{Median} & \textbf{Median} \\
& \textbf{Volume}& \textbf{Est. Rev.} & \textbf{Builder Tips} & \textbf{Est. PnL} & \textbf{Trade Rev.} & \textbf{Trade PnL} & \textbf{Profit Margin} \\
& [USD]  & [USD] & [USD] & [USD] & [USD] & [USD] & \\
\midrule
\midrule
Wintermute & 74.8B & 71.4M  & 47.1M & 24.3M & 11.36 & 1.65 & 20.9\% \\
SCP & 63.5B & 71.1M & 57.4M  & 13.7M & 7.34 & 0.86 & 18.5\% \\
Kayle & 41.0B & 28.3M  & 16.0M & 12.3M & 6.05 & 1.15 & 27.9\% \\
Galio & 25.9B & 9.9M  & 4.5M   & 5.4M & 24.93 & 7.10 & 31.7\% \\
Shen & 12.3B & 18.1M   & 6.3M  & 11.8M & 8.42 & 3.77 & 53.2\% \\
Taric & 5.0B & 4.4M   & 1.8M   & 2.6M & 22.92 & 9.28 & 43.3\% \\
Lucian & 3.9B & 10.9M  & 3.9M   & 7.0M & 14.74 & 5.83 & 45.8\% \\
Riven & 3.1B & 3.1M   & 1.3M   & 1.9M & 31.42 & 8.16 & 42.5\% \\
Thresh & 2.7B & 3.2M   & 1.0M   & 2.2M & 34.12 & 13.65 & 59.7\% \\
Ahri & 1.7B & 541.4K & 399.0K  & 162.4K & 24.08 & 4.30 & 22.8\% \\
Darius & 1.1B & 497.0K  & 331.3K &  165.6K & 29.62 & 6.62 & 27.9\% \\
Karma & 674.8M & 3.0M  & 517.5K  & 2.5M & 22.82 & 15.02 & 71.2\% \\
Senna & 596.3M & 503.9K & 268.1K  & 235.8K & 5.22 & 1.06 & 27.2\% \\
Maokai & 588.4M & 4.4M   & 1.1M  & 3.3M & 52.40 & 35.13 & 69.7\% \\
Zed & 580.2M & 3.1M &  1.1M  & 2.1M & 13.99 & 4.83 & 41.8\% \\
Graves & 229.3M & 173.4K & 353.1K & -179.7K & 8.77 & -19.34 & -272.1\% \\
Caitlyn & 96.1M & 253.7K & 67.6K & 186.1K & 69.02 & 39.62 & 68.8\% \\
Akali & 49.4M & 378.6K & 100.5K & 278.1K & 22.20 & 12.37 & 59.4\% \\
Poppy & 35.5M & 371.7K & 171.0K & 200.7K & 25.00 & 13.43 & 55.6\% \\
\bottomrule
\end{tabular}
}
\end{table}

\begin{figure}[ht!]
    \centering
    \includegraphics[width=1\linewidth]{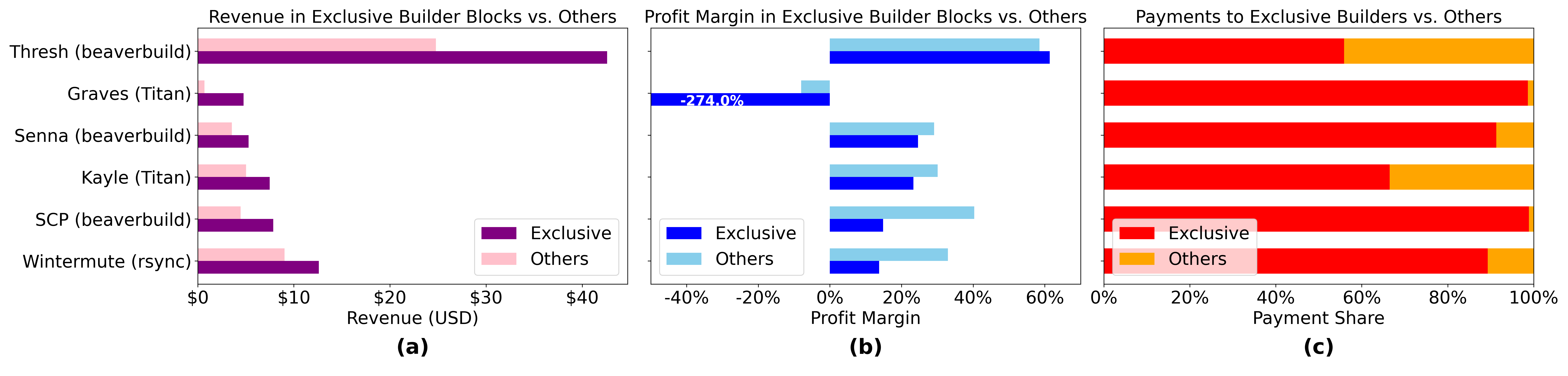}
    \caption{ \textbf{a)} Revenue (median) of exclusive searchers' trades in their respective exclusive builders' blocks vs. in other blocks. \textbf{b)} Profit margin (median) of exclusive searchers' trades in their respective exclusive builders' blocks vs. in other blocks. \textbf{c)} Percentage of total payments from exclusive searchers to their exclusive builder vs. to other builders. The y-axis is shared between three plots and represents the exclusive pair of searcher and builder.}
    \label{fig:exclusive_profitability}
\end{figure}

\Cref{fig:exclusive_profitability} offers deeper insights into these dynamics between exclusive searchers and their corresponding exclusive builders:
\begin{enumerate}[topsep=2pt, itemsep=2pt, parsep=2pt]
    \item In Figure~\ref{fig:exclusive_profitability}a, we observe that exclusive searchers consistently send higher-revenue trades to their exclusive builder, compared to other builders.
    \item In Figure~\ref{fig:exclusive_profitability}b, for most exclusive searchers, their trades in their respective exclusive builder's block consistently have lower median profit margins, indicating a higher proportion of revenue shared with the exclusive builder than other builders. For instance, both \texttt{Wintermute} and \texttt{SCP} transfer nearly 90\% of their arbitrage revenue to their integrated builder \texttt{rsync} and \texttt{beaverbuild}, retaining lower profit margins slightly above 10\%. While most \texttt{Grave}'s trades are not profitable, their trades in their exclusive builder \texttt{Titan}'s blocks have a median profit margin of -274.0\%.
    \item In Figure~\ref{fig:exclusive_profitability}c, as a result of submitting higher revenue trades and sharing greater revenue proportion, exclusive searchers favor their exclusive builder with substantially higher total payments than the amount paid to other builders.
\end{enumerate}

Both \texttt{Wintermute} and \texttt{SCP} transfer nearly 90\% of their arbitrage revenue directly to their integrated builder to gain an edge in the block building auction and seek a prioritized execution for their trades. \texttt{Graves} sends their trade exclusively to \texttt{Titan} with payments much higher than their revenue. While \texttt{Kayle} splits its trade volume almost evenly between \texttt{beaverbuild} and \texttt{Titan}, the above economics point to a much higher integration with \texttt{Titan}: trades landed in \texttt{Titan} blocks earn noticeably higher revenue but carry a markedly lower profit margin, and \texttt{Kayle}'s total payments to Titan are roughly double those paid to the rest builders combined. Similar patterns are also evident for \texttt{Senna} with \texttt{beaverbuild}. 

\texttt{Thresh}, by contrast, appears only loosely affiliated with \texttt{beaverbuild} despite sending 60\% of its volume there. The median margin \texttt{Thresh} keeps in \texttt{beaverbuild} blocks still exceeds 50\%—in fact slightly higher than in other blocks—and its total payments to \texttt{beaverbuild} are only modestly larger than to builders, which is reasonable considering the success of \texttt{beaverbuild} as a builder \cite{mevboostpics}. This indicates that \texttt{Thresh} simply routes most of their trades to \texttt{beaverbuild} but is less likely to engage in an exclusive arrangement.

In summary, neutral searchers distribute trades across several builders and pay only the lowest necessary tips for block inclusion, thereby retaining the largest possible share of revenue. Exclusive searchers—likely bound by an exclusive orderflow deal with the builder—often direct their profitable trades to a single builder, willingly sharing a substantially larger fraction of their revenue and accepting slim or even negative PnL at the trade level. 

\subsection{Impact of Exclusive CEX-DEX Flow on Market Structures}
\label{sec:exclusive_builder_performance}
Prior research has highlighted the strategic advantage builders gain by accessing high-value exclusive CEX-DEX flow \cite{gupta2023centralizing,nonatomic,whowinsandwhy,wu2024strategic,yang2024decentralization}. Here, we extend this discussion by closely examining interactions between exclusive searcher-builder pairs and their impact on both the searcher and builder markets.

\begin{figure}[t]
    \centering
    \begin{subfigure}{0.5\textwidth}
    \captionsetup{justification=centering}
    \includegraphics[width=\linewidth]{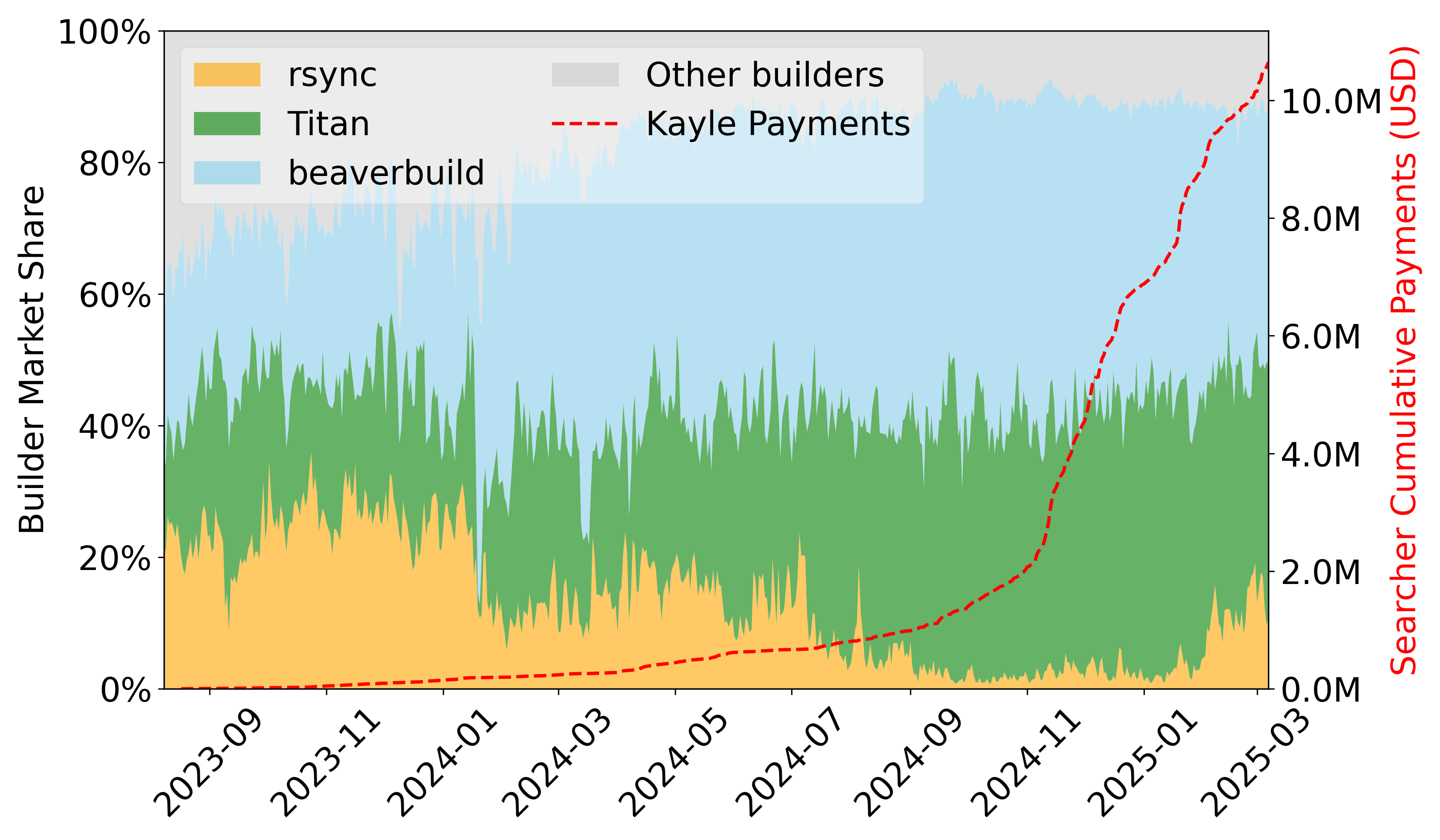}
    \caption{}
    \label{fig:kayle_Graves_titan}
    \end{subfigure}%
    \begin{subfigure}{0.5\textwidth}
    \includegraphics[width=\linewidth]{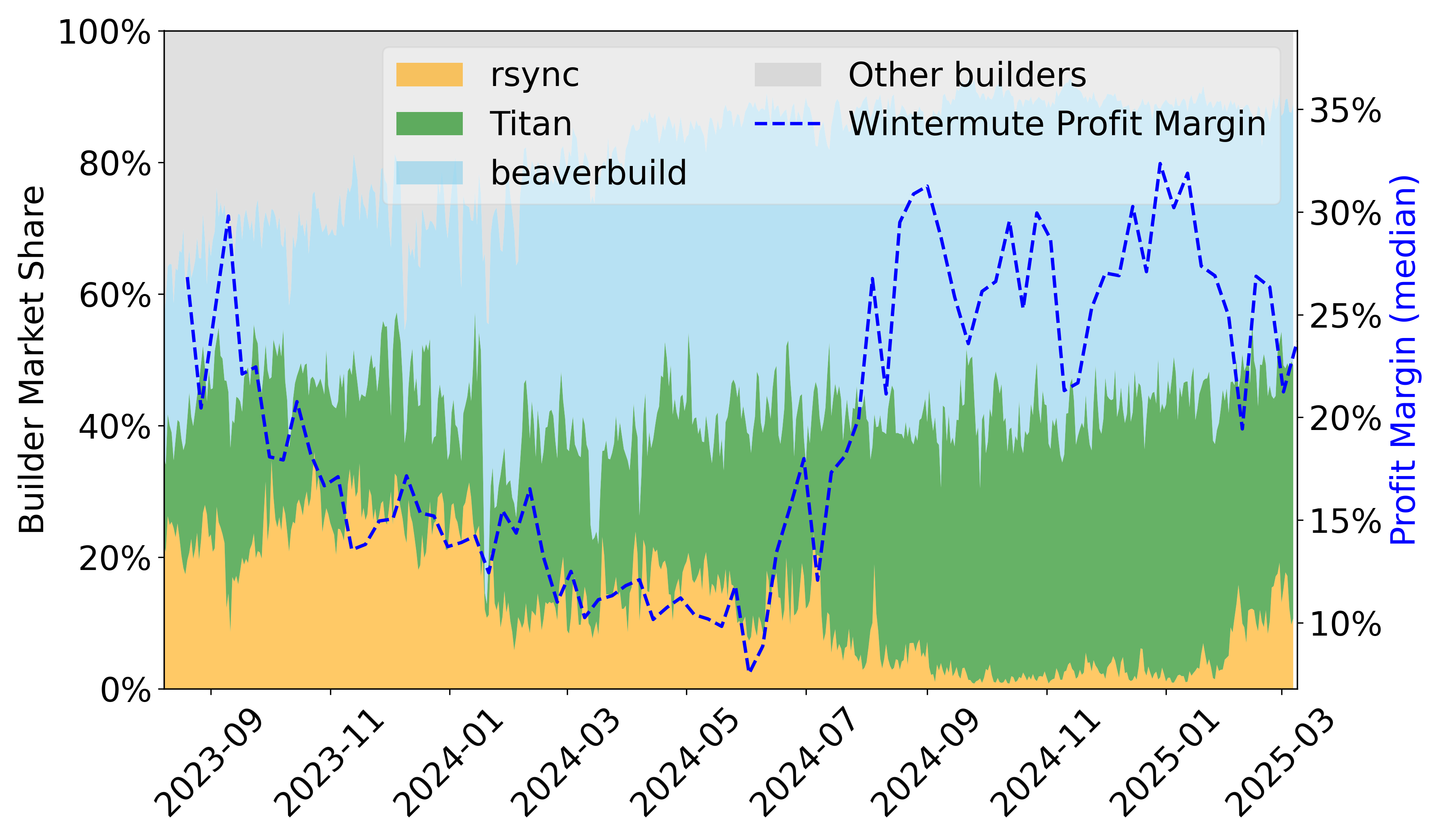}
    \captionsetup{justification=centering}
    \caption{}
    \label{fig:wm_margin_vs_share}
    \end{subfigure}
    \caption{\textbf{a)} \texttt{Titan} market share and \texttt{Kayle}'s cumulative payments to \texttt{Titan}. The left y-axis shows the builder market share, and the right y-axis shows \texttt{Kayle}' cumulative payments to \texttt{Titan}. \textbf{b)} \texttt{rsync} market share and \texttt{Wintermute} profit margin. The left y-axis shows the builder market share, and the right y-axis shows \texttt{Wintermute}'s median profit margin.}
\end{figure}

As previously described, \texttt{Kayle}, an exclusive searcher affiliated with builder \texttt{Titan}, quickly ascended to top-tier searcher status with the third-largest CEX-DEX volume since June 2024. Simultaneously, \Cref{fig:kayle_Graves_titan} shows \texttt{Titan}'s market share grew by 15\% till September 2024, coinciding with a marked surge in cumulative payments from \texttt{Kayle}. To quantitatively assess this relationship, we conduct a correlation analysis between \texttt{Kayle}'s daily CEX-DEX volume share and \texttt{Titan}'s daily market share between June 1 to September 30, 2024 (122 days). 

\begin{figure}[t]
    \centering
    \includegraphics[width=1\linewidth]{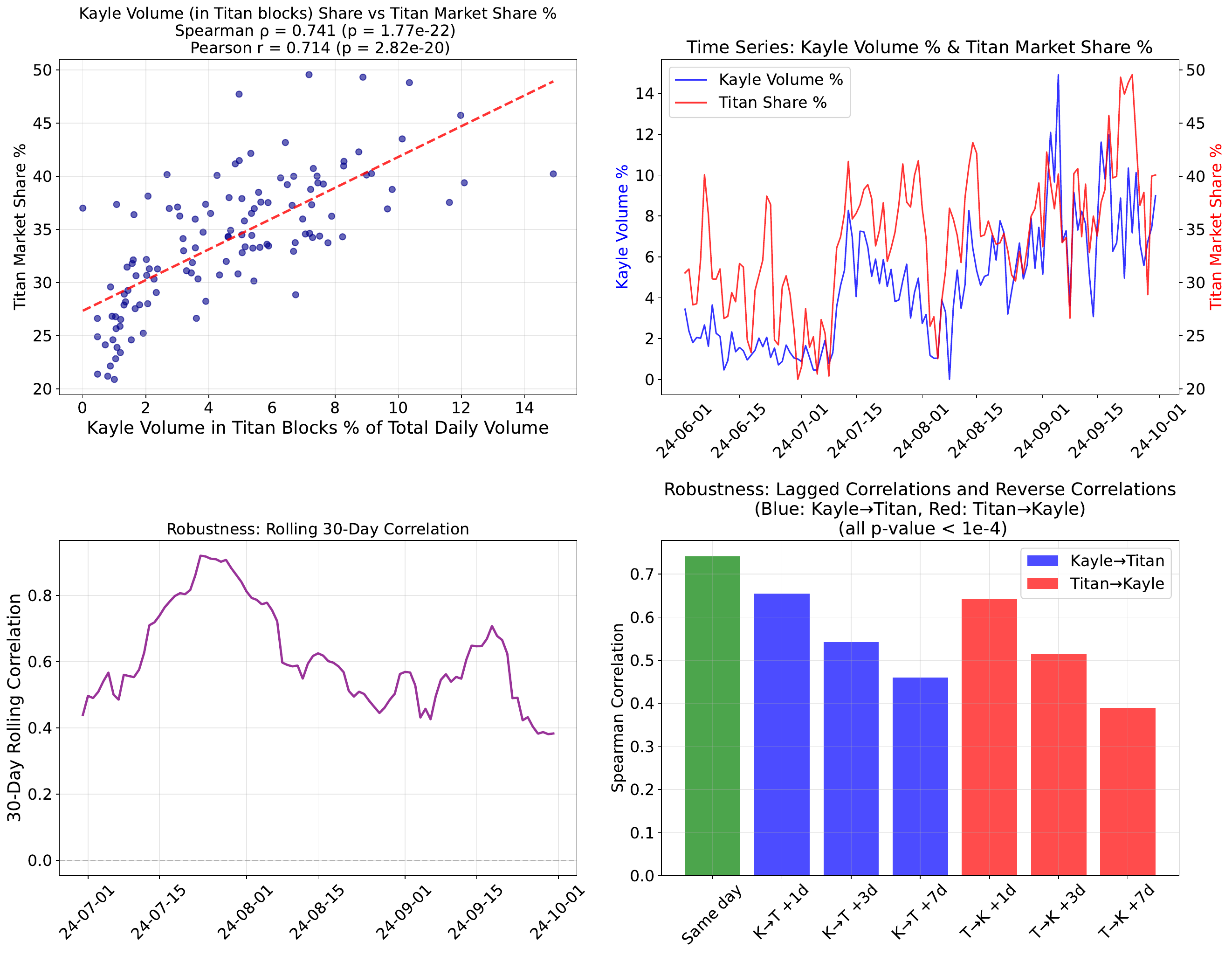} %
    \caption{Correlation analysis of \texttt{Kayle}'s daily volume share in \texttt{Titan}'s blocks (as a percentage of total daily CEX-DEX volume) and \texttt{Titan}'s daily market share. Using percentage metrics effectively controls for overall fluctuations in daily total CEX-DEX volume.}
    \label{fig:kayle_titan_corr}
\end{figure}

We find a strong and robust mutual correlation between \texttt{Kayle}'s daily volume share in \texttt{Titan's} blocks and \texttt{Titan}'s daily market share during this period. Specifically, \Cref{fig:kayle_titan_corr} (top-left panel) shows a significant contemporaneous correlation (Spearman’s $\rho = 0.74$, $p < 0.0001$), indicating that higher proportion of \texttt{Kayle}'s trade volume included in \texttt{Titan}'s blocks strongly coincide with higher share for \texttt{Titan} on the same day. The aligned trajectories shown in the time series comparison (top-right panel) further confirm their synchronized behavior. Robustness checks using rolling 30-day correlations confirm the consistency and significance of this relationship across the period analyzed (bottom-left panel).

We further investigate the lagged and reverse correlations at intervals of 1, 3, and 7 days (bottom-right panel). The results indicate a notable predictive component: higher daily volume share from \texttt{Kayle} to \texttt{Titan} positively correlates with increased market share for \texttt{Titan} in subsequent days. This effect is most pronounced at a 1-day lag (Spearman’s $\rho = 0.655,  p < 0.0001$), gradually decreasing over one week. One plausible explanation is that larger \texttt{Kayle} flow in one day brings larger tip revenue for Titan and enlarges \texttt{Titan}'s builder surplus, enhancing \texttt{Titan}'s bidding power in subsequent days. Moreover, we identify a reverse effect of similar magnitude, suggesting a strong feedback loop: higher market share for \texttt{Titan} subsequently increases \texttt{Kayle}'s market capture and flow directed back to \texttt{Titan}. Indeed, we observe a substantial surge in CEX-DEX volume share included in \texttt{Titan}'s blocks since June 2024 (cf. \Cref{appendix:mev_compare}, \Cref{fig:volume_hhi_builder}). 

This dynamic played a major role in accelerating centralization within the searcher market during this period and beyond (cf. \Cref{appendix:mev_compare}, \Cref{fig:volume_hhi}), and had significant repercussions for competing builders, particularly \texttt{rsync}. As a result of \texttt{Titan}'s expansion, in \Cref{fig:wm_margin_vs_share}, we observe \texttt{rsync}'s market share fell below 5\% by September 2024, effectively retreating from competition in the builder market. At the same time, \texttt{Wintermute}'s median profit margin quickly rebounded, indicating a strategic shift away from aggressively supporting \texttt{rsync} towards maximizing searcher profitability. Interestingly, \texttt{beaverbuild}'s market position and \texttt{SCP}'s profitability remained robust despite similar competitive pressures, likely buffered by the timely emergence of another exclusive searcher, \texttt{Senna} (See \Cref{appendix:impact_cexdex} for results not presented here).

\subparagraph*{Interpretation and Implications.}
Combined with the analysis presented in \Cref{sec:searcher_profitability}, these findings provide strong quantitative evidence supporting a mutually reinforcing exclusivity partnership between \texttt{Kayle} and \texttt{Titan}. Definitive evidence of exclusivity would require relay-level data inspection. Indeed, preliminary evidence provided by \cite{canidio2025immutableethereum}, where the author analyzed both winning and non-winning block contents submitted to the Agnostic Relay over a 38-slot window from slot 10534387 to 10534424, confirms that \texttt{Kayle}'s trades were exclusively seen in \texttt{Titan}'s blocks. 

Searchers with exclusive builder partnerships, despite paying higher per-trade costs and retaining lower margins, likely gain prioritized inclusion, potentially capturing a greater share of arbitrage opportunities. Builders, in turn, leverage exclusive high-value flow to strengthen their competitive positions in the block auctions, further empowering their searcher partners. While explicit off-chain rebates and agreements remain opaque, the observed patterns—higher payments, lower searcher margins, and increased market share—strongly indicate such arrangements. 

Collectively, these results underscore the significant centralizing pressure exerted by exclusive partnerships and vertical integration within the Ethereum searcher and builder markets.

\section{Integrated Builder Profits Correction}
\Cref{sec:searcher_profitability} shows \texttt{SCP} and \texttt{Wintermute} only retain 10-15\% of their arbitrage revenue and transfer nearly 90\% to their integrated builders. However, \texttt{beaverbuild} and \texttt{rsync} are unlikely to pass this entire amount as their bid to the proposer; part of the searcher payments may be pocketed as builder surplus upon winning the block. Previous research measures only the builders' on-chain profits, and therefore omits any margin retained at their integrated searcher level \cite{whowinsandwhy, yang2024decentralization}. In this section, we correct this oversight by combining builders' on-chain profits with the searcher-level PnL estimates. This reconciliation provides deeper insight into the profitability of integrated searcher-builder entities.

We collect the Ultra Sound bid adjustments data \cite{bidadjustment} and MEV-Boost payloads data \cite{relayscan}, and analyze builder profitability from block 17866488 to block 21998438 spanning the period from August 8, 2023, to March 8, 2025. \Cref{tab:builder_profits} summarizes the profitability of \texttt{beaverbuild}, \texttt{rsync}, and \texttt{Titan}, considering integrated searcher profits (if any). ``Builder Margin" presents the cumulative mean of profit margin if only builder profit for each block is considered, while ``Aggregated Profit Margin" presents the cumulative mean of profit margin if considering both builder profit and searcher PnL for each block. Note that the ``Total Searcher PnL" presented in this table only considers the searcher's trades in the respective integrated builder's blocks, excluding those in other blocks, to focus solely on their performance as builders. Recall, given that the searcher PnL is an upper-bound estimtate, the aggretated profit and margin are also upper-bound estimates. We detail the calculation method in \Cref{appendix:integrated_profit_cal}.

\begin{table}[t]
\centering
\caption{Builder profitability summary.}
\label{tab:builder_profits}

\resizebox{\textwidth}{!}{%
\begin{tabular}{lcccccccc}
\toprule
\textbf{Builder-Searcher} & \textbf{Total} & \textbf{Total} & \textbf{Total} &  \textbf{Builder} & \textbf{Total} & \textbf{Aggregated} & \textbf{Agg. Profit} \\
& \textbf{Blocks} & \textbf{Bid Value} & \textbf{Builder Profit} & \textbf{Margin} & \textbf{Searcher PnL} & \textbf{Profit} & \textbf{Margin}\\
& [\#] & [\#] & [USD] & & [USD] & [USD] & \\
\midrule
\midrule
beaverbuild - SCP & 1,603,780 & 402.87M & 66.68M & 6.69\% &11.05M & 77.73M & 7.92\% \\
rsync - Wintermute & 543,564 & 156.78M & 9.65M & -2.24\% & 14.77M & 24.42M & 27.06\% \\
Titan & 1,206,806 & 266.80M & 92.44M & 5.85\% & N/A & 92.44M & 5.85\% \\
\bottomrule
\end{tabular}
}
\end{table}

Results show that \texttt{beaverbuild}'s builder profit is roughly 6 times their searcher profit, and the profit margin only increases by about 1\% by incorporating their searcher profit. This confirms that while \texttt{SCP} transfers nearly 90\% of their revenue to their builder, a sizable share never reaches the proposer and is instead retained by their builder.

Since \texttt{rsync} is less competitive as a builder, their builder profit is lower even with a slightly negative builder margin. However, if incorporating their searcher profit, we find higher profit retained at their searcher level than \texttt{SCP}, and the blocks they win are exceptionally lucrative with a 27.06\% profit margin. This indicates that they compete selectively for the most profitable blocks and mainly profit as a searcher instead of as a builder.

Finally, \texttt{Titan}, despite not being vertically integrated with any searcher, earns the highest total profit. This is strong evidence that they have exclusive deals with high-value orderflow providers (e.g., CEX-DEX searcher \texttt{Kayle} and Banana Gun Telegram Bot \cite{whowinsandwhy,yang2024decentralization}), and rebate part of their builder profit back to them.

We further examine builder subsidization behavior in MEV-Boost auctions. Prior studies measured subsidies with builders’ on-chain profits only, suggesting builders subsidize blocks to maintain market presence, potentially offset by exclusive orderflow provider profits \cite{whowinsandwhy, yang2024decentralization, titanfrontierpost}. We refine this measure by including integrated searcher profits for \texttt{beaverbuild} and \texttt{rsync}. We define a block as ``subsidized" only if both the builder's on-chain profit and aggregated profit are negative. If the builder profit is positive, we do not classify the block as subsidized, even if the aggregated profit is negative. \Cref{tab:builder_subsidy} summarizes the results. After accounting for integrated searcher profits, the apparent subsidy burden for both builders shrinks by roughly 15\% fewer blocks and slightly lower outflows. This suggests that the majority of block subsidizing behavior by \texttt{beaverbuild} and \texttt{rsync} are not directly driven by immediate coverage from their searcher profits in that block.

\begin{table}[t]
\centering
\caption{Builder subsidy summary.}
\label{tab:builder_subsidy}

\resizebox{\textwidth}{!}{%
\begin{tabular}{lccccccc}
\toprule
\textbf{Builder} & \textbf{Subsidized Blocks} & \textbf{Subsidized Blocks} & \textbf{Builder Subsidy} & \textbf{Builder Subsidy} \\
& Before Correction & After Correction & Before Correction & After Correction  \\
& [\#] & [\#] & [USD] & [USD] \\
\midrule
\midrule
beaverbuild & 498,388 & 426,335 & -2.31M & -2.13M \\
rsync & 179,925 & 153,713 & -3.89M & -3.77M \\
\bottomrule
\end{tabular}
}
\end{table}

\section{Discussion}

\subparagraph*{Limitations of Our Approach.}
Although our heuristics expand the coverage of identified CEX-DEX arbitrage trades compared to prior methodologies \cite{whowinsandwhy,nonatomic}, certain trades may remain undetected, resulting in potentially underestimating searcher revenues and PnL. Given the resources available, our methodology represents the best feasible solution at present. 

Our revenue estimation method—based on empirically derived optimal execution horizons across historical trades for each searcher—introduces approximation errors by assuming uniform hedge execution timing across all trades for each searcher and neglecting the price impact of hedging, particularly relevant for low-liquidity tokens. Despite these limitations, our approach is grounded in microstructure literature and captures dominant execution patterns for each searcher, offering consistent upper-bound estimates that allow meaningful comparisons among searchers.

Finally, we do not account for potential off-chain rebates or Orderflow Auction refunds, which might underestimate actual searcher PnL and overstate builder profits, especially for exclusive searcher-builder partnerships.

\subparagraph*{Centralization in the searcher market.}
Previous studies have discussed the impact of vertical integration, exclusive orderflow, and builder market centralization, highlighting concerns such as censorship and proposer loss \cite{whowinsandwhy,yang2024decentralization,nonatomic}. Our analysis pushes the centralization narrative one step upstream. Critically, we are the first to provide robust quantitative estimates of the value extracted through CEX-DEX arbitrages and shed empirical light on profit-sharing dynamics previously understood only qualitatively. 

It appears that, just as arbitrages in traditional finance have been concentrated among a few specialist firms \cite{limitsofarbitrage}—often because they purchase retail order flow (i.e., payment for order flow) or control clearing licenses \cite{duffie2011does,hu2022competition}—arbitrages at the TradFi and DeFi interface are similarly centralized by searcher-builder orderflow arrangements and vertical integration.


Specifically, the two largest CEX-DEX searchers by volume and revenue—\texttt{SCP} and \texttt{Wintermute}—\\are each vertically integrated with leading block builders, \texttt{beaverbuild} and \texttt{rsync}, respectively. This vertical integration grants them a decisive strategic advantage in securing block inclusion for their arbitrage transactions. 
The third best searcher \texttt{Kayle} emerged as a significant market participant beginning around June 2024, likely following an exclusive arrangement with the second largest builder \texttt{Titan}. Collectively, these three searchers dominate approximately three-quarters of the total arbitrage volume and value extracted, reinforcing further centralization in both the builder and searcher markets.

In contrast, smaller searchers, despite maintaining higher margins per trade, have seen their overall volume and revenue share diminish over time. Given the already high entry barriers to CEX-DEX arbitrage, this further centralization significantly elevates barriers to entry, discouraging market participation by smaller entities and fostering economies of scale.

Recent research also highlights centralization concerns in DEX solver markets, where notably, \texttt{Wintermute} and \texttt{SCP} similarly dominate the solver markets on prominent platforms like CoWSwap, Uniswap X, and 1inchFusion \cite{yuminaga2025executionwelfaresolverbaseddexes}. Taken together, these observations suggest that the entire MEV supply chain—from solver to searcher to builder—is becoming increasingly consolidated around a few dominant market participants.

\subparagraph*{Mitigations and Challenges.} One proposed mitigation discussed in prior works \cite{lvr,lvr2,nonatomic,fritsch2024measuringarbitragelossesprofitability} involves shorter block times (e.g., EIP-7782 \cite{7782}) to reduce price movements between blocks and limit overall arbitrage opportunities. While this may lower the absolute value extractable from CEX-DEX arbitrages, it is less clear that shorter block times would meaningfully shift the market structure, given the sustained advantage held by dominant, vertically integrated players.

Mechanisms designed to promote decentralization in the downstream builder market, such as Orderflow Auctions \cite{mevblocker, mevshare} and 
BuilderNet \cite{buildernet}, could enhance decentralization at the builder level. However, their impact on the upstream searcher market remains uncertain and warrants further study.

Lastly, mechanisms aimed at capturing, redistributing, or burning MEV at the application \cite{zhang2024rediswapmevredistributionmechanism,angstrom} or protocol level \cite{mevsmooth,mevburn,burian2024mevcapturedecentralizationexecution} could reduce centralization pressures. Nevertheless, these approaches rely on accurate MEV oracles \cite{yang2024decentralization}, which remain an open technical challenge.

\section{Conclusion}
In this work, we shed light on value extraction, profitability, and market dynamics associated with CEX-DEX arbitrage on Ethereum. Our findings highlight increasing centralization as three major searchers affiliated with top builders dominate CEX-DEX arbitrage opportunities, compressing smaller participants despite their higher per-trade margins. Exclusive searcher-builder arrangements further amplify these centralization pressures both downstream and upstream of the MEV supply chain, underscoring critical economic and strategic considerations for Ethereum’s decentralization guarantees.

\section*{Acknowledgments}
The work is partially supported by The Latest in DeFi Research (TLDR) Fellowship 2024 funded by Uniswap Foundation. We appreciate Gaussian Process, Kevin Pang, Bill Zhang, Tom Zhao, and Romain Butteaud for helpful discussions on the methodology. We thank Burak \"Oz and Max Resnick for insightful comments on this work. We also thank Alexander Tesfamichael from Ultra Sound Money and \v{Z}an Knafelc from Eden Network for their help with data collection.

\bibliography{bib}
\newpage    
\appendix
\section{Data Collection}
\subsection{Searcher labels and contract addresses}
\label{appendix:searcher_label}
The searcher contract addresses and their corresponding labels presented in Tables~\ref{tab:searcher_addresses} and \ref{tab:searcher_addresses_2} are primarily derived from prior datasets and labeling efforts \cite{botlist,loldash}. We further refine and update this list by tracing historical contract activity and ownership through Arkham \cite{arkham} to identify newer addresses associated with labeled searchers. Despite our best efforts to ensure completeness, it is possible that some relevant contract addresses belonging to the listed entities may have been omitted. 

\begin{table}[h]
\centering
\caption{Searcher labels and contract addresses.}
\label{tab:searcher_addresses}
\renewcommand{\arraystretch}{1.0}
\begin{tabular}{|l|l|}
\hline
\textbf{Searcher Label} & \textbf{Contract Address} \\
\hline
Wintermute & 0x0000006daea1723962647b7e189d311d757fb793 \\
 & 0x00000000ae347930bd1e7b0f35588b92280f9e75 \\
 & 0x0087bb802d9c0e343f00510000729031ce00bf27 \\
 & 0xaf0b0000f0210d0f421f0009c72406703b50506b \\
 & 0x280027dd00ee0050d3f9d168efd6b40090009246 \\
 & 0x51C72848c68a965f66FA7a88855F9f7784502a7F \\
 & 0x3b55732f6d3997a7d44a041b8496e1a60712a35f \\
 & 0xec6fc9be2d5e505b40a2df8b0622cd25333823db \\\hline
SCP & 0xa69babef1ca67a37ffaf7a485dfff3382056e78c \\
 & 0x56178a0d5f301baf6cf3e1cd53d9863437345bf9 \\
 & 0xa57bd00134b2850b2a1c55860c9e9ea100fdd6cf \\
 & 0x4Cb18386e5d1F34dC6EEA834bf3534A970a3f8e7 \\
 & 0x5050e08626c499411B5D0E0b5AF0E83d3fD82EDF \\\hline
Kayle & 0xbc2c6cd5013585ac720160efcb1feced30837177 \\
 & 0x593AC83229D3099C57cC721540c4d535a7fbBFd6 \\
 & 0x9def7cde171841a9f0724124ca0b01a622d749e4 \\
 & 0x2e5ca1238654ad4adc4c60b34664b656da17d4da \\
 & 0xfbd4cdb413e45a52e2c8312f670e9ce67e794c37 \\
 & 0x68d3A973E7272EB388022a5C6518d9b2a2e66fBf \\\hline
Galio & 0xe8cfad4c75a5e1caf939fd80afcf837dde340a69 \\
 & 0x1bf621aa9cee3f6154881c25041bb39aed4ca7cc \\
 & 0x5dc62cea20b0e7c3607adcc61a885ff9369dbc60 \\\hline
Shen & 0x6f1cdbbb4d53d226cf4b917bf768b94acbab6168 \\
 & 0xbfef411d9ae30c5b471d529c838f1abb7b65d67f \\
 & 0xeff6cb8b614999d130e537751ee99724d01aa167 \\\hline
Taric & 0x2D722C96f79d149dD21e9eF36F93fc12906CE9f8 \\
 & 0x767C8bB1574BEE5D4FE35E27e0003c89D43C5121 \\\hline
Lucian & 0x98c3d3183c4b8a650614ad179a1a98be0a8d6b8e \\\hline
Thresh & 0xef97b8a6cbb72feeccf5bc5e897078e9e53ee0a4 \\
 & 0x817648d73fd85c802cfde8a29eba9f68b783ca60 \\\hline
Ahri & 0xd7f3fbe8c72a961a5515203eada59750437fa762 \\
 & 0x2deae6ce94d65ac1de19a1fc4bb160c4e02c92ef \\\hline
\end{tabular}
\end{table}

\begin{table}[H]
\centering
\caption{Searcher labels and contract addresses.}
\label{tab:searcher_addresses_2}
\begin{tabular}{|l|l|}
\hline
\textbf{Searcher Label} & \textbf{Contract Address} \\
\hline
Riven & 0x053f661abf26d086194540f20f312e0d90a61302 \\
 & 0x752e87b5f1397e171D5383cec3D4C51A8D3C114B \\
 & 0xaAFb85ad4a412dd8adC49611496a7695A22f4aeb \\
 & 0x32801aB1957Aaad1c65289B51603373802B4e8BB \\
 & 0xD198fBE60C49D4789525fC54567447518C7D2a11 \\
 & 0xb6613cc55866e282638006455390207c1d485be9 \\
 & 0xbeb5fd030ffb0fbc95d68113c1c796eff65526d7 \\
 & 0x4c405bc9dc26435a48fe6a637b6b08eb78b9da5 \\\hline
Darius & 0x28e261390adaa654f29dbe268109baf06e9b4cc4 \\\hline
Karma & 0x807cf9a772d5a3f9cefbc1192e939d62f0d9bd38 \\\hline
Bard & 0x360e051a25ca6decd2f0e91ea4c179a96c0e565e \\\hline
Senna & 0xd4bc53434c5e12cb41381a556c3c47e1a86e80e3 \\
 & 0x203e2349666c08a538266afaed434b388e01a657 \\\hline
Maokai & 0x000000000dfde7deaf24138722987c9a6991e2d4 \\\hline
Zed & 0xfbeedcfe378866dab6abbafd8b2986f5c1768737 \\
 & 0x99999999D116Ffa7D76590De2f427d8e15AEb0b8 \\\hline
Jinx & 0xd249942f6d417cbfdcb792b1229353b66c790726 \\\hline
Tristana & 0x4a137fd5e7a256ef08a7de531a17d0be0cc7b6b6 \\\hline
Graves & 0x33565e5e5bb57cce2c606e16b99f435a80adc674 \\
 & 0xc22d4ca2362c78b0f7e7c370484f9e191eb656cb \\\hline
Lux & 0x966d8c1f61bae657d577077abfbd7d896c09e242 \\
 & 0xa0b18CdF5F395D98c061B753d8dedB28c7Aee450 \\
 & 0xc6feCDF760Af24095cDEd954dE7d81aB49f8Bae1 \\
 & 0x12ff0e28318e53a6f91d42cf607963076af6c03f \\
 & 0x15dc6e110423d97339105e6c377ce08191527e95 \\
 & 0x6d1d1ebe7da598194293784252659e862d55b52c \\\hline
Caitlyn & 0x70c66f3ce5a5387a70e2773d054eff572525c6f4\\ 
 & 0x05f016765c6c601fd05a10dba1abe21a04f924a5 \\\hline
Akali & 0xcfd4176f7975c70f800d87aeaca316270521595a \\
 & 0x9EA3cda5c2Adf0370454b9Ee28786a068227b1a4 \\
 & 0x70e86223507724bf2c51fe3ac2cc78c67bfad366 \\
 & 0x73a8a6f5d9762ea5f1de193ec19cdf476c7e86b1 \\\hline
Poppy & 0xe6ae75be7c9317af842b8f2c2cd6dc7f49f17184\\\hline 
\end{tabular}
\end{table}

\subsection{Other Data Sources}
In \Cref{tab:data_sources}, we summarize the other data sources used in the paper.

\begin{table}[H]
\centering
\caption{Data sources.}
\label{tab:data_sources}

\begin{tabular}{lr}
\toprule
\textbf{Data} & \textbf{Source} \\
\midrule
\midrule
DEX trades & Dune dex.trades table \cite{dunedextrades} \\
Binance Spot historical quotes & Tardis.dev \cite{tardisdata}\\
Binance-listed ERC-20 token contract addresses & Etherscan \cite{etherscan}, CoinMarketCap \cite{coinmarketcap}\\
MEV-Boost bids and payloads & relayscan.io \cite{relayscan}\\
Ultra Sound bid adjustments & Ultra Sound relay API \cite{bidadjustment}\\
\bottomrule
\end{tabular}

\end{table}

\section{Searcher gross return patterns}
\label{appendix:revenue_dist}
We here present the gross return distribution of the remaining 17 searchers in \Cref{fig:gross_returns_10s_14searchers}, and summarize the optimal execution horizon and pattern category of each searcher in \Cref{tab:pattern_sum}.

\begin{table}[h]
\centering
\caption{Summary of searchers' optimal execution horizons and patterns.}
\label{tab:pattern_sum}

\renewcommand{\arraystretch}{1.0}
\begin{tabular}{|l|l|l|}
\hline
\textbf{Searcher Label} & \textbf{Optimal Execution Horizon} &\textbf{Pattern} \\
\hline
Wintermute  & 1.5s & Pattern 1 \\\hline
SCP & 0.5s & Pattern 1 \\\hline
Kayle & 1.0s & Pattern 1 \\\hline
Galio & 0.5s & Pattern 1 \\\hline
Shen & 1.5s & Pattern 1 \\\hline
Taric & 1.0s & Pattern 2 \\\hline
Lucian & 1.0s & Pattern 2 \\\hline
Riven & 1.5s & Pattern 1 \\\hline
Thresh & 2.0s & Pattern 1 \\\hline
Ahri & 2.0s & Pattern 1 \\\hline
Darius & 0.5s & Pattern 1 \\\hline
Karma & 2.0s & Pattern 2 \\\hline
Bard & N/A & Pattern 3 \\\hline
Senna & 0.5s & Pattern 1 \\\hline
Maokai & 1.0s & Pattern 2 \\\hline
Zed & 1.5s & Pattern 2 \\\hline
Jinx & N/A & Pattern 3 \\\hline
Tristana & N/A & Pattern 3 \\\hline
Graves & 1.5s & Pattern 1 \\\hline
Lux & N/A & Pattern 3 \\\hline
Caitlyn & 1.0s & Pattern 1 \\\hline
Akali & 1.0s & Pattern 2 \\\hline
Poppy & 1.5s & Pattern 2 \\\hline
\end{tabular}
\end{table}

\begin{figure}[H]
  \centering
  \captionsetup{justification=centering}
  \begin{subfigure}[b]{0.32\linewidth}
    \centering
    \includegraphics[width=\linewidth]{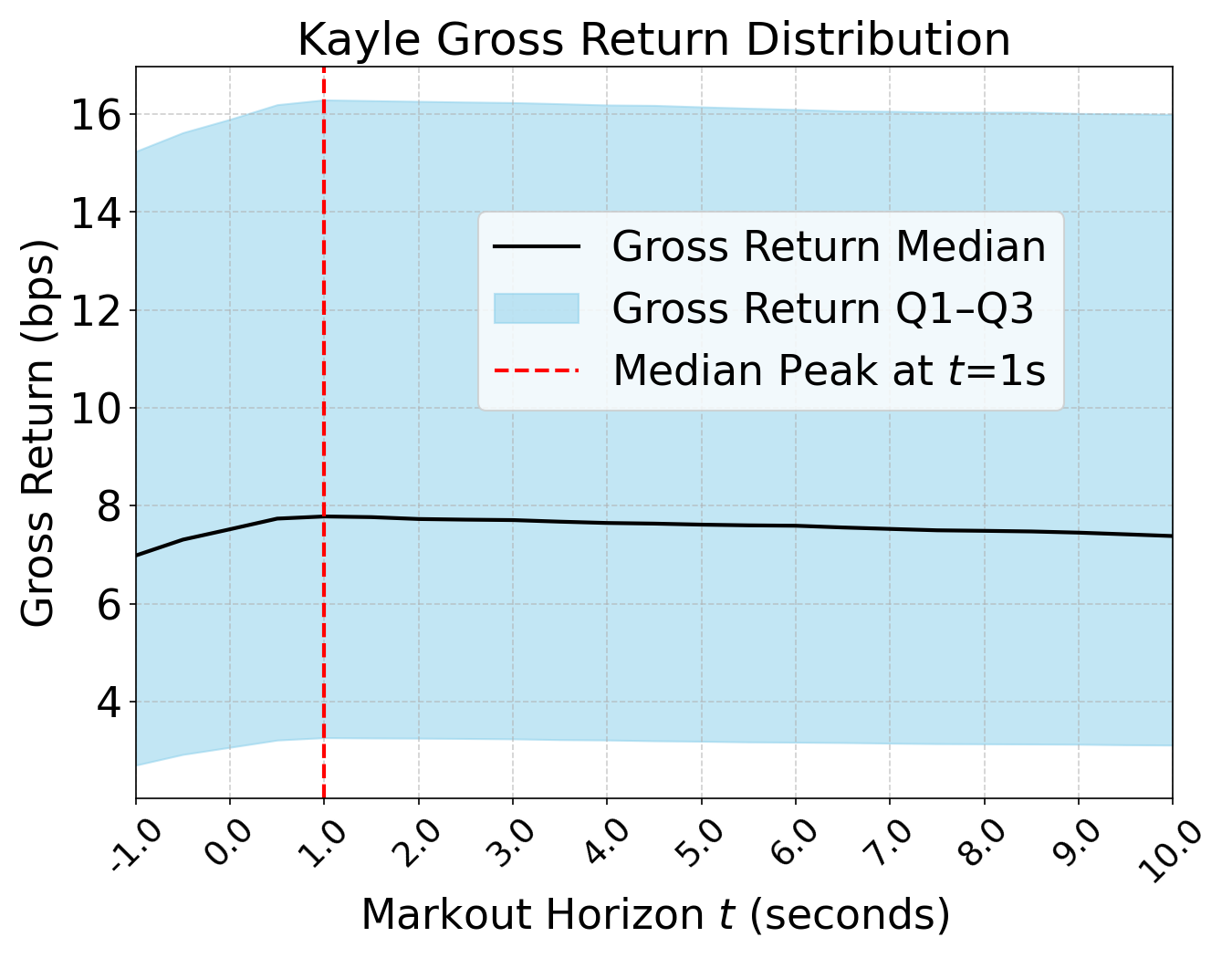}
    \caption{\texttt{Kayle} (Pattern 1)}
    \label{fig:kayle_return}
  \end{subfigure}\hfill
  \begin{subfigure}[b]{0.32\linewidth}
    \centering
    \includegraphics[width=\linewidth]{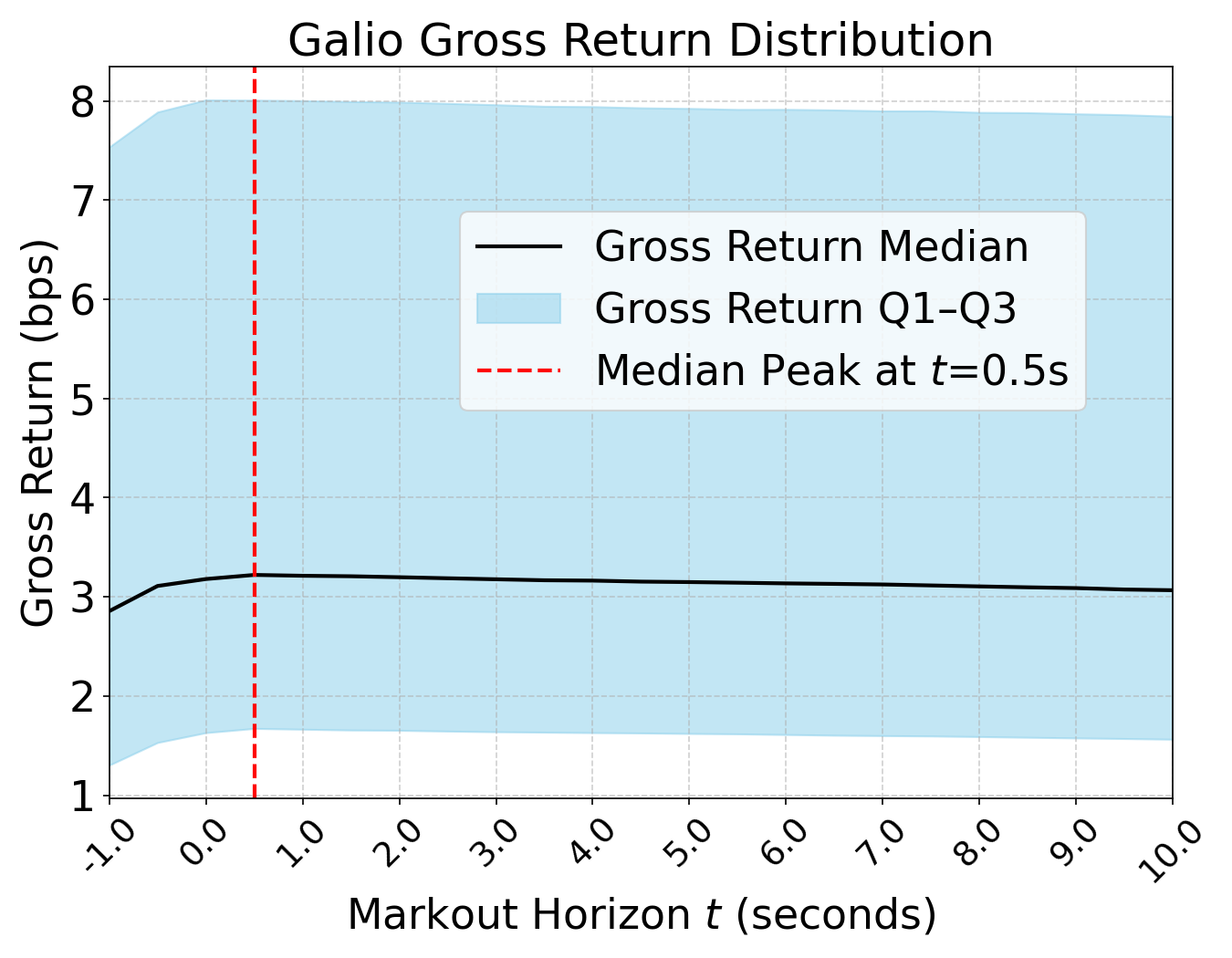}
    \caption{\texttt{Galio} (Pattern 1)}
    \label{fig:galio_return}
  \end{subfigure}\hfill
  \begin{subfigure}[b]{0.32\linewidth}
    \centering
    \includegraphics[width=\linewidth]{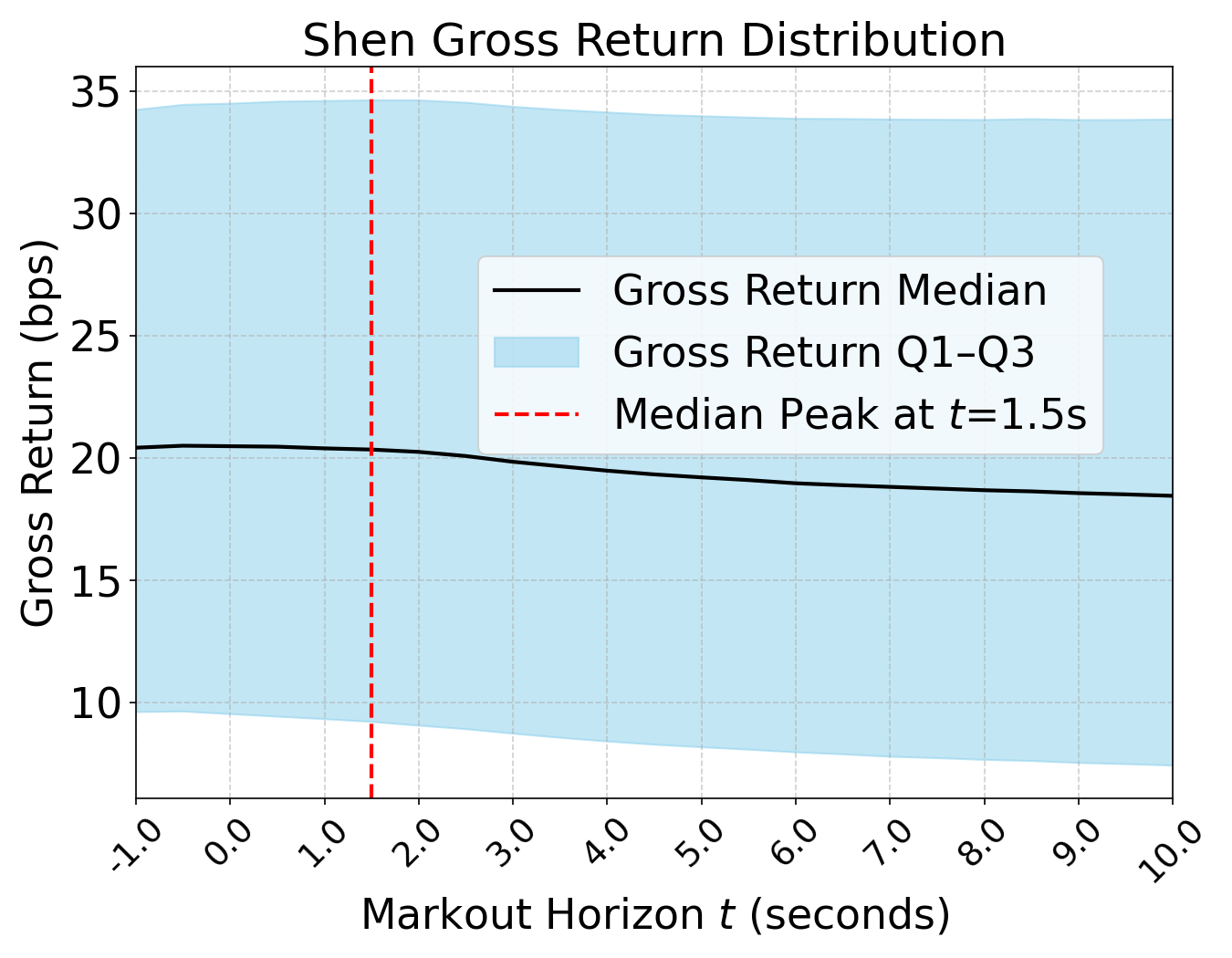}
    \caption{\texttt{Shen} (Pattern 1)}
    \label{fig:shen_return}
  \end{subfigure}

  \vspace{1em}
  
  \begin{subfigure}[b]{0.32\linewidth}
    \centering
    \includegraphics[width=\linewidth]{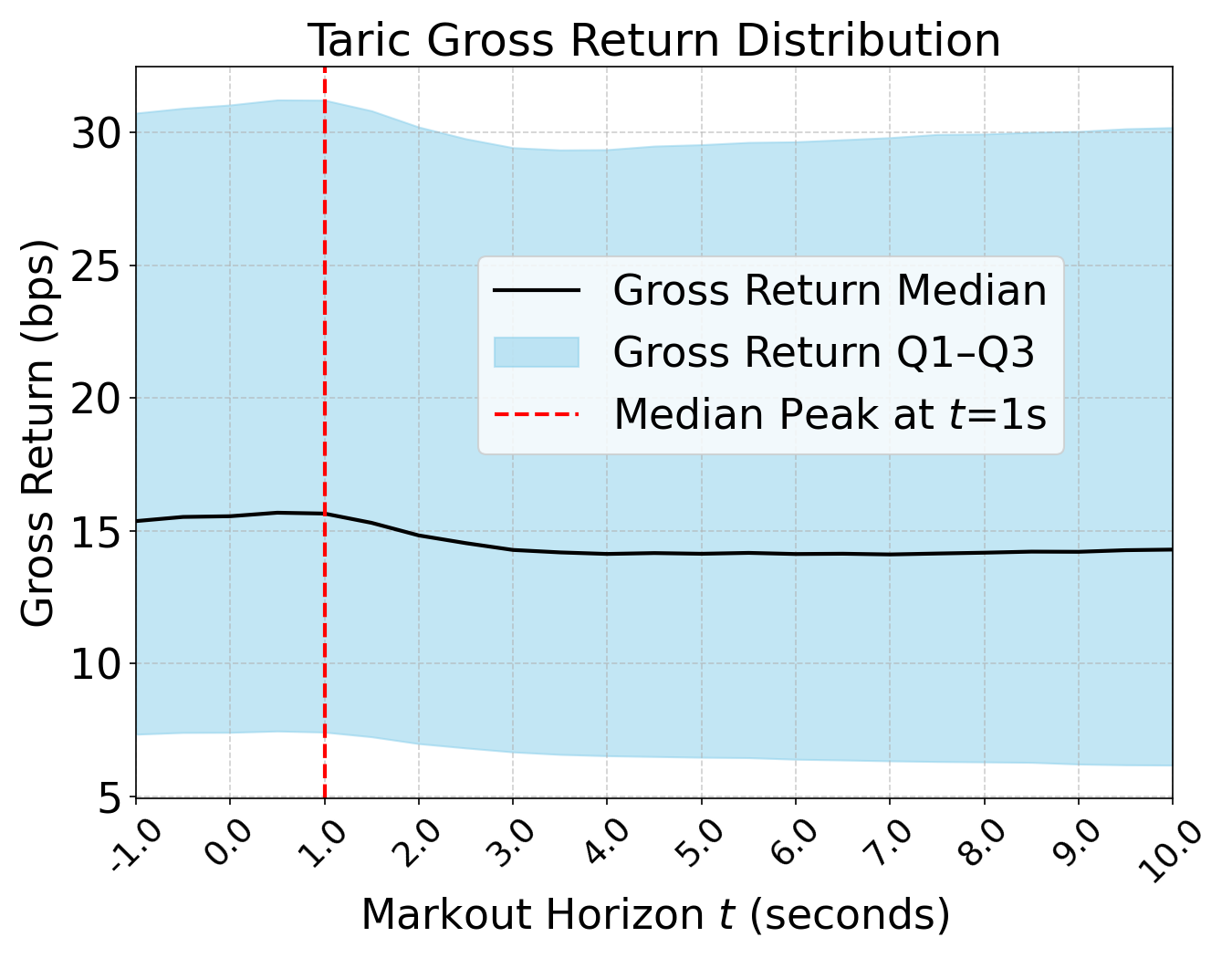}
    \caption{\texttt{Taric} (Pattern 2)}
    \label{fig:taric_return}
  \end{subfigure}\hfill
  \begin{subfigure}[b]{0.32\linewidth}
    \centering
    \includegraphics[width=\linewidth]{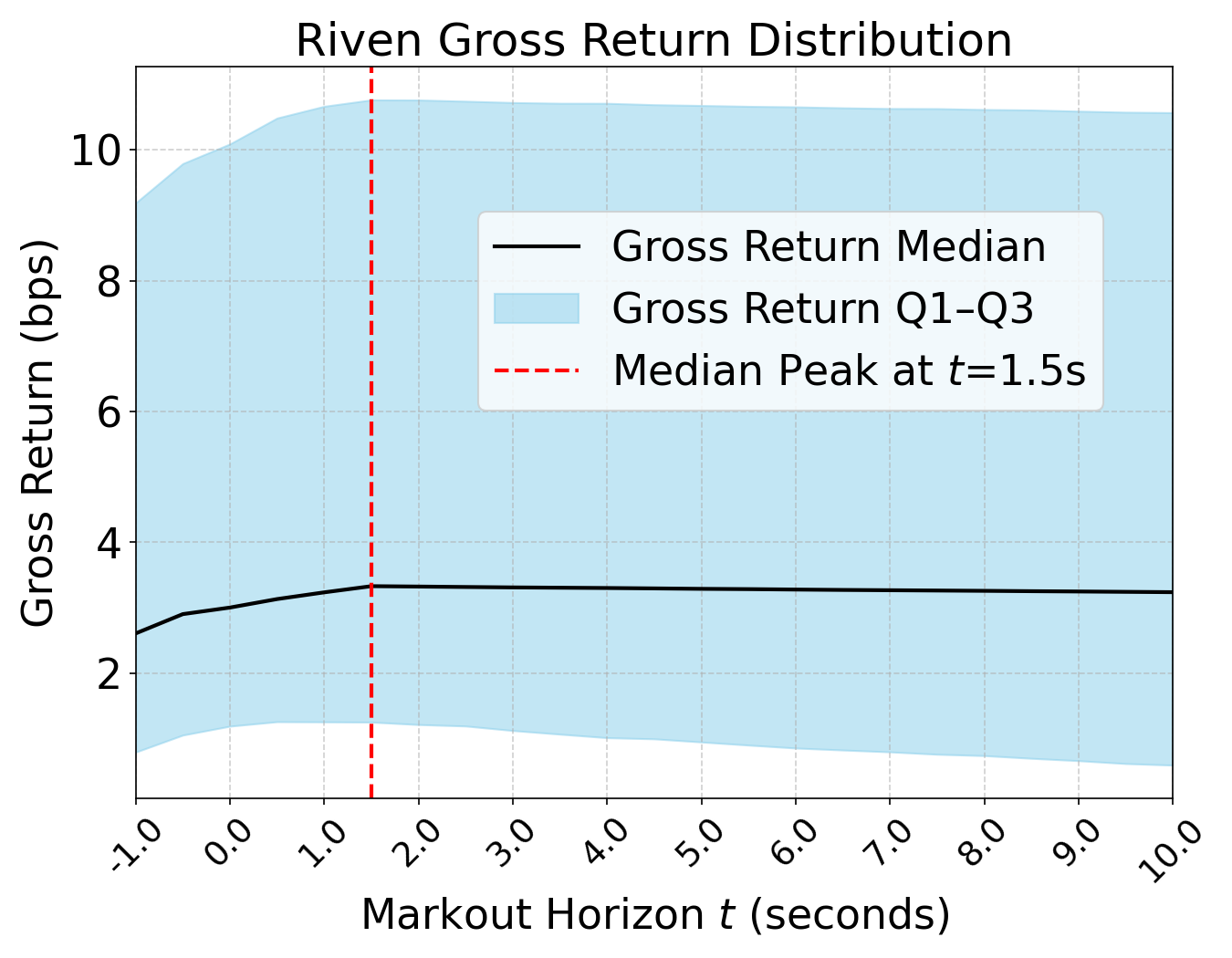}
    \caption{\texttt{Riven} (Pattern 1)}
    \label{fig:riven_return}
  \end{subfigure}\hfill
  \begin{subfigure}[b]{0.32\linewidth}
    \centering
    \includegraphics[width=\linewidth]{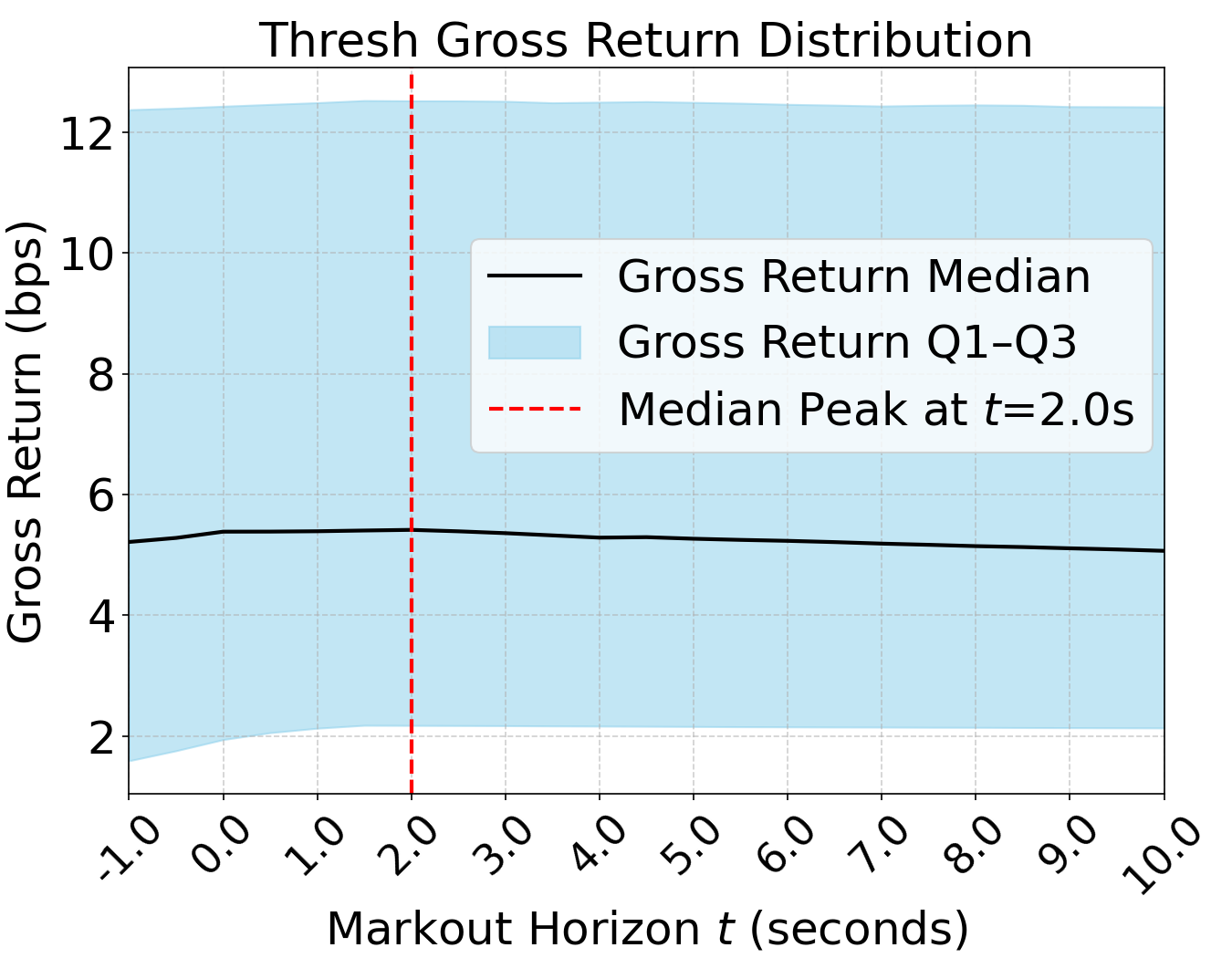}
    \caption{\texttt{Thresh} (Pattern 1)}
    \label{fig:thresh_return}
  \end{subfigure}

  \vspace{1em}

  \begin{subfigure}[b]{0.32\linewidth}
    \centering
    \includegraphics[width=\linewidth]{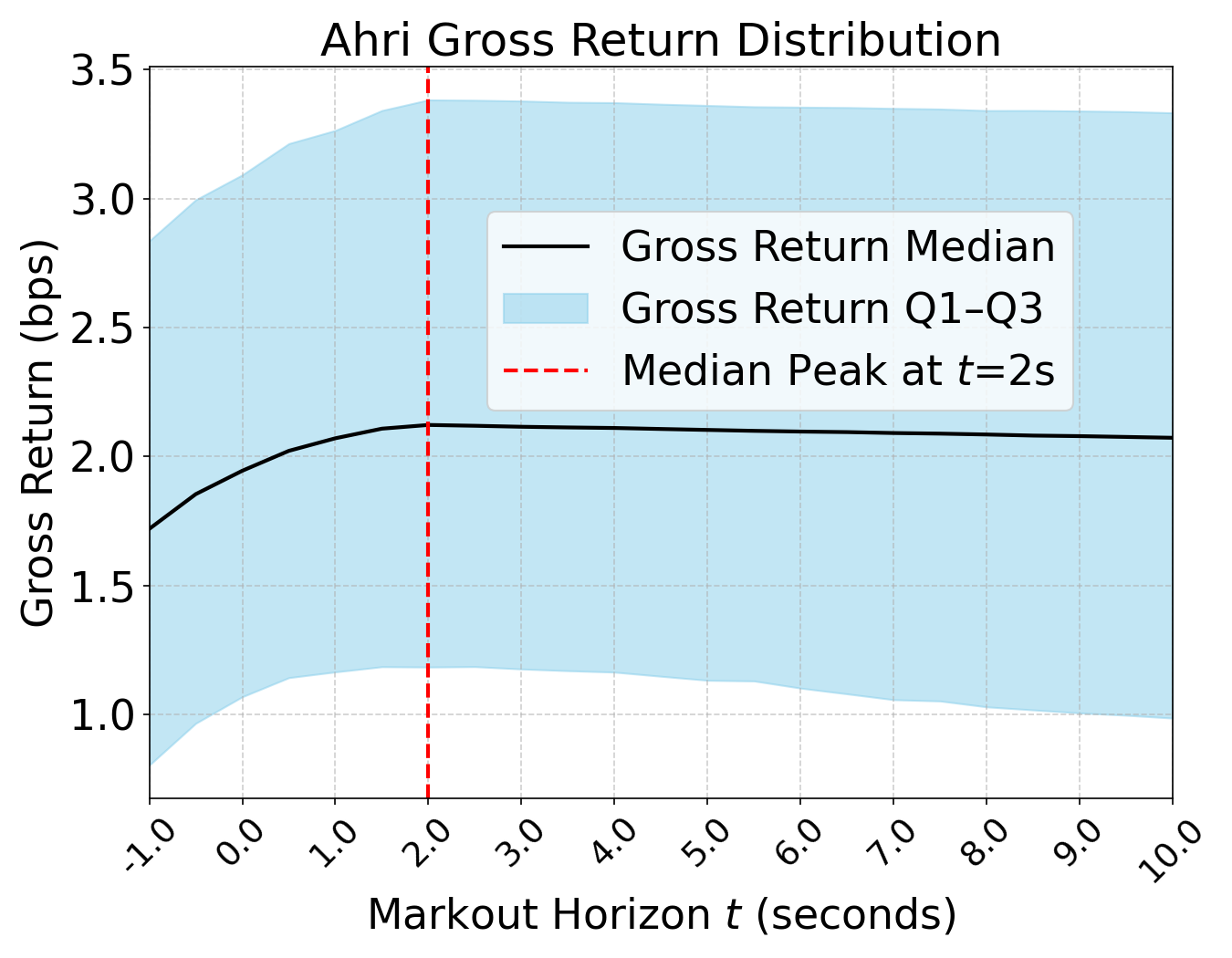}
    \caption{\texttt{Ahri} (Pattern 1)}
    \label{fig:ahri_return}
  \end{subfigure}\hfill
  \begin{subfigure}[b]{0.32\linewidth}
    \centering
    \includegraphics[width=\linewidth]{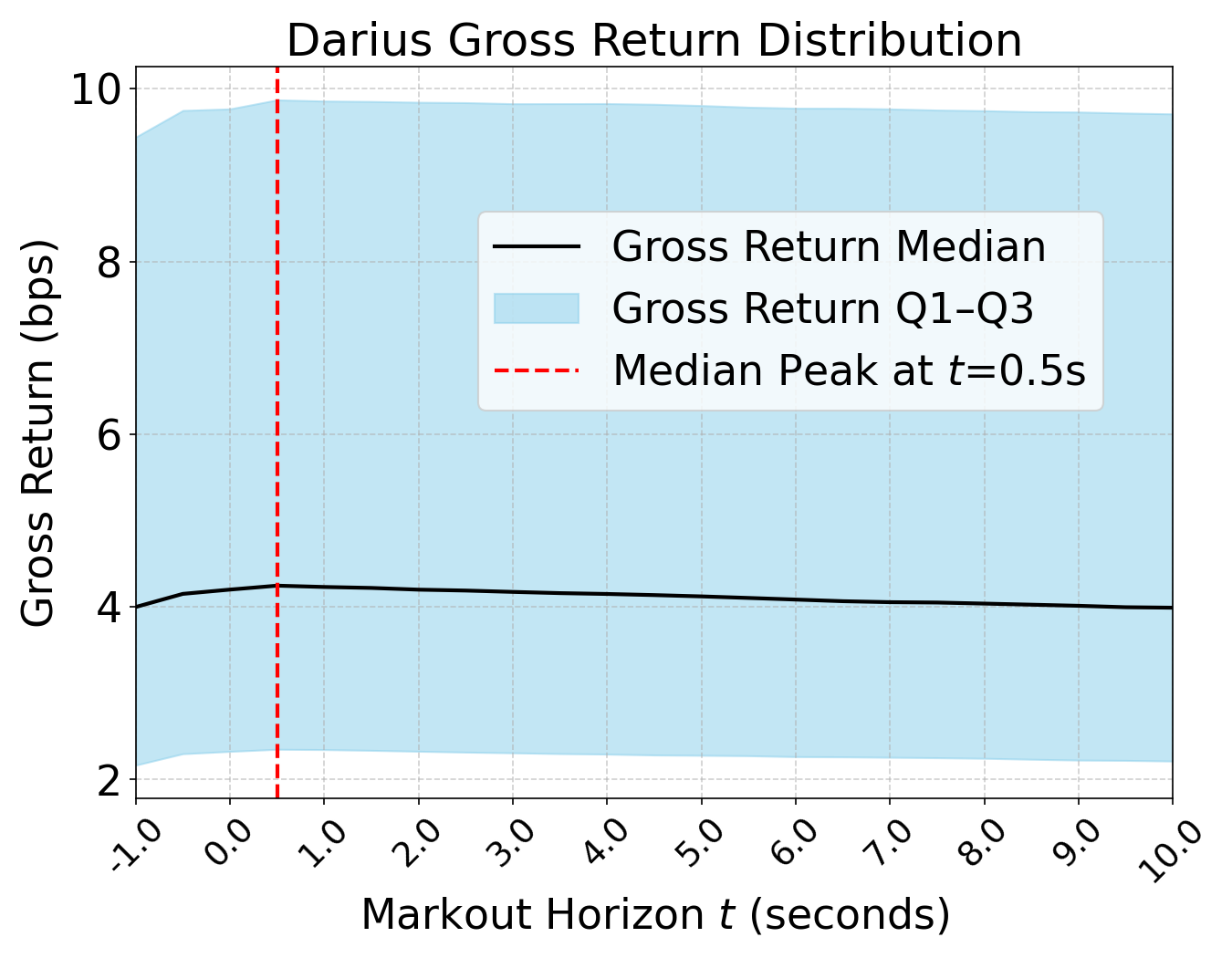}
    \caption{\texttt{Darius} (Pattern 1)}
    \label{fig:darius_return}
  \end{subfigure}\hfill
  \begin{subfigure}[b]{0.32\linewidth}
    \centering
    \includegraphics[width=\linewidth]{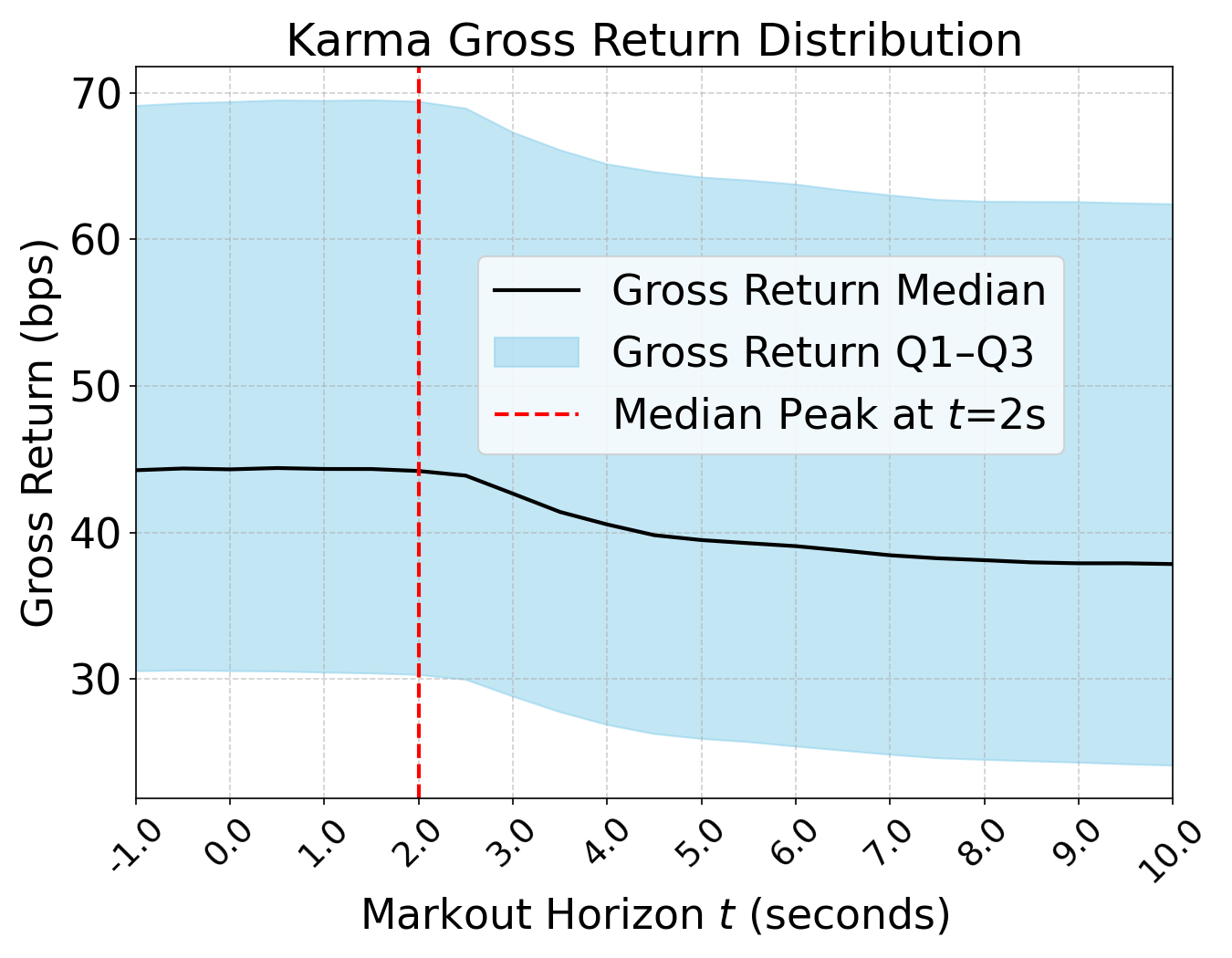}
    \caption{\texttt{Karma} (Pattern 2)}
    \label{fig:bard_return}
  \end{subfigure}

  \vspace{1em}
    
  \begin{subfigure}[b]{0.32\linewidth}
    \centering
    \includegraphics[width=\linewidth]{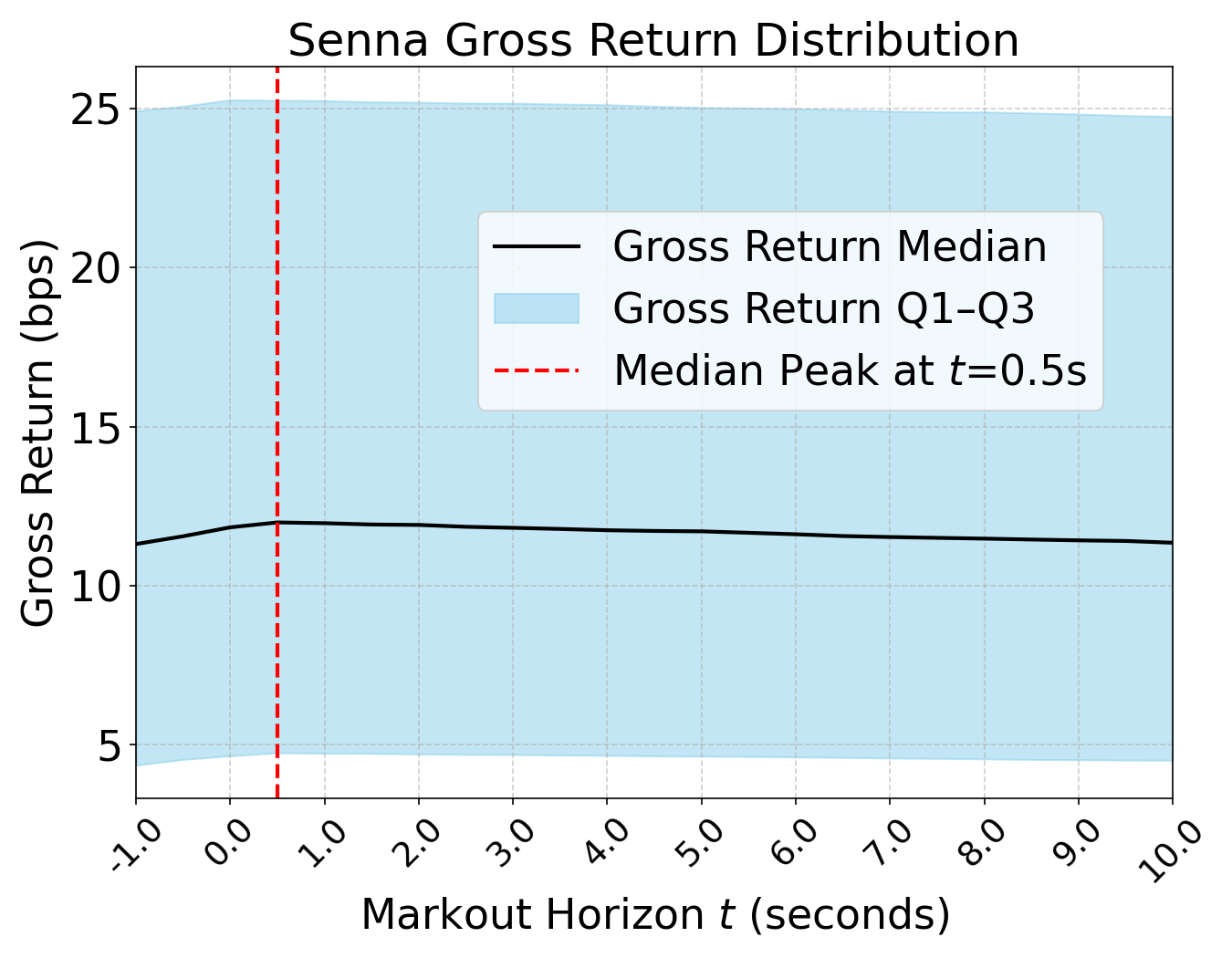}
    \caption{\texttt{Senna} (Pattern 1)}
    \label{fig:senna_return}
  \end{subfigure}\hfill
  \begin{subfigure}[b]{0.32\linewidth}
    \centering
    \includegraphics[width=\linewidth]{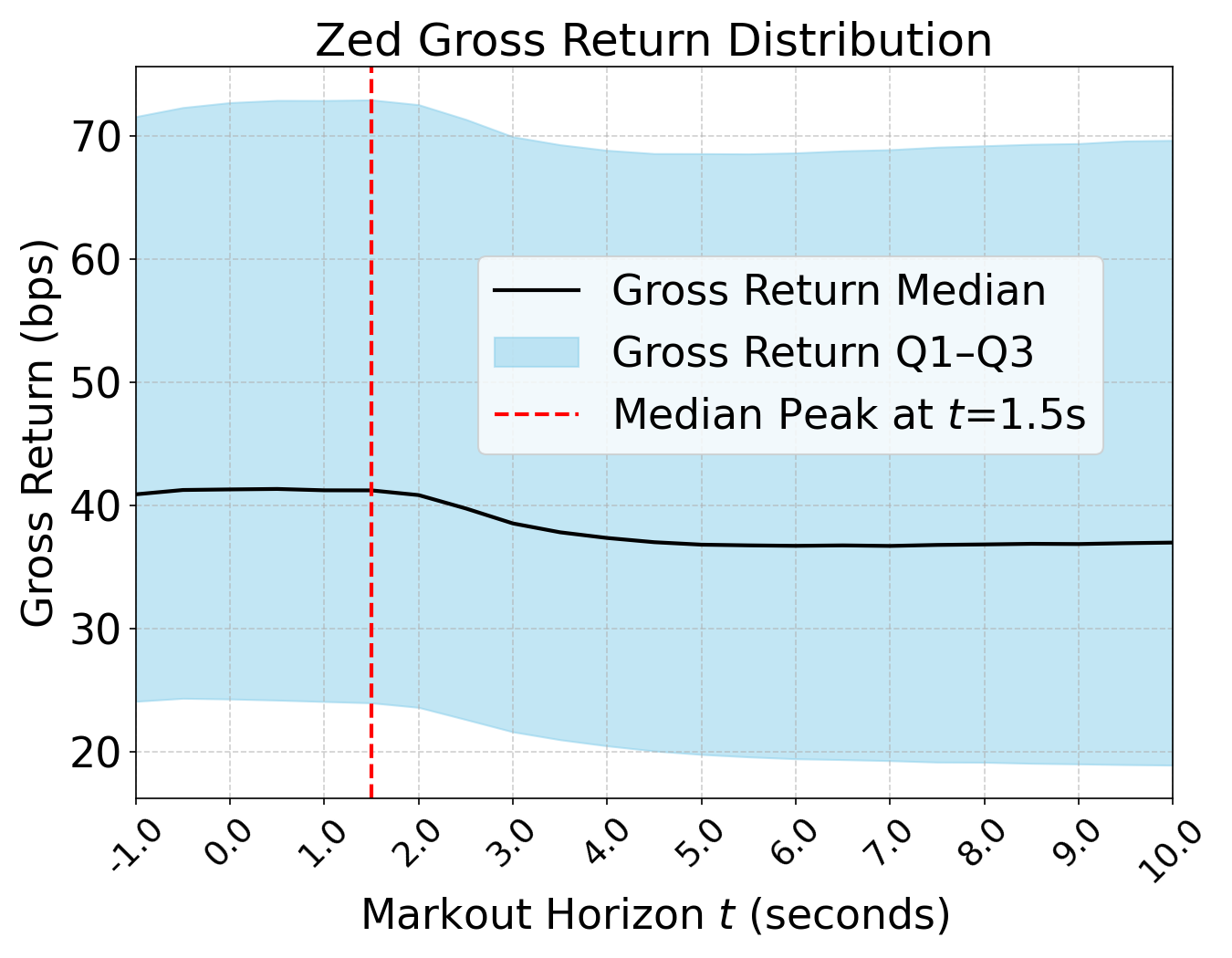}
    \caption{\texttt{Zed} (Pattern 2)}
    \label{fig:maokai_return}
  \end{subfigure}\hfill
  \begin{subfigure}[b]{0.32\linewidth}
    \centering
    \includegraphics[width=\linewidth]{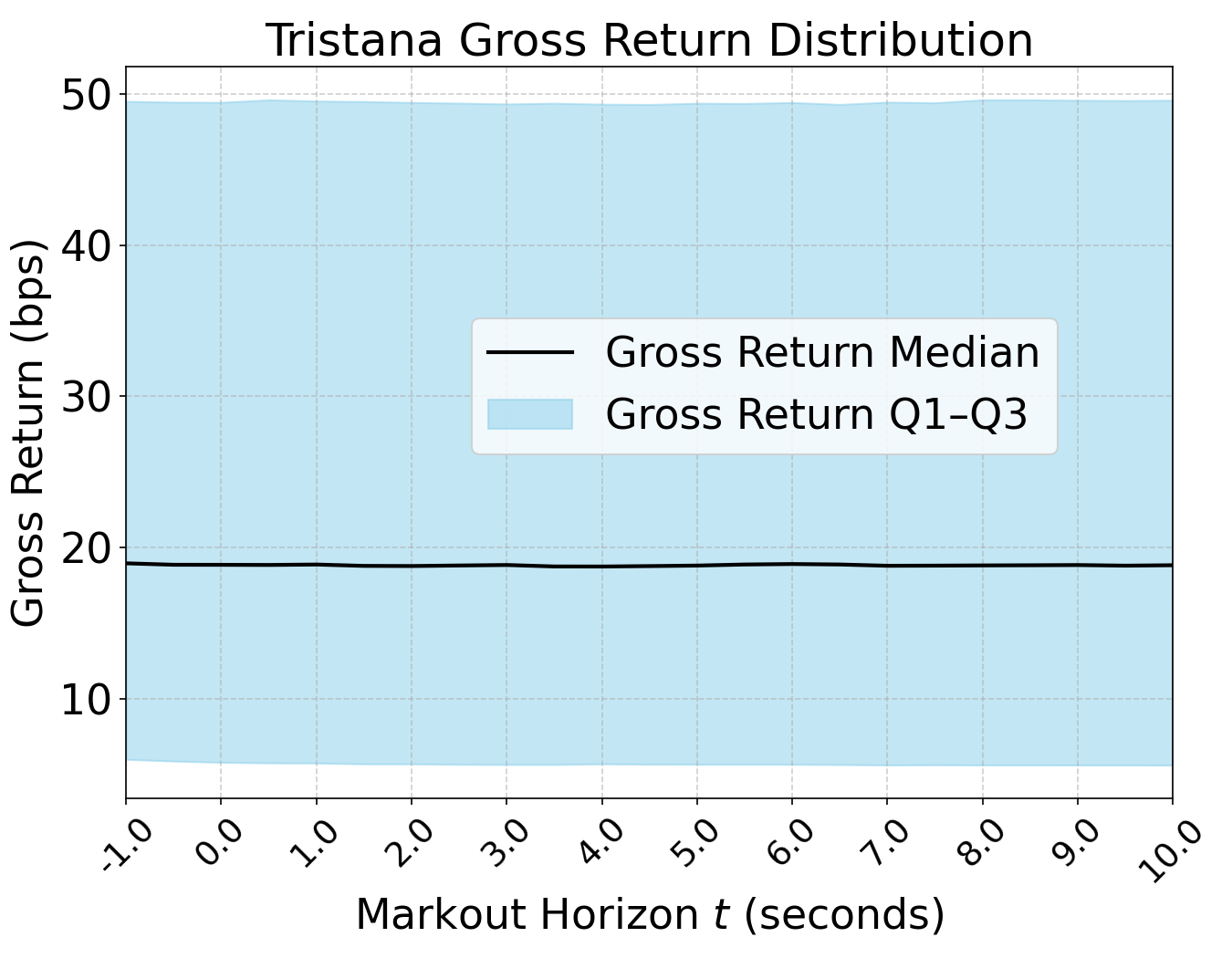}
    \caption{\texttt{Tristana} (Pattern 3)}
    \label{fig:tristana_return}
  \end{subfigure}

  \vspace{1em}

\end{figure}

\begin{figure}[H]

  \ContinuedFloat
  \centering
  \captionsetup{justification=centering}


  \vspace{1em}




  \vspace{1em}

  \begin{subfigure}[b]{0.32\linewidth}
    \centering
    \includegraphics[width=\linewidth]{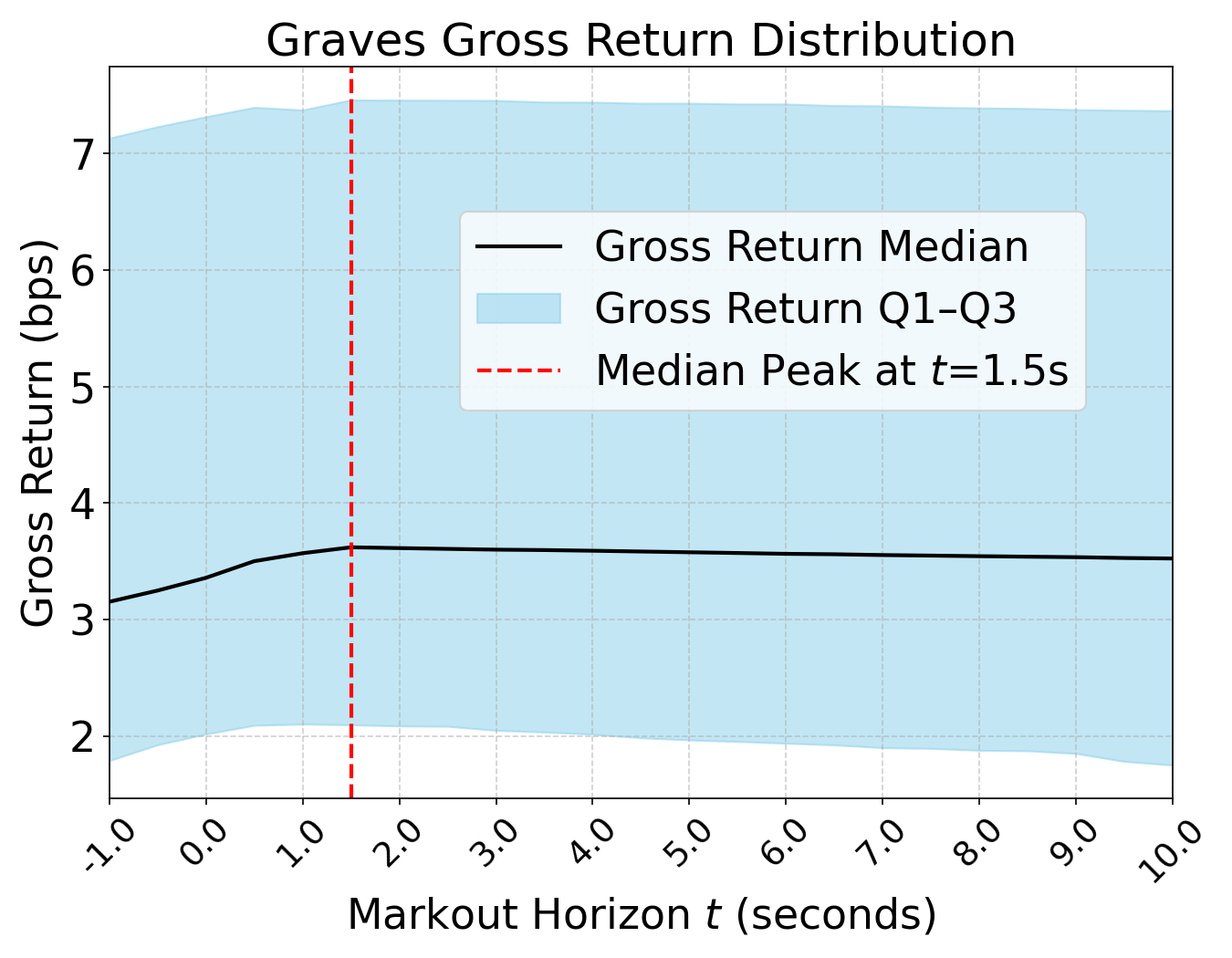}
    \caption{\texttt{Graves} (Pattern 1)}
    \label{fig:graves_return}
  \end{subfigure}\hfill
  \begin{subfigure}[b]{0.32\linewidth}
    \centering
    \includegraphics[width=\linewidth]{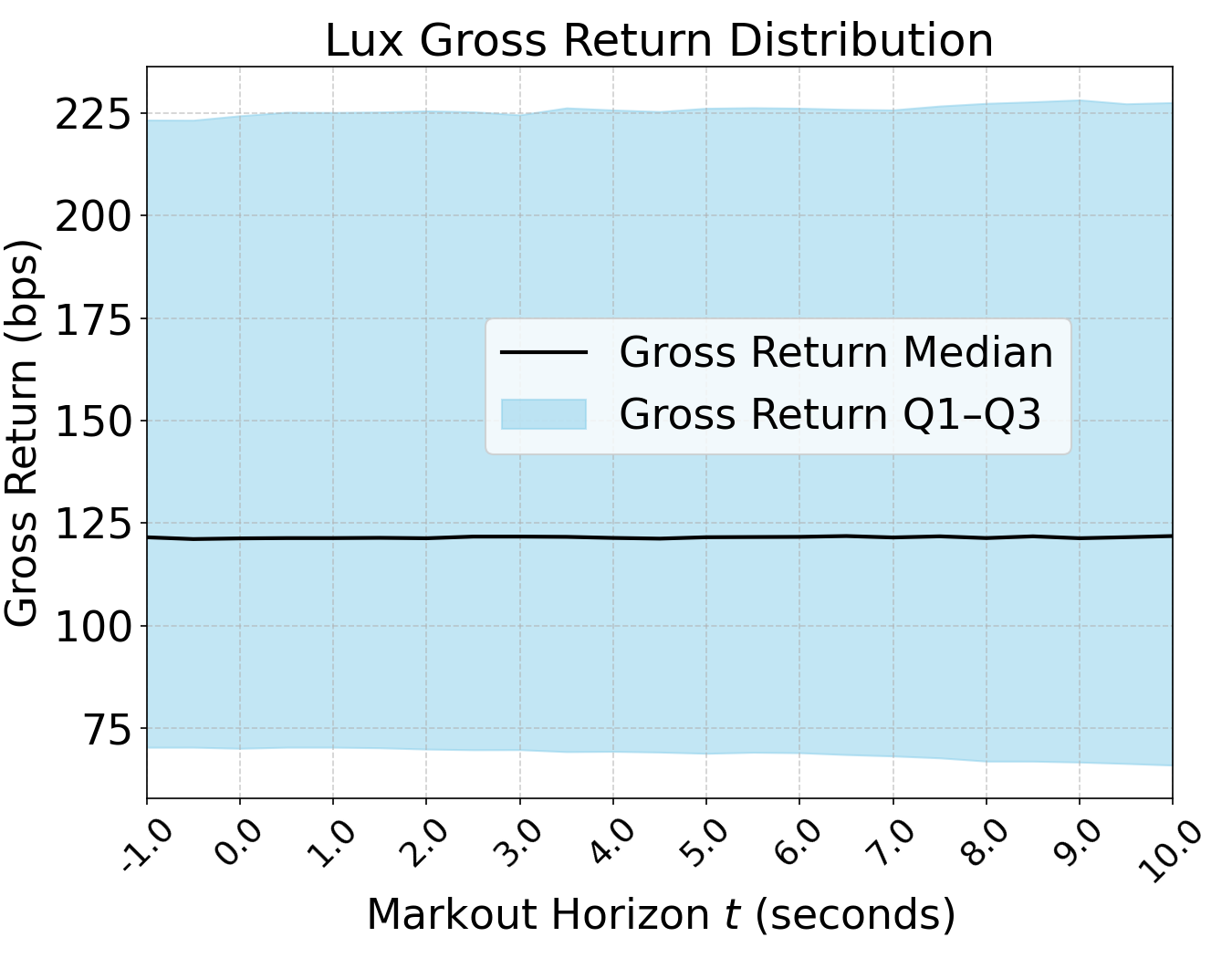}
    \caption{\texttt{Lux} (Pattern 3)}
    \label{fig:lux_return}
  \end{subfigure}
  \begin{subfigure}[b]{0.32\linewidth}
    \centering
    \includegraphics[width=\linewidth]{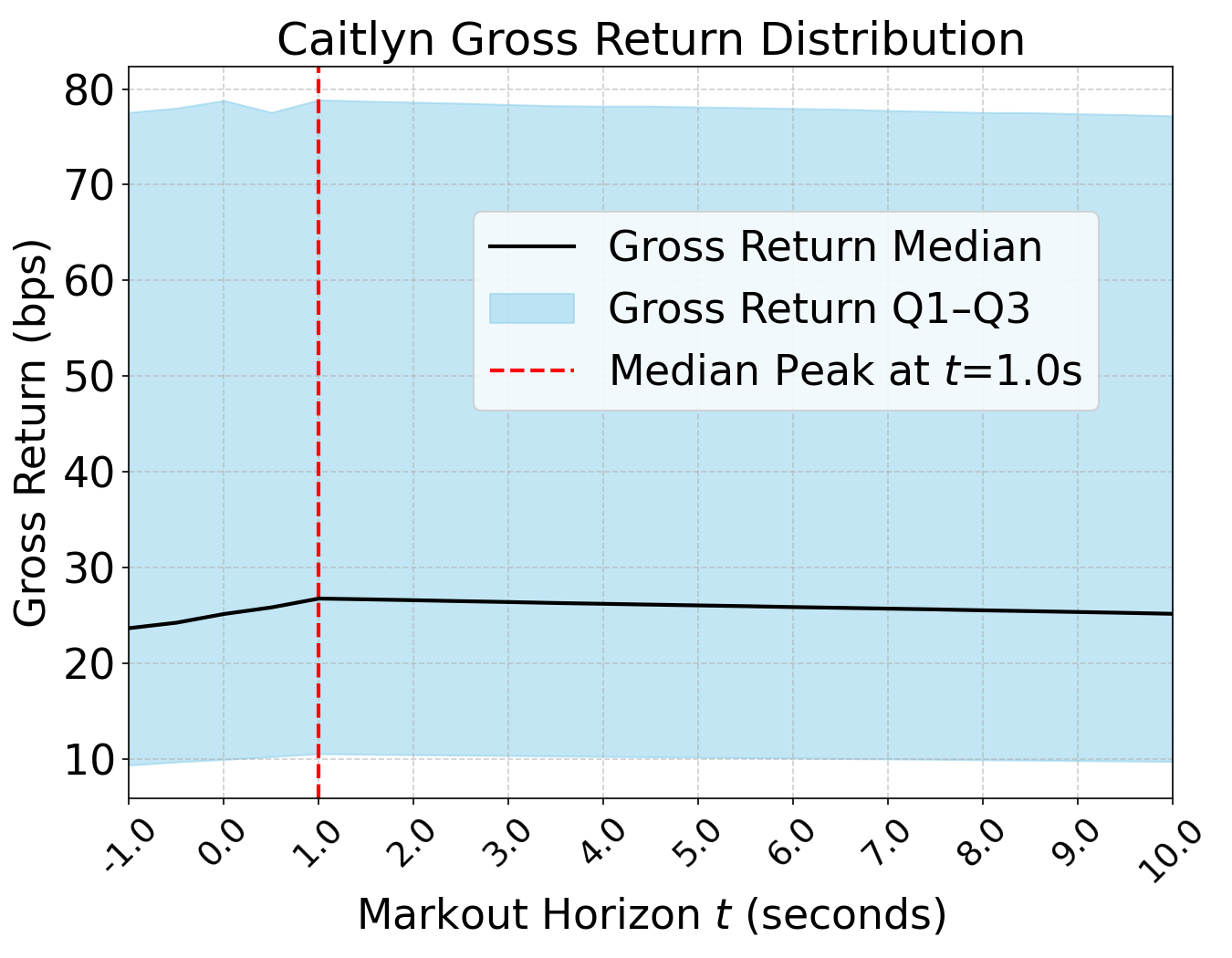}
    \caption{\texttt{Caitlyn} (Pattern 1)}
    \label{fig:akali_return}
  \end{subfigure}\hfill

  \vspace{1em}  
  
  \begin{subfigure}[b]{0.32\linewidth}
    \centering
    \includegraphics[width=\linewidth]{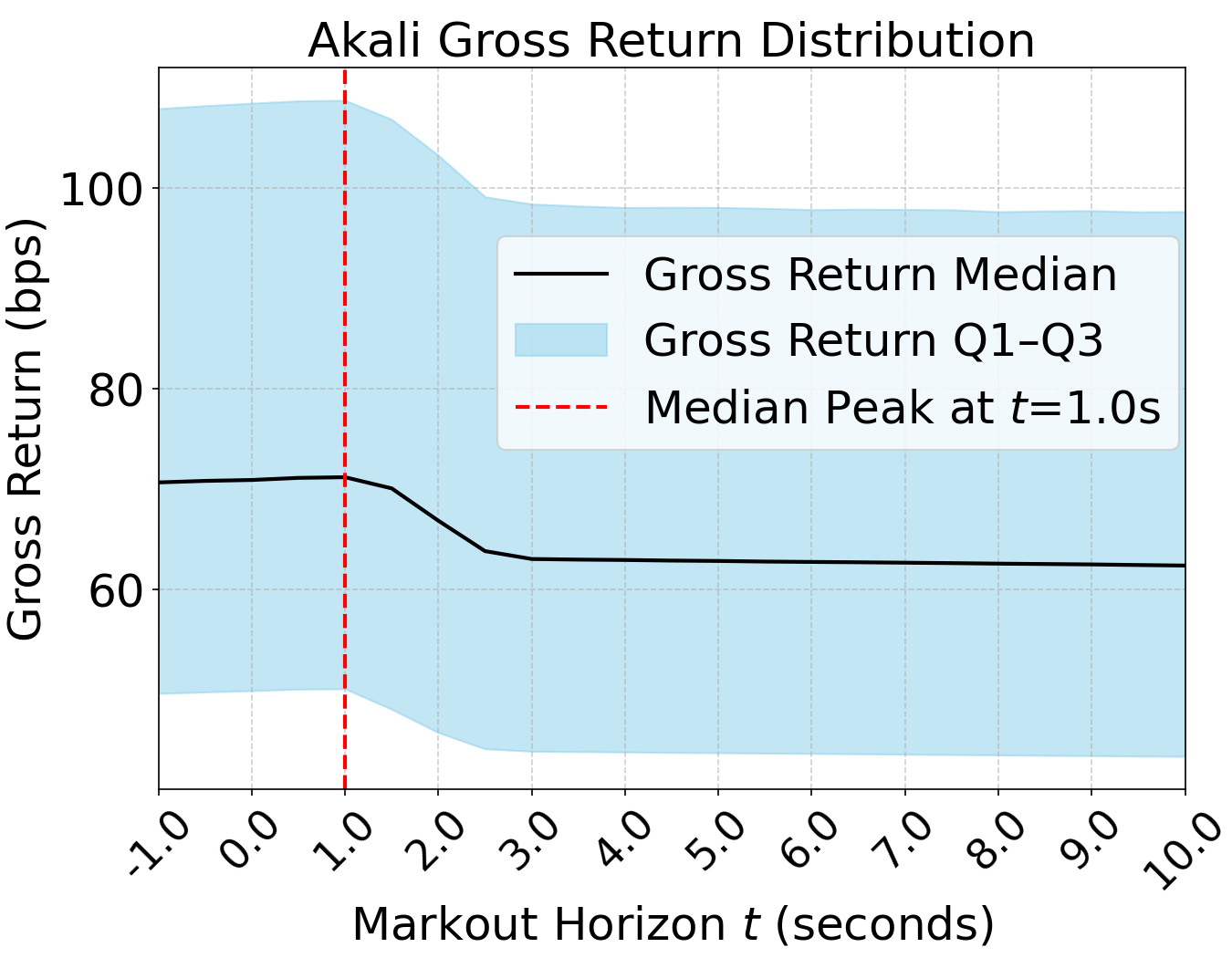}
    \caption{\texttt{Akali} (Pattern 2)}
    \label{fig:cat_return}
  \end{subfigure}\hfill
  \begin{subfigure}[b]{0.32\linewidth}
    \centering
    \includegraphics[width=\linewidth]{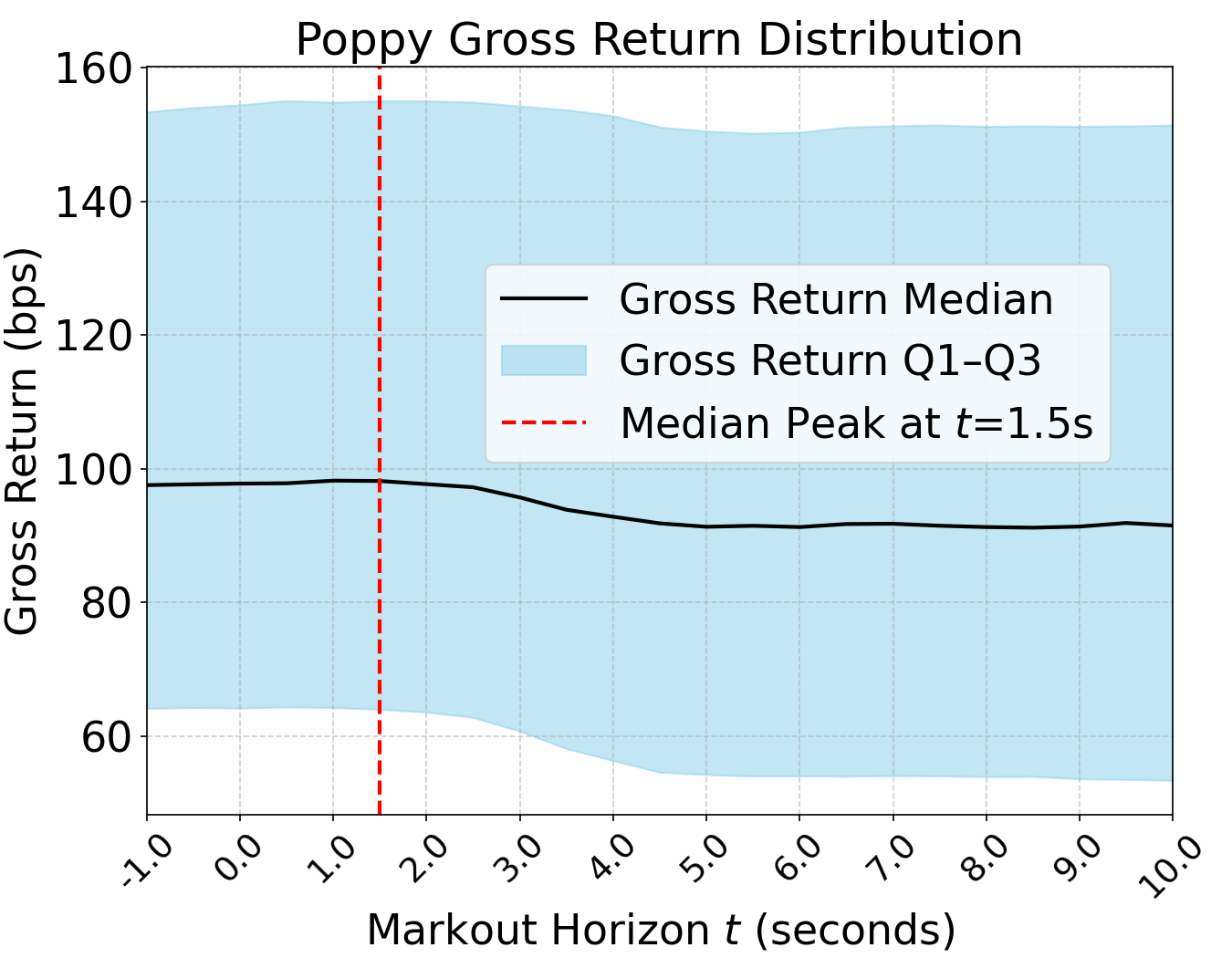}
    \caption{\texttt{Poppy} (Pattern 2)}
    \label{fig:poppy_return}
  \end{subfigure}
  
  \caption{The gross return distribution and pattern of the remaining searchers.}
  \label{fig:gross_returns_10s_14searchers}
\end{figure}

\section{Token pairs}
In \Cref{tab:token_full}, we present the count and volume of token pairs traded by each searcher for all 23 labeled searchers. Notably, Pattern 1 searchers typically concentrate on trading Major-Major pairs, whereas Pattern 2 and 3 searchers' trades more often involve ALT tokens.
\label{appendix:token}

\begin{table}[H]
\centering
\caption{Count and volume of token pairs traded by each searcher.}
\label{tab:token_full}
\resizebox{\textwidth}{!}{%
\begin{tabular}{@{}l | c | cc | cc | cc@{}}
\toprule
Searcher     & Pattern 
  & \multicolumn{2}{c|}{Major–Major} 
  & \multicolumn{2}{c|}{Major–ALT} 
  & \multicolumn{2}{c}{ALT–ALT} \\
             & 
  & \# (\%)       & USD (\%) 
  & \# (\%)       & USD (\%) 
  & \# (\%)       & USD (\%) \\
\midrule
\midrule
Wintermute   & 1 
  & 977,048 (54.2\%)   & 67.7B (89.3\%)  
  & 794,181 (44.0\%)   &  7.9B (10.4\%)  
  &  32,741 (1.8\%)    & 242.5M (0.3\%) \\

SCP          & 1 
  & 908,414 (43.2\%)   & 57.1B (88.1\%)  
  & 1,190,421 (56.7\%) &  7.7B (11.9\%)  
  &   2,246 (0.1\%)    & 11.7M (0.0\%)  \\

Kayle        & 1 
  & 680,494 (46.3\%)   & 33.7B (81.2\%)  
  & 781,030 (53.1\%)   &  7.7B (18.7\%)  
  &   8,184 (0.6\%)    & 52.9M (0.1\%)  \\

Galio        & 1 
  & 161,238 (66.2\%)   & 22.6B (87.0\%)  
  &  81,391 (33.4\%)   &  3.4B (12.9\%)  
  &     940 (0.4\%)    & 17.4M (0.1\%)  \\

Shen         & 1 
  & 190,502 (27.2\%)   &  9.0B (72.1\%)  
  & 502,718 (71.9\%)   &  3.5B (27.8\%)  
  &   6,276 (0.9\%)    &  9.8M (0.1\%)  \\

Taric        & 2 
  &  70,496 (39.8\%)   &  3.6B (71.2\%)  
  & 100,805 (56.9\%)   &  1.4B (28.3\%)  
  &   5,767 (3.3\%)    & 27.2M (0.5\%)  \\

Lucian       & 2 
  &  14,458 (2.7\%)    &  1.3B (32.6\%)  
  & 512,152 (97.0\%)   &  2.6B (67.2\%)  
  &   1,268 (0.2\%)    &  4.6M (0.1\%)  \\

Riven        & 1 
  & 17,429 (78.6\%)    &  3.0B (96.1\%)  
  &  4,489 (20.2\%)    & 121.3M (3.9\%)  
  &    261 (1.2\%)     &  2.4M (0.1\%)  \\

Thresh       & 1 
  & 25,808 (97.7\%)    &  2.6B (98.8\%)  
  &   619 (2.3\%)      &  30.9M (1.2\%)  
  &     0 (0.0\%)      &   0 (0.0\%)    \\

Ahri         & 1 
  & 11,846 (100.0\%)   &  1.7B (100.0\%) 
  &     0 (0.0\%)      &   0 (0.0\%)    
  &     0 (0.0\%)      &   0 (0.0\%)    \\

Darius       & 1 
  &  8,756 (73.6\%)    &  1.0B (91.2\%)  
  &  3,135 (26.4\%)    &  97.0M (8.8\%)  
  &     0 (0.0\%)      &   0 (0.0\%)    \\

Karma        & 2 
  &  4,226 (7.5\%)     & 274.7M (40.6\%) 
  & 48,633 (86.1\%)    & 388.3M (57.4\%) 
  &  3,618 (6.4\%)     & 13.3M (2.0\%)  \\

Bard         & 3 
  &    263 (0.1\%)     &  24.7M (3.8\%)  
  & 170,074 (96.4\%)   & 618.8M (94.6\%) 
  &  6,141 (3.5\%)     & 10.5M (1.6\%)  \\

Senna         & 1 
  & 19,786 (36.9\%)    & 368.9M (60.5\%) 
  & 32,625 (60.8\%)    & 236.9M (38.9\%) 
  &  1,208 (2.3\%)     &  3.7M (0.6\%)  \\

Maokai       & 2 
  &    579 (2.1\%)     & 186.0M (31.4\%) 
  & 26,066 (94.7\%)    & 399.2M (67.3\%) 
  &    893 (3.2\%)     &  7.9M (1.3\%)  \\

Zed          & 2 
  &     13 (0.0\%)     & 319.7K (0.1\%)  
  & 95,724 (100.0\%)   & 584.9M (99.9\%) 
  &     0 (0.0\%)      &   0 (0.0\%)    \\

Jinx         & 3 
  &    479 (1.2\%)     &  8.9M (2.8\%)  
  & 39,073 (98.7\%)    & 306.0M (97.1\%) 
  &    16 (0.0\%)      & 159.1K (0.1\%) \\

Tristana     & 3 
  &  4,680 (27.9\%)    & 212.9M (73.1\%) 
  & 11,609 (69.1\%)    &  77.7M (26.7\%) 
  &   506 (3.0\%)      & 756.2K (0.3\%)  \\

Graves        & 1 
  &  8,532 (93.3\%)    & 251.5M (99.6\%) 
  &   612 (6.7\%)      & 978.1K (0.4\%)  
  &     0 (0.0\%)      &   0 (0.0\%)    \\

Lux          & 3 
  &     15 (0.1\%)     & 298.9K (0.3\%)  
  & 11,300 (99.9\%)    & 118.5M (99.7\%) 
  &     0 (0.0\%)      &   0 (0.0\%)    \\

Caitlyn          & 1 
  &     729 (54.2\%)     & 86.5M (88.3\%)  
  & 616 (45.8\%)    & 11.5M (11.7\%) 
  &     0 (0.0\%)      &   0 (0.0\%)    \\

Akali        & 2
  &     0 (0.0\%)     & 0 (0\%)  
  & 9771 (100.0\%)    & 49.4M (100.0\%) 
  &     0 (0.0\%)      &   0 (0.0\%)    \\

Poppy         & 2 
  &     0 (0.0\%)     & 0 (0.0\%)  
  & 8115 (100.0\%)    & 35.6M (100.0\%) 
  &     0 (0.0\%)      &   0 (0.0\%)    \\

\bottomrule
\end{tabular}%
}
\end{table}

\section{Other metrics of the CEX-DEX market}
\label{appendix:mev_compare}

\begin{figure}[h]
    \centering
    \includegraphics[width=0.49\linewidth]{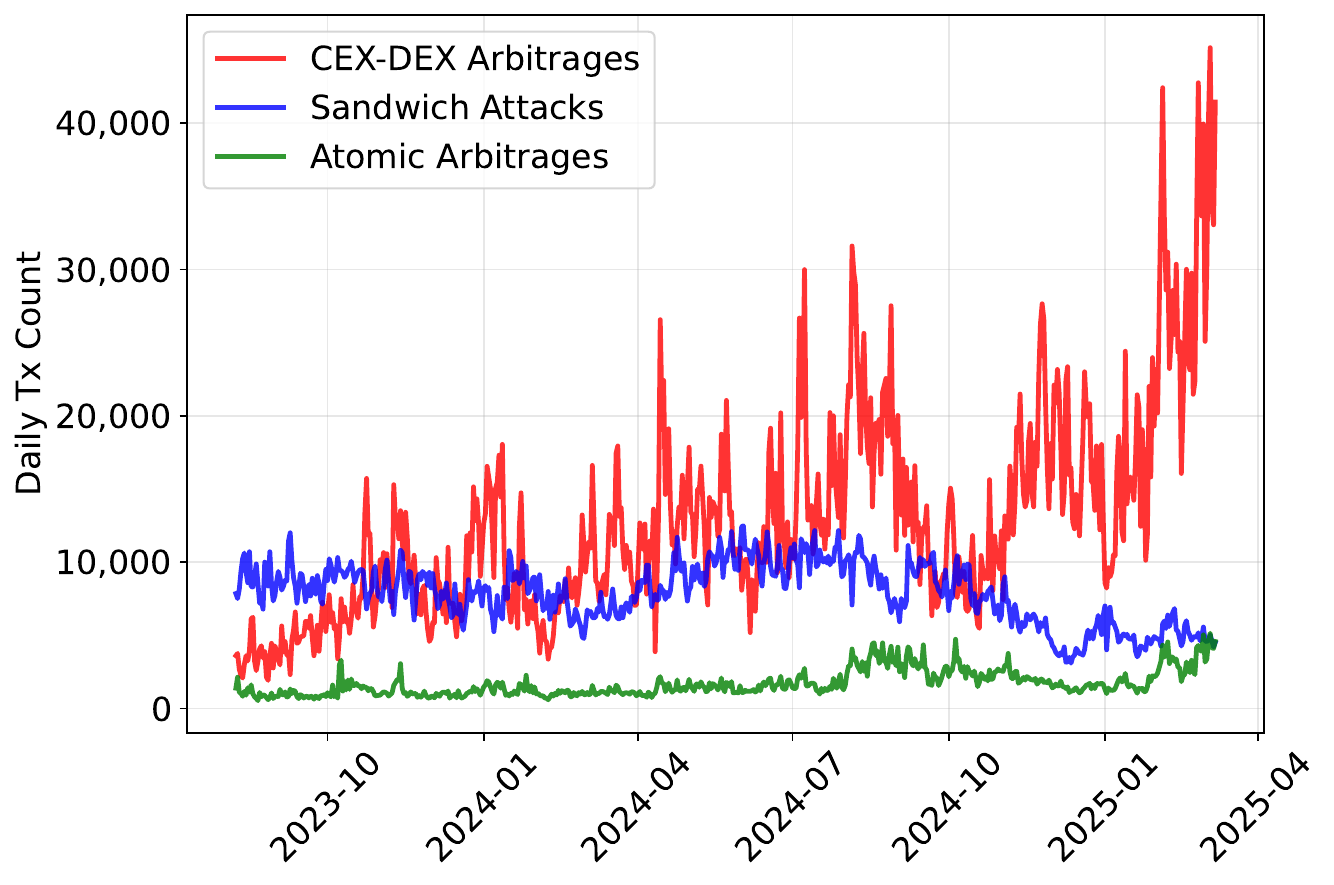} %
    \includegraphics[width=0.49\linewidth]{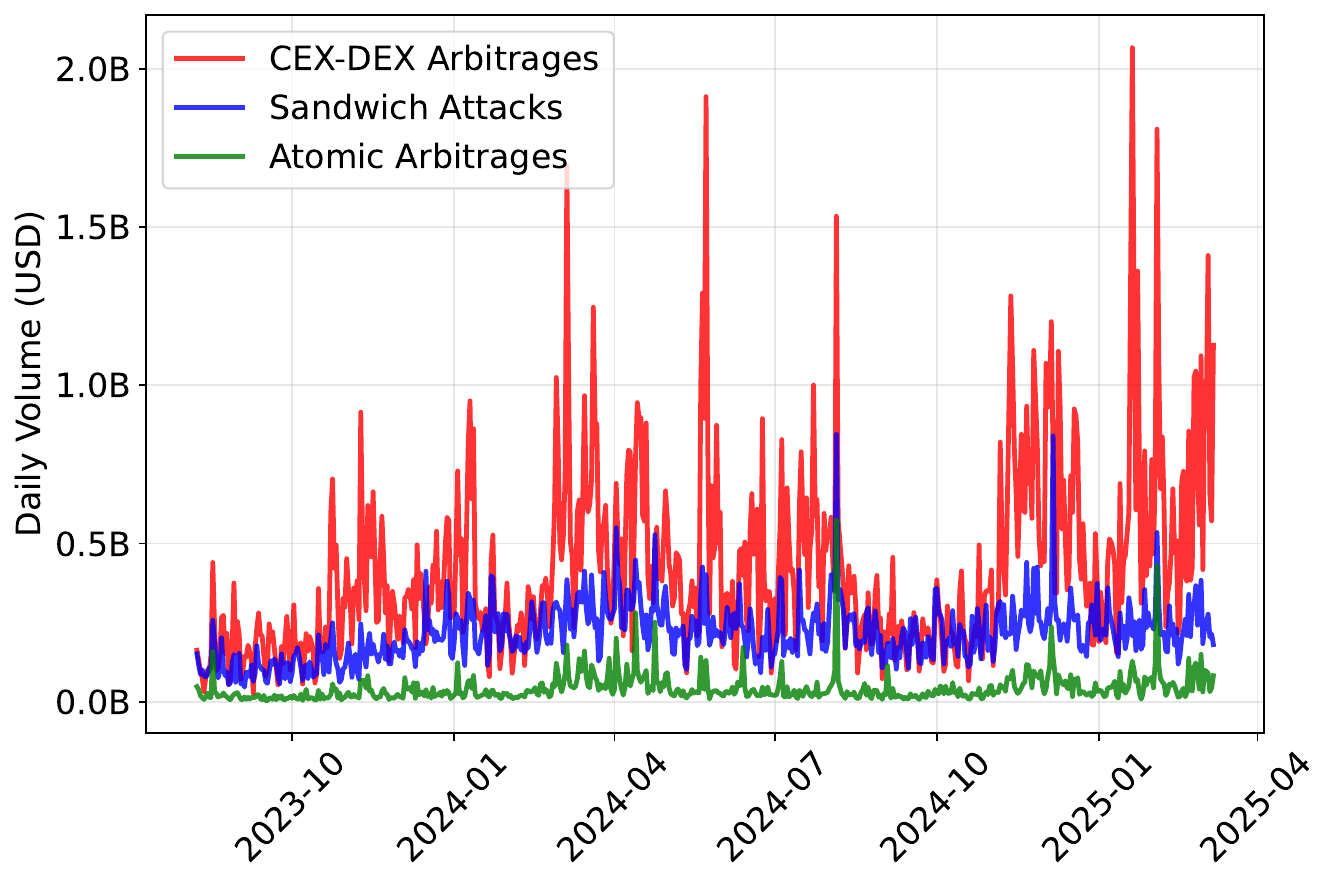} %
    \caption{Daily DEX trade count and volume from atomic arbitrages, sandwich attacks, and CEX-DEX arbitrages by 19 labeled searchers during the period of August 2023 to March 2025.}
    \label{fig:mev_compare}
\end{figure}

In \Cref{fig:mev_compare}, we present the comparison between CEX-DEX arbitrages executed by 19 labeled searchers and other MEV transaction types, such as atomic arbitrages and sandwich attacks during the observed period of August 2023 to March 2025. The data for atomic arbitrages and sandwich attacks are collected from \cite{hildobbyatomic,hildobbysandwich}.

\begin{figure}[h]
    \centering
    \includegraphics[width=1\linewidth]{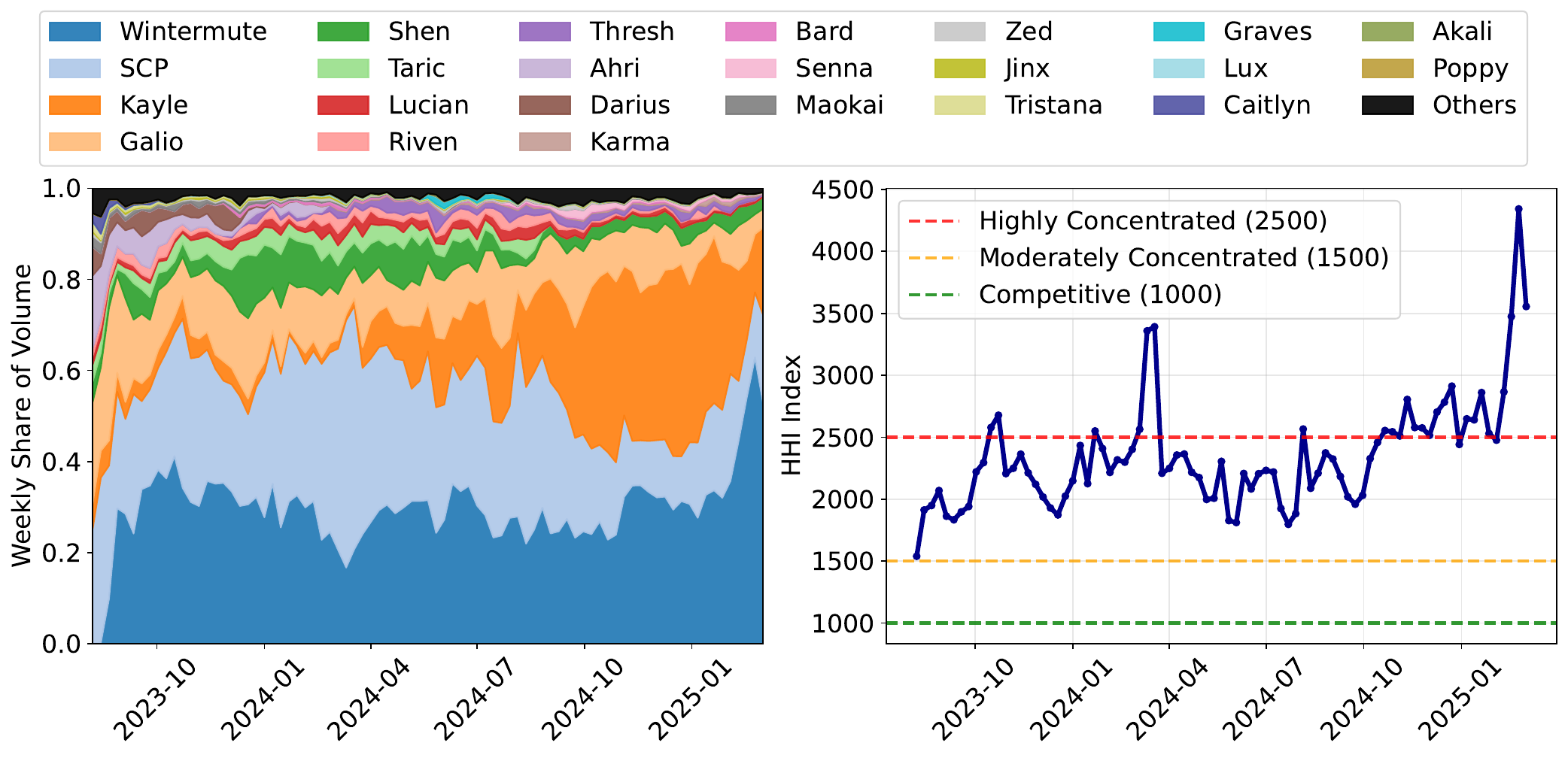} %
    \caption{Weekly share of CEX-DEX volume by all the searchers, including Pattern 3 and other searchers, and HHI index of weekly CEX-DEX volume.}
    \label{fig:volume_hhi}
\end{figure}

In \Cref{fig:volume_hhi}, we present the weekly share of CEX-DEX arbitrage volume by all searchers, including Pattern 3 unlabeled searchers and other searchers, and the HHI index of weekly CEX-DEX trade volume. Data shows that the volume is concentrated on a few leading searchers starting from October 2024, contributing to the market being highly centralized.

\begin{figure}[h]
    \centering
    \includegraphics[width=0.5\linewidth]{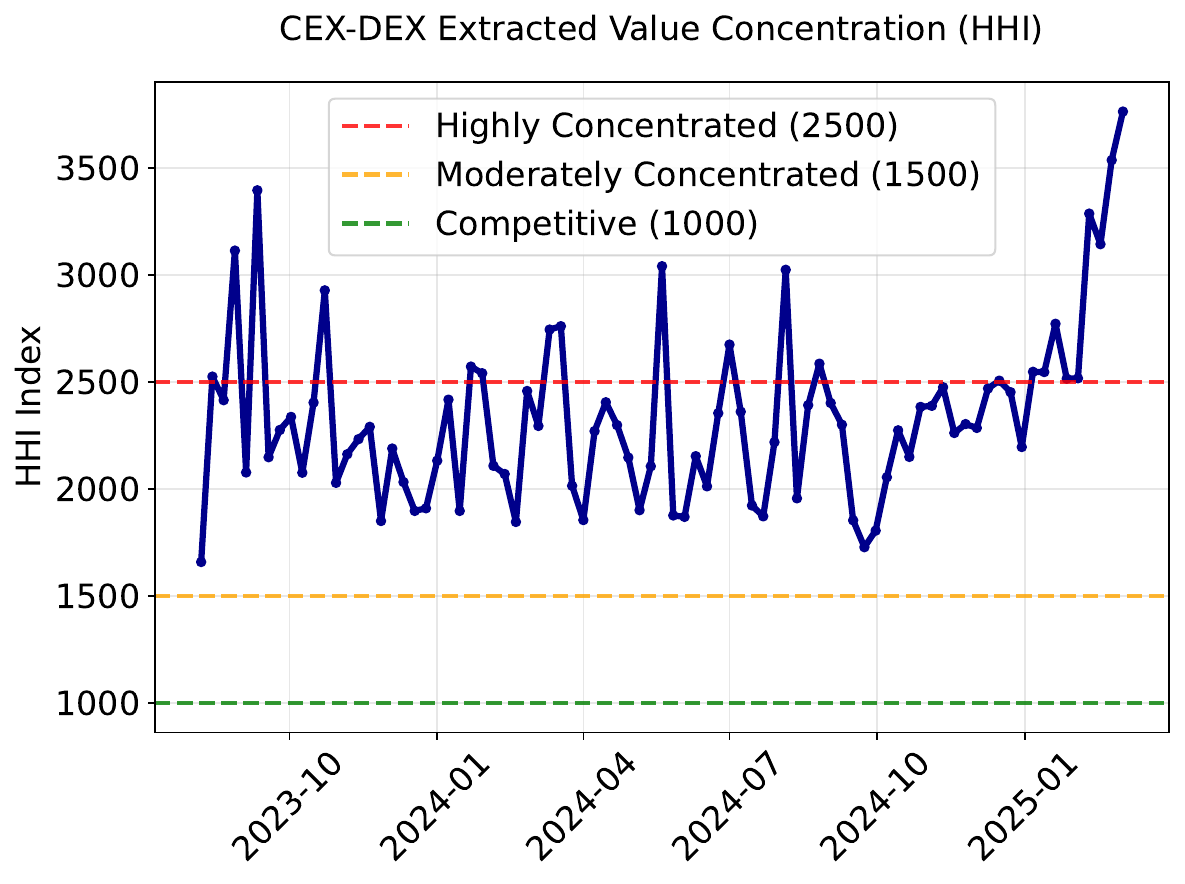} %
    \caption{HHI Index of weekly extracted value from CEX-DEX arbitrages by 19 labeled searchers.}
    \label{fig:ev_hhi}
\end{figure}

In \Cref{fig:ev_hhi}, we present the HHI Index of weekly extracted value from CEX-DEX arbitrages by 19 labeled searchers (cf. \Cref{sec:ev}, \Cref{fig:ev}). Data shows that the extracted value from CEX-DEX arbitrages starts to become increasingly concentrated since October 2024 and has become highly centralized to a few leading searchers by Q1 2025.

\begin{figure}[h]
    \centering
    \includegraphics[width=1\linewidth]{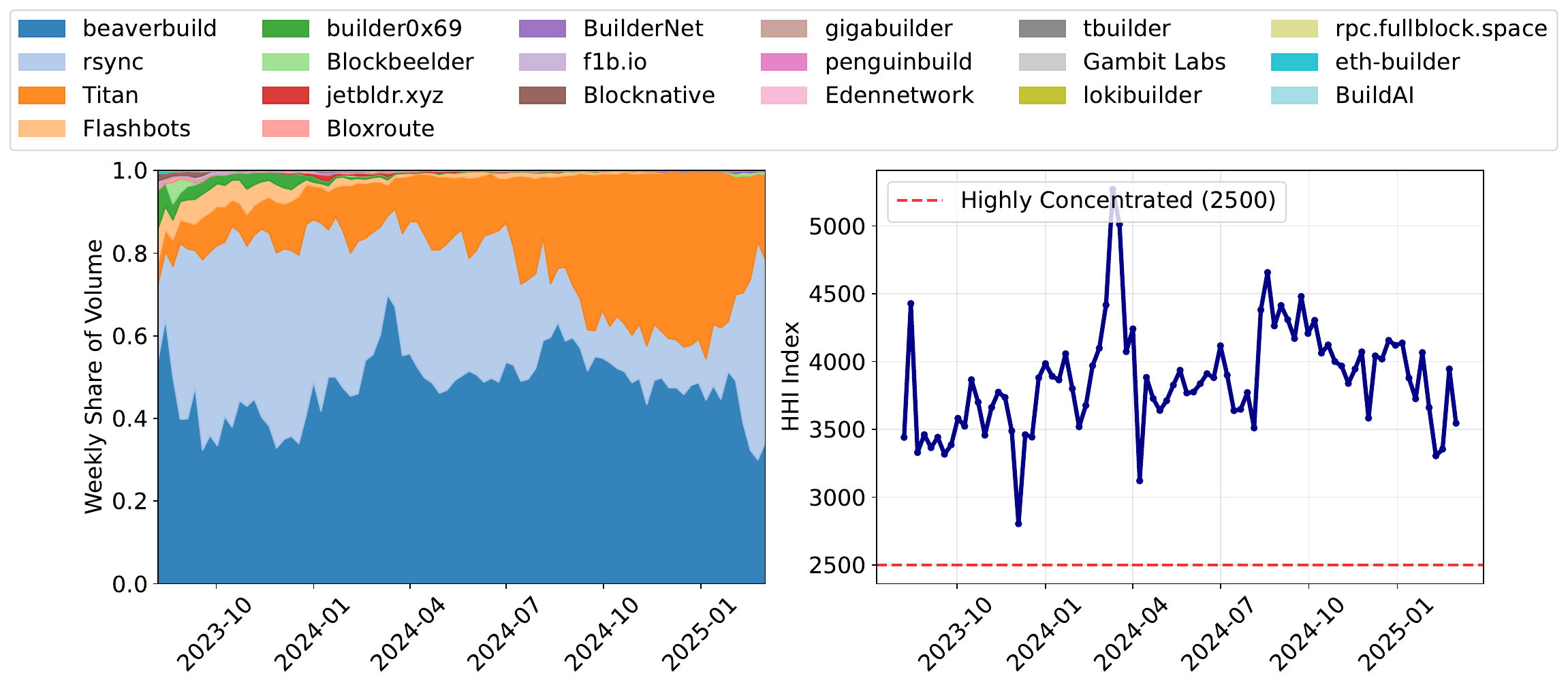} %
    \caption{Weekly share of CEX-DEX volume in Top 20 builders' blocks and the HHI index.}
    \label{fig:volume_hhi_builder}
\end{figure}

In \Cref{fig:volume_hhi_builder}, we present the weekly share of CEX-DEX volume in the top 20 builders' blocks and the HHI index. We observe a significant increase of CEX-DEX volume share included in \texttt{Titan}'s blocks between June 2024 and September 2024.

\section{Additional impact of exclusive CEX-DEX flow}
\label{appendix:impact_cexdex}

In this section, we present our additional findings about the impact of exclusive CEX-DEX flow that are not presented in \Cref{sec:exclusive_builder_performance}.

\begin{figure}[h]
    \centering
    \begin{subfigure}{0.5\textwidth}
    \captionsetup{justification=centering}
    \includegraphics[width=\linewidth]{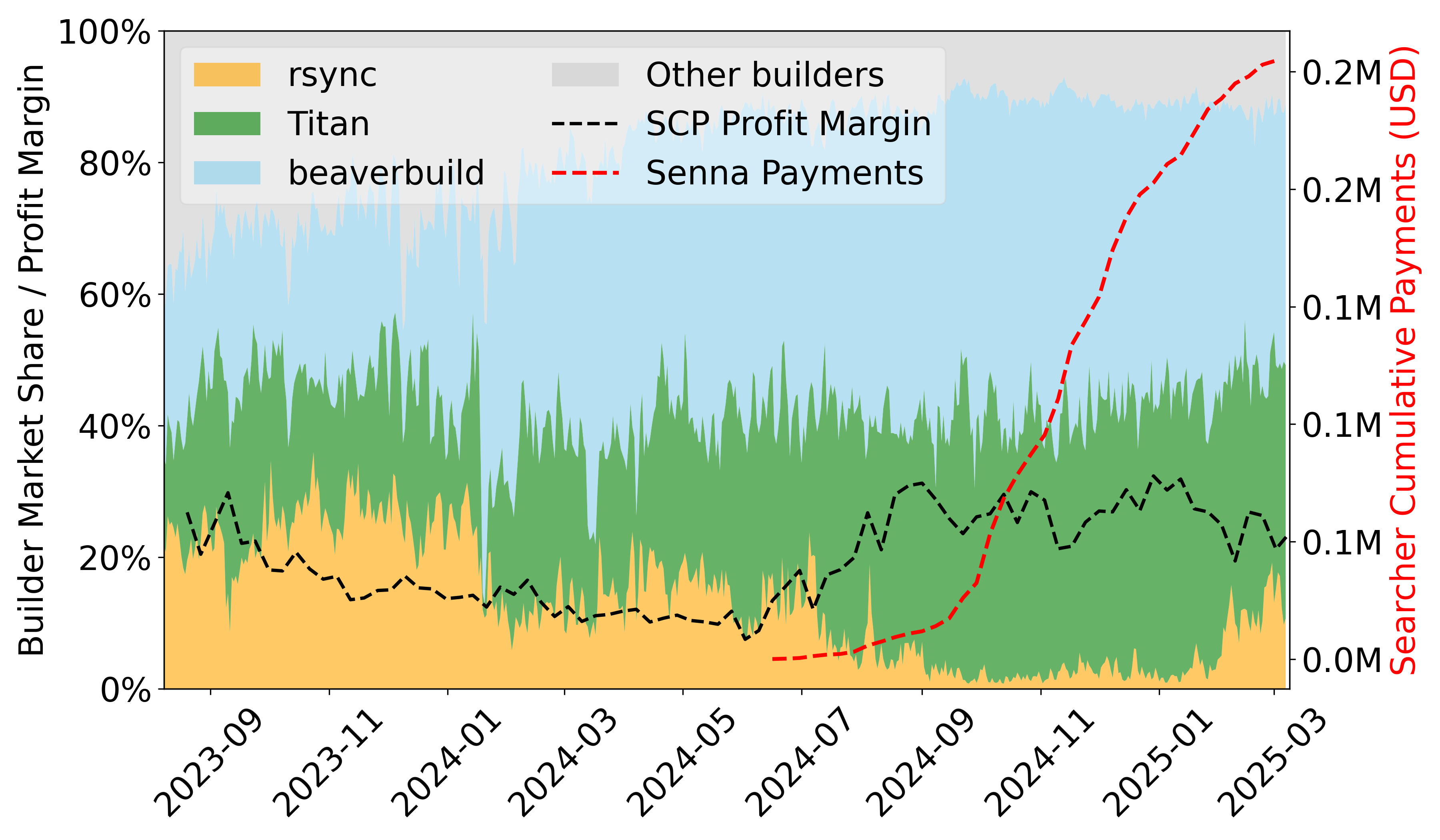}
    \caption{}
    \label{fig:senna_beaver}
    \end{subfigure}%
    \begin{subfigure}{0.5\textwidth}
    \includegraphics[width=\linewidth]{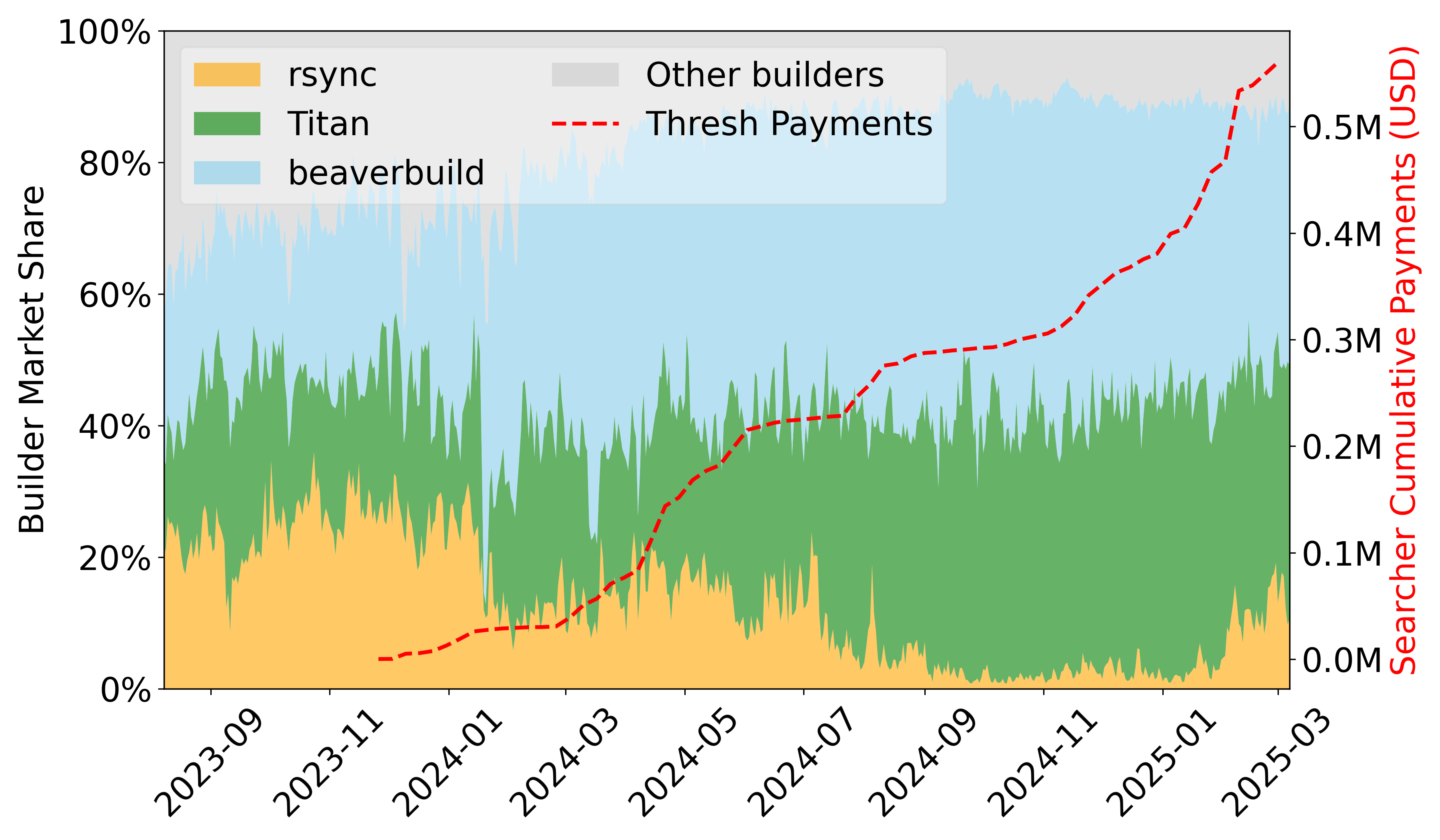}
    \captionsetup{justification=centering}
    \caption{}
    \label{fig:thresh_beaver}
    \end{subfigure}
    \caption{\textbf{a)} \texttt{beaverbuild} market share, \texttt{SCP} median profit margin, and \texttt{Senna}'s cumulative payments to \texttt{beaverbuild}. The left y-axis shows the builder market share and \texttt{SCP} median profit margin, and the right y-axis shows \texttt{Senna}'s cumulative payments to \texttt{beaverbuild}. \textbf{b)} \texttt{beaverbuild} market share and \texttt{Thresh} cumulative payments to \texttt{beaverbuild}. The left y-axis shows the builder market share, and the right y-axis shows \texttt{Thresh}'s cumulative payments to \texttt{beaverbuild}.}
\end{figure}

As is illustrated in \Cref{fig:senna_beaver}, unlike \texttt{rsync} and \texttt{Wintermute}, the market share of \texttt{beaverbuild} remained robust and \texttt{SCP}'s profit margin even slightly increased during \texttt{Titan}'s expansion between June 2024 and September 2024. A potential factor could be the concurrent operation of \texttt{Senna}, an exclusive searcher affiliated with \texttt{beaverbuild}, whose considerable payments potentially helped stabilize \texttt{beaverbuild}'s market position and support \texttt{SCP}'s profitability.

We also observe another shift in the builder market happened earlier between February 2024 and April 2024. As shown in \Cref{fig:thresh_beaver}, beginning in February 2024, \texttt{Titan} market share began rising, reportedly through an exclusive deal with the Banana Gun Telegram bot \cite{whowinsandwhy,yang2024decentralization}. Around the same time, \texttt{beaverbuild} similarly steadily expanded its market share by 15\%. This timing is consistent with the operation of \texttt{Thresh} as an exclusive searcher submitting most of their high-value trades with substantial payments directly to \texttt{beaverbuild}, as we observe their cumulative payments significantly increases around this time. As a result, \texttt{rsync}'s share declined from 25\% to below 15\%.

In response to the high-value flow captured by \texttt{beaverbuild} and \texttt{Titan}, \texttt{Wintermute} appeared to engage defensively by sacrificing profitability to support its integrated builder, \texttt{rsync}. \Cref{fig:wm_margin_vs_share} shows a steady decline in \texttt{Wintermute}’s median profit margin during this interval, consistent with the searcher ceding a larger share of their arbitrage revenue to \texttt{rsync} so the builder could bid more aggressively and remain competitive in the builder market. 

By June 2024, we notice that \texttt{Wintermute} median profit margin was already below 10\%. As a result, during \texttt{Titan}'s expansion supported by \texttt{Kayle} from June, \texttt{Wintermute} could not further transfer more arbitrage revenue to \texttt{rsync} to maintain their builder market share as before. This likely triggered \texttt{rsync}'s retreat from the builder market.

\section{Heuristics for identifying CEX-DEX transactions}
\label{appendix:heuristics}
\subsection{Limitations}
Despite employing a comprehensive set of heuristics, our detection methodology still has some inherent constraints and makes certain tradeoffs to ensure the cleanliness of our dataset given the resources we have:
\begin{enumerate}[topsep=2pt, itemsep=2pt, parsep=2pt]
    \item We only include transactions involving tokens listed on Binance. Consequently, our analysis excludes arbitrage transactions involving tokens listed exclusively on other CEXes. This choice likely results in an underestimation of the total revenue and profitability (PnL) of some searchers. Since Binance consistently ranks as the CEX with the highest trading volumes \cite{coinmarketcap}, our dataset maintains sufficient coverage.
    
    \item We only consider transactions in which at least one swap executed is the first trade in its direction within the corresponding DEX pool in that block. This heuristic helps filter out retail or unrelated trades but may also exclude valid arbitrage trades that occur after other swaps. While this contributes to potential underestimation of searchers' total revenues and PnL, relaxing this heuristic would significantly increase noise in our dataset and compromise the accuracy of our analysis.
    
    \item Finally, cross-chain arbitrage transactions (i.e., arbitrages executed between decentralized exchanges on different blockchains) might occasionally be misclassified as CEX-DEX arbitrages. Although we have verified that known cross-chain arbitrageur addresses identified in prior research \cite{burakcrosschainarb} are not included in our dataset, some cross-chain arbitrage transactions may still be inadvertently captured. Nonetheless, this does not materially affect our conclusions on the revenue and profitability of the explicitly labeled CEX-DEX searchers.
    
\end{enumerate}

\subsection{Removed heuristics from previous works}
We remove two heuristics in previous works \cite{nonatomic,whowinsandwhy}, and we here detail the explanations.

\begin{enumerate}[topsep=2pt, itemsep=2pt, parsep=2pt]
    \item ``The transaction consumes less than 400,000 gas." We expanded our analysis to include transactions involving multiple token swaps and ERC-20 token transfers, which inherently consume more gas than single-swap transactions. Consequently, the previous gas consumption limit of 400,000 units no longer applies.\footnote{Example transaction hash: \href{https://etherscan.io/tx/0x5bdcacc0feee0fd4685fb9a0ba6820cc2ea01046cd9403783c80e6f91314e91f}{\texttt{0x5bdcacc0feee0fd4685fb9a0ba6820cc2ea01046cd9403783c80e6f91314e91f}}.}
    \item ``The transaction includes a coinbase transfer or a priority fee of at least 1 GWei." Theoretically, integrated searchers do not need to send explicit tips for inclusion in the block built by their integrated builders. Therefore, this heuristic is removed to capture trades by such integrated searchers accurately.
\end{enumerate}

\subsection{Manually removed contract addresses}
Due to limitations in existing datasets and identification frameworks \cite{orderflowart, zeromev, dextradingbot}, we manually removed several contract addresses associated with frontends, routers, solvers, and atomic MEV bots not captured by our heuristics. Below, we detail these addresses along with the rationale for their exclusion:

\begin{itemize}[topsep=2pt, itemsep=2pt, parsep=2pt]
    \item \href{https://etherscan.io/address/0x00000000003b3cc22af3ae1eac0440bcee416b40}{\texttt{0x00000000003b3cc22af3ae1eac0440bcee416b40}} and \href{https://etherscan.io/address/0x000000d40b595b94918a28b27d1e2c66f43a51d3}{\texttt{0x000000d40b595b94918a28b27d1e2\\c66f43a51d3}}: sandwich MEV bots not labeled by \cite{zeromev,dextradingbot}.
    \item \href{https://intel.arkm.com/explorer/address/0x5FF137D4b0FDCD49DcA30c7CF57E578a026d2789}{\texttt{0x5FF137D4b0FDCD49DcA30c7CF57E578a026d2789}}: an ERC-4337 Account Abstraction frontend/router integrating \texttt{1inch}'s API. This contract was omitted from \cite{orderflowart} and not captured by our Heuristic 5 due to the sender being a smart contract rather than regular user accounts.
    
    \item \href{https://intel.arkm.com/explorer/address/0xA8C1C98aAF99A5DfC907d61b892b2aD624901185}{\texttt{0xA8C1C98aAF99A5DfC907d61b892b2aD624901185}}: \texttt{Rizzsolver} on \texttt{1inch} run by \texttt{Wintermute} not captured by \cite{orderflowart}. We verified that nearly all transactions from this address are included in the DEX aggregator trades table on Dune \cite{dexaggre}, excluding a few direct token transfers. Therefore, no genuine CEX-DEX arbitrage transactions executed by \texttt{Wintermute} through this contract were removed.
    \item \href{https://etherscan.io/address/0xfcb51642a2a33eafefd79c236480e295ccbd4a44}{\texttt{0xfcb51642a2a33eafefd79c236480e295ccbd4a44}}: confirmed as a Uniswap X filler (solver) not listed in \cite{orderflowart}, verified through \cite{uniswapxfillers}.
\end{itemize}

\section{Integrated searcher-builder profitability calculation}
\label{appendix:integrated_profit_cal}
Builder on-chain profit from winning the block in the MEV-Boost auction consist of tips and coinbase transfers received from the transactions included in the block, net of the bid paid to the proposer and any OFA refund to the users \cite{whowinsandwhy}. For most blocks, these profits can be directly tracked via the changes in their \texttt{block.coinbase} ETH balance before and after winning the block \cite{relayscan}. However, this method becomes inaccurate when the builder uses an EOA to pay the proposer. This situation frequently happens for \texttt{beaverbuild}, \texttt{Titan}, and \texttt{rsync}, who utilize the Ultra Sound bid adjustment feature, introduced from block 18719819 on Dec 5, 2023 \cite{bidadjustment}. This mechanism allows Ultra Sound relay to adjusts the winning bid value to be the second-highest bid on any relay plus 1 WEI, issue the proposer payment using the builder's EOA, and refund a proportion of the delta (if any) back to the builder. Before 2024-03-05 05:00 UTC, Ultra Sound relay refunded 100\% of this delta, while after this timestamp, the refund rate decreased to 50\%, with the remainder kept by the relay.

Formally, consider block $k$ built by builder $j$. Let the builder's observed on-chain profit from winning block $k$ in the MEV-Boost auction be $BP_{j,k}$, and the \texttt{block.coinbase} balance difference of the builder address before and after winning block $k$ be $\Delta C_{j,k}$. We further let $\delta_k$ be the bid adjustment delta reported by Ultra Sound relay if the builder uses the bid adjustment feature. The builder $j$'s on-chain profit $BP_{j,k}$ can be given by:
\begin{equation}
BP_{j,k} = 
\begin{cases}
\Delta C_{j,k} & \text{without Ultra Sound bid adjustment}, \\
\Delta C_{j,k} - b_{j,k} + r \cdot \delta_k & \text{with Ultra Sound bid adjustment},
\end{cases}    
\end{equation}
where $b_{j,k}$ is the builder's original bid value, and $r$ is the fraction of adjustment delta refunded to the builder: $r = 1$ before 2024-03-05 05:00 UTC and $r = 0.5$ thereafter.\footnote{For calculation, we convert the price unit of the builder's on-chain profits from a block from ETH to USD using the ETH-USDT mid-price at the corresponding \texttt{slot time}.}

Next, consider the integrated searcher associated with builder $j$. Let  $\mathcal{I}_{j,k}$ be the set of CEX-DEX arbitrage transactions from that searcher included in block $k$. Using the PnL estimates per trade from \Cref{sec:pnl_definition}, the searcher's retained profit and the aggregated profit captured by integrated searcher-builder entity $j$ for block $k$ is:
\[
SP_{j,k} = \sum_{i \in \mathcal{I}_{j,k}} \widehat{\operatorname{PnL}}_{i,j}, \text{ and }P_{j,k} = BP_{j,k} + SP_{j,k}. 
\]
The aggregated profit margin is thus:

\[\text{PM}_{j,k} = \frac{P_{j,k}}{P_{j,k}+b_{j,k}-r \cdot \delta_k}. 
\]

\section{Searcher transaction count details}
\label{appendix:tx_count}
We here present details of the transaction count examined in our work. Out of all 8,723,233 transactions detected by our heuristics, 163,148 transactions are first removed because the CEX prices of the tokens involved are not recorded on Tardis.dev for certain dates.  683,539 transactions are treated as ``inventory adjustment trades" and removed from revenue estimation as their markout revenue fails to cover the base fees throughout the examined interval. 176,159 transactions from \texttt{Bard}, 36,444 transactions from \texttt{Jinx}, 16,212 transactions from \texttt{Tristana}, and 11,118 transitions from \texttt{Lux}, and 433,053 transactions from all other unlabeled searchers, including a total of 805 EOAs and bot contract addresses, are removed from revenue estimation. In \Cref{tab:trade_count}, we summarize the trade count for 19 labeled searchers, including their total trades, inventory adjustment trades, geneiue arbitrage trades, profitable trades ($\widehat{\operatorname{PnL}}_{i} \geq 0$), and unprofitable trades ($\widehat{\operatorname{PnL}}_{i} < 0$).

\begin{table}[H]
\centering
\caption{Summary of trade count for 19 labeled searchers.}
\label{tab:trade_count}
\begin{tabular}{lccccc}
\toprule
\textbf{Searcher} & \textbf{Total} & \textbf{Inventory Adj.} & \textbf{Arbitrage} & \textbf{Profitable} & \textbf{Unprofitable} \\
 & \textbf{Trades} & \textbf{Trades} & \textbf{Trades} & \textbf{Trades} & \textbf{Trades} \\
 & [\#] & [\#] & [\#] & [\#] & [\#]\\
\midrule
\midrule
Wintermute & 1,916,577 & 153,671 & 1,762,906  & 1,297,128 & 465,778 \\
SCP        & 2,188,986 & 142,086 & 2,046,900  & 1,579,632 & 467,268 \\
Kayle      & 1,525,397 &  90,970 & 1,434,427  & 1,131,090 & 303,337 \\
Galio      &   257,377 &  15,858 & 241,519  &   196,208 &  45,311 \\
Shen       &   707,491 &  12,658 & 694,833  &   649,458 &  45,375 \\
Taric      &   179,277 &   4,053 & 175,224  &   156,569 &  18,655 \\
Lucian     &   528,422 &   4,502 & 523,920  &   498,731 &  25,189 \\
Riven      &    23,699 &   2,536 & 21,163  &    14,586 &   6,577 \\
Thresh     &    31,622 &   5,904 & 25,718  &    20,061 &   5,657 \\
Ahri       &    12,161 &     879 & 11,282  &     7,611 &   3,671 \\
Darius     &    12,415 &     700 & 11,715  &     9,312 &   2,403 \\
Karma      &    58,085 &   1,730 & 56,355  &    55,883 &     472 \\
Senna      &    55,910 &   5,523 & 50,387  &    44,399 &   5,988 \\
Maokai     &    27,976 &     649 & 27,327  &    26,345 &     982 \\
Zed        &    96,954 &   2,649 & 94,305  &    86,205 &   8,100 \\
Graves     &    12,173 &   5,750 & 6,423  &       925 &   5,498 \\
Caitlyn    &     1,559 &     247 & 1,312  &     1,184 &    128 \\
Akali      &     9,803 &     55  & 9,748  &     9,595 &     153 \\
Poppy      &     8,131 &     35  & 8,096  &     7,865 &     231 \\
\bottomrule
\end{tabular}
\end{table}

\end{document}